\documentclass[aps,preprint,showpacs,showkeys,nofootinbib,superscriptaddress,floatfix]{revtex4}%
\usepackage{amsfonts}%
\usepackage{amsmath}%
\usepackage{amssymb}%
\usepackage{graphicx}
\usepackage{color}
\providecommand{\U}[1]{\protect\rule{.1in}{.1in}}

\begin{document}
\preprint{ }
\title[Isotope Shift Measurements of Lithium Isotopes]
{Isotope Shift Measurements of Stable and Short-Lived Lithium Isotopes for
Nuclear Charge Radii Determination }
\author{W.~N\"{o}rtersh\"{a}user}
\affiliation {GSI Helmholtzzentrum f\"ur Schwerionenforschung GmbH, 64291 Darmstadt, Germany}
\affiliation{Institut f\"ur Kernchemie, Universit\"at Mainz, 55099 Mainz, Germany}

\author{R.~S\'anchez}
\affiliation {GSI Helmholtzzentrum f\"ur Schwerionenforschung GmbH, 64291 Darmstadt, Germany}
\affiliation{Institut f\"ur Kernchemie, Universit\"at Mainz, 55099 Mainz, Germany}

\author{G.~Ewald}
\affiliation {GSI Helmholtzzentrum f\"ur Schwerionenforschung GmbH, 64291 Darmstadt, Germany}

\author{A. Dax}
\altaffiliation[Current address ]{CERN, CH-1211 Geneva 23, Switzerland.}
\affiliation {GSI Helmholtzzentrum f\"ur Schwerionenforschung GmbH, 64291 Darmstadt, Germany}

\author{J.~Behr}
\affiliation {TRIUMF, Vancouver, BC, Canada V6T 2A3}

\author{P.~Bricault}
\affiliation {TRIUMF, Vancouver, BC, Canada V6T 2A3}

\author{B.~A.~Bushaw}
\affiliation {Pacific Northwest National Laboratory, Richland, WA 99352, USA}

\author{J.~Dilling}
\affiliation {TRIUMF, Vancouver, BC, Canada V6T 2A3}

\author{M.~Dombsky}
\affiliation {TRIUMF, Vancouver, BC, Canada V6T 2A3}

\author{G.~W.~F.~Drake}
\affiliation {Department of Physics, University of Windsor, Windsor, Ontario, Canada, N9B 3P4}

\author{S.~G\"otte}
\affiliation {GSI Helmholtzzentrum f\"ur Schwerionenforschung GmbH, 64291 Darmstadt, Germany}

\author{H.-J.~Kluge}
\affiliation {GSI Helmholtzzentrum f\"ur Schwerionenforschung GmbH, 64291 Darmstadt, Germany}

\author{Th.~K\"uhl}
\affiliation {GSI Helmholtzzentrum f\"ur Schwerionenforschung GmbH, 64291 Darmstadt, Germany}

\author{J.~Lassen}
\affiliation {TRIUMF, Vancouver, BC, Canada V6T 2A3}

\author{C.~D.~P.~Levy}
\affiliation {TRIUMF, Vancouver, BC, Canada V6T 2A3}

\author{K.~Pachucki}
\affiliation {Faculty of Physics, University of Warsaw, 00-681 Warsaw, Poland}

\author{M.~Pearson}
\affiliation {TRIUMF, Vancouver, BC, Canada V6T 2A3}

\author{M.~Puchalski}
\affiliation{Faculty of Chemistry, Adam Mickiewicz University, Grunwaldzka 6, 60-780 Pozna\'n, Poland}

\author{A.~Wojtaszek}
\altaffiliation[Current address ]{Instytut Fizyki, Jan Kochanowski University, PL-25-406, Kielce, Poland.}
\affiliation {GSI Helmholtzzentrum f\"ur Schwerionenforschung GmbH, 64291 Darmstadt, Germany}

\author{Z.-C.~Yan}
\affiliation {Department of Physics, University of New Brunswick, Fredericton, New Brunswick, Canada E3B 5A3}

\author{C.~Zimmermann}
\affiliation {Physikalisches Institut, Universit\"at T\"ubingen, 72076 T\"ubingen, Germany}

\keywords{lithium, laser spectroscopy, nuclear charge radius, halo nucleus}
\pacs{21.10.Gv, 21.10.Ft, 27.20.+h, 32.10.Fn}

\begin{abstract}
Changes in the mean square nuclear charge radii along the lithium isotopic chain were determined using a
combination of precise isotope shift measurements and theoretical
atomic structure calculations. Nuclear charge radii of light
elements are of high interest due to the appearance of the nuclear
halo phenomenon in this region of the nuclear chart. During the past
years we have developed a new laser spectroscopic approach to
determine the charge radii of lithium isotopes which combines
high sensitivity, speed, and accuracy to measure the extremely small
field shift of an 8 ms lifetime isotope with production rates on the order of
only 10\,000 atoms/s. The method was applied to all bound isotopes
of lithium including the two-neutron halo isotope $^{11}$Li at the on-line isotope separators at GSI, Darmstadt, Germany and at TRIUMF, Vancouver, Canada. We
describe the laser spectroscopic method in detail, present updated and
improved values from theory and experiment, and discuss the results.

\end{abstract}
\volumeyear{2010}
\volumenumber{number}
\issuenumber{number}
\eid{identifier}
\date[Date text]{date}
\received[Received text]{date}

\revised[Revised text]{date}

\accepted[Accepted text]{date}

\published[Published text]{date}

\startpage{1}
\maketitle

\section{Introduction}

Laser spectroscopy of lithium isotopes has recently attracted much
interest. This is for two reasons: First, as a three-electron system
it can be used to test the fundamental theoretical description of
few-electron systems at high accuracy. Second, the extraction of
nuclear charge radii from isotope shifts for very light elements
became possible by combining high-accuracy measurements with atomic
theory. These charge radii are of special interest since one of the
lithium isotopes, $^{11}$Li, is the best investigated halo nucleus,
but a nuclear-model-independent value of its charge radius was not
known until recently. Our ToPLiS\footnote{{\bf T}w{\bf o}-{\bf
P}hoton {\bf Li}thium {\bf S}pectroscopy} Collaboration succeeded in
measuring the charge radii of all lithium isotopes
\cite{Ewald04,Ewald05,Sanchez06}. Here we will describe the
experimental setup in detail and discuss the results, including
updated and improved values from theory and experiment.

Laser spectroscopy has considerably contributed to our knowledge of
ground-state properties of short-lived isotopes. Information on
nuclear spins, charge radii, magnetic dipole moments and
spectroscopic electric quadrupole moments can be extracted from
isotope shift and hyperfine structure measurements in atomic
transitions. A particular strength of these methods is that they
provide nuclear-model-independent data. This field has been
regularly reviewed, see, {\it e.g.},
\cite{Otten89,Billowes95,Neugart02,Kluge03} within the last decades.
While nuclear moments can be obtained with laser spectroscopy from
the lightest to the heaviest elements, it was so far not possible to
determine nuclear charge radii for short-lived isotopes lighter than
neon \cite{Geithner00,Geithner05}.

The reason is that the isotope shift, in which the charge radius
information is encoded, has two contributions: One is the change in
nuclear mass between two isotopes (mass shift, MS) and the second
one is the difference in the charge distribution inside the nucleus
(volume shift or field shift, FS). The MS by far dominates the isotope
shift in light elements and then decreases rapidly approximately
with increasing mass number $A$ by $A^{-2}$. By contrast, the FS is more than 10\,000 times
smaller than the MS in the case of the lithium isotopes but
increases in proportion to $Z^{2}A^{-1/3}$ with the nuclear charge number
$Z$ and by far exceeds the MS in heavy elements. Separating the tiny
FS in the light elements is a complicated task and could only be
performed on very simple and stable atoms or ions containing not more than two electrons. The general approach is a
very accurate calculation of parts contributing to the mass shift
and the comparison with the experimentally observed isotope shift,
as first demonstrated in 1993 for the case of helium by Drake
\cite{Drake93}. The difference between these values is then
attributed to the change in the nuclear charge radius. Consequently,
relative accuracy better than 10$^{-5}$ must be reached in
experiment as well as in atomic structure calculations. Here, the correlations in the atom between more than two electrons were an unresolved task. Therefore until a few years ago, this was only possible for one- and two-electron
systems, and this approach was used for the stable isotopes of
hydrogen \cite{SchmidtKaler95,Huber98}, helium
\cite{Drake93,Shiner95}, and Li$^+$ \cite{Riis1994}.

The extension of such measurements to short-lived isotopes was a
challenge. The excitation and detection scheme used for measuring
isotope shifts in the lithium isotope chain must provide both high
efficiency, because $^{11}$Li is produced with production rates of only a
few thousand atoms/s, and high resolution and accuracy for
extraction of the tiny field shift contribution. Furthermore, it
must be a rapid technique since $^{11}$Li is the shortest-lived
isotope ($T_{1/2}=8$~ms) that has ever been addressed with cw
lasers as required for the necessary accuracy\footnote{Only some radioisotopes of americium, investigated by the group of H. Backe \cite{Backe2000}, have even shorter half lives but here broad-band pulsed lasers could be applied.} . To probe the nuclear charge radius, a transition out of the
atomic ground state of lithium seems to be the natural choice.
Resonance spectroscopy in the triplet system of the Li$^+$ system
\cite{Riis1994} could in principle be performed as well, but the
preparation of such a metastable state is usually quite inefficient
and therefore not appropriate. Spectroscopy on trapped ions or atoms
has proved to provide the required sensitivity and accuracy for such
measurements \cite{Wang04,Nakamura2006}; however, the $^{11}$Li
lifetime is much too short to allow sufficient time for trapping and
cooling. The  $2s\;^2{\rm S}_{1/2} \rightarrow 2p\;^2{\rm P}_J$
transitions in Li were used previously for several investigations of
$^{8,9,11}$Li in collinear laser spectroscopy by optical pumping and
$\beta$-NMR detection
\cite{Arnold87,Arnold92,Arnold94,Borremans05,Neugart08}. In these
cases the lithium ions, provided by the ISOLDE radioactive beam
facility at CERN, were neutralized in flight in a charge exchange
cell and then used as a fast atomic beam. However, the Doppler shift
of the transition frequency due to the beam velocity of the accelerated ions,
and the limited knowledge of the acceleration voltage with an uncertainty of typically
$10^{-4}$ is prohibitive for isotope shift measurements. Recently,
such measurements became practicable for Be$^+$ ions by using a
frequency comb and simultaneous spectroscopy in parallel and
antiparallel direction \cite{Noertershaeuser09,Zakova10} in order to
eliminate voltage uncertainties. Those studies are still limited by
statistical and systematic uncertainties to an accuracy of about
1~MHz, which is not sufficient for lithium isotopes. Hence, in the
experiment discussed here, the Li$^+$ ions are stopped and
neutralized before spectroscopy is performed on
the atomic cloud. To avoid large Doppler broadening, the $2s\;^2{\rm
S}_{1/2} \rightarrow 3s\;^2{\rm S}_{1/2}$ two-photon transition is
studied, which is to first order Doppler-free.

After a series of short letters \cite{Bushaw2003,Ewald04,Ewald05,Sanchez06}
in which we reported results on the charge radii of all isotopes, we
describe here in detail the experimental technique as well as the
theoretical calculations and discuss all possible sources of
systematic uncertainties that were further studied. The paper is
organized as follows: In Section~\ref{sec:Theory} we give a summary of the theoretical
mass shift calculations that were performed and steadily improved
during the last years, including relativistic, QED and nuclear
structure corrections. Afterwards, the experimental setup is
described, including radioactive beam production, separation and the
transformation into a thermal atomic beam. Resonance ionization mass
spectroscopy is combined with two-photon spectroscopy to reach the
required accuracy and efficiency simultaneously. The corresponding
results and the change in nuclear charge radius for the pair of stable isotopes $^{6,7}$Li are presented in Section~\ref{sec:Results}. Results for the stable isotopes are compared with other experimental data obtained by optical spectroscopy of the lithium ion or atom to verify the consistency of the data.
The paper closes with the presentation of the extracted changes in the mean square (ms) nuclear charge radii
$\delta \left\langle r^{2}\right\rangle^{6,A}$ relative to the
reference isotope $^6$Li. These
values can be extracted
independently from further information in a model-independent way \cite{Otten89}.

Most interesting for nuclear-structure studies are absolute nuclear charge or nuclear matter radii. To determine charge radii
from our isotope shift measurements, we need the nuclear charge
radius of at least one reference isotope obtained by a different technique. A
nuclear charge radius determination solely based on optical
measurements of the absolute transition frequency and atomic theory
calculations is thus far only possible for hydrogen \cite{Udem97}.
Experimental data are now in principle also available for the lithium
isotopes \cite{Sanchez2009}; however, atomic theory for three-electron atoms is not yet
able to achieve absolute transition frequencies sufficiently
accurate to extract the $10^{-9}$ contribution of the
finite nuclear size effect. The choice of the reference isotope and
the respective charge radii together with an evaluation of
theoretical results will be discussed in a following publication
\cite{Noertershaeuser2010}.

\section{Theory}
\label{sec:Theory}
To extract nuclear charge radii from isotope shift measurements of the lightest
elements, calculations of the mass-dependent part of the isotope shift at the
highest level that is currently achievable are required. In this Section we
will present the latest results for the lithium isotopes. The calculation
starts with constructing the solution of the nonrelativistic Schr\"odinger
equation using variational calculations in Hylleraas coordinates. These
solutions are then the basis for the calculation of relativistic and quantum
electrodynamic contributions. Finally, nuclear structure corrections are
included. The finite nuclear size effect is actually the part that will be
extracted from experiment. But the proportionality coefficient between the
change in the mean square nuclear charge radius and the extracted nuclear
volume effect in the isotope shift has to be provided by theory. Additionally,
as with many halo nuclei, the $^{11}$Li nucleus possesses strong electric dipole transitions
to low-lying states, making it a so-called ``soft-dipole''  with a relatively large
nuclear polarizability compared with less exotic nuclei. The influence of this
polarizability on the atomic electron levels was recently evaluated \cite{Puchalski2006} and leads
to a significant contribution. In the following Sections, the details of the
atomic structure calculations will be explained.

\subsection{General Approach}
In order to extract the rms nuclear charge radius from the measured isotope
shift, we begin by writing the isotope shift for an atomic transition
$a\rightarrow b$ between isotopes $A$ and $A^\prime$ in the form
\begin{eqnarray}
\delta\nu_{a\rightarrow b}^{A,A^\prime} & = & \nu^{A^\prime}_{a\rightarrow b} - \nu^A_{a\rightarrow b} 
\label{eq:IS-definition}
\\
& = & \delta\nu_{a\rightarrow b}^{(0)}(A,A^\prime) +
C_{a\rightarrow b}[\bar{r}_c^2(A^\prime) -\bar{r}_c^2(A)] \nonumber \\
\delta\nu_{\rm IS}^{A,A^\prime} & = & \delta\nu_{\rm MS}^{A,A^\prime} + C_{A,A^\prime}
\cdot \delta \left\langle r^{2}\right\rangle^{A,A^\prime},
\label{eq:IsotopeShift}
\end{eqnarray}
where $\bar{r}_c(A)$ and $\bar{r}_c(A^\prime)$ are the rms charge radii for the
two isotopes, and the remaining term $\delta\nu_{a\rightarrow b}
^{(0)}(A,A^\prime)$ comes from the mass dependence of the atomic energy levels.
Hence, it is called the mass shift (MS). As will be seen, the coefficient
$C_{a\rightarrow b}$ is nearly independent of the isotopes involved, but in the
case of lithium, relativistic corrections to the wave function at the origin
and leading recoil corrections should be included \cite{Puchalski2010}. The main
challenge is to calculate the term $\delta\nu_{a\rightarrow
b}^{(0)}(A,A^\prime)$ to  sufficient accuracy. Equation (\ref{eq:IS-definition}) 
can be written in the form (\ref{eq:IsotopeShift}) that usually appears in experimental papers,
where the particular transition index is suppressed. Here, the mass-shift  and the
$\delta \left\langle r^{2}\right\rangle^{A,A^\prime}$ dependence of the field
shift are clearly stated. In the following, we will always refer to the $2s\;
^2{\rm S}_{1/2} \rightarrow 3s\; ^2{\rm S}_{1/2} $ two-photon transition. 

To give an overview of the contributions  to $\delta\nu_{\rm MS}^{A,A^\prime}$,
it is convenient to arrange them in the form of a double series expansion
in powers of $\alpha \simeq 1/137.036$ and the electron reduced-mass-to-nuclear-mass 
ratio $\mu/M \simeq 10^{-4}$, where $\mu= mM/(m+M)$. Table \ref{table1}
summarizes the various contributions, including the QED corrections and the
finite nuclear size term.  Since all the lower-order terms can now be
calculated to very high accuracy, including the QED terms of order
$\alpha^3$\,Ry, the dominant source of uncertainty comes from the QED
corrections of order $\alpha^4$\,Ry or higher.  For the isotope shift, the QED
terms independent of $\mu/M$ cancel, and so it is only the radiative recoil
terms of order $\alpha^4\mu/M \simeq 10^{-12}$Ry ($\sim$10 kHz) that
contribute to the uncertainty.  Since this is much less than the finite nuclear
size correction of about 1 MHz, the comparison between theory and experiment
clearly provides a means to determine the nuclear size. This is the key point
to keep in mind when considering the theoretical contributions to the isotope
shift.
\begin{table}
\begin{center}
\caption{Contributions to the energy in units of the Rydberg
constant and their leading orders of magnitude in
terms of $Z$, $\mu/M \sim 10^{-4}$, and $\alpha \sim 10^{-2}$. $\alpha_{\rm d,nuc}$ is the nuclear dipole polarizability and $a_0$ the Bohr atomic radius.}
\label{table1}
\begin{tabular}{ll}
\hline
Contribution & Magnitude\\
\hline
Nonrelativistic energy & $Z^2$\\
Mass polarization      & $Z^2\mu/M$\\
Second-order mass polarization & $Z^2(\mu/M)^2$\\
Relativistic corrections& $Z^4\alpha^2$\\
Relativistic recoil    & $Z^4\alpha^2\mu/M$\\
Anomalous magnetic moment & $Z^4\alpha^3$\\
Hyperfine structure       & $Z^3g_I\mu_0^2$\\
Lamb shift             & $Z^4\alpha^3\ln\alpha + \cdots$\\
Radiative recoil       & $Z^4\alpha^3(\ln\alpha)\mu/M$\\
Finite nuclear size    & $Z^4\langle \bar{r}_c/a_0\rangle^2$\\
Finite size recoil     & $Z^4\,\mu/M \, \langle r_c/a_0 \rangle^2 $ \\
Nuclear polarization   & $Z^3e^2 \alpha_{\rm d,nuc}/(\alpha a_0)$ \\
\hline
\end{tabular}
\end{center}
\end{table}

\subsection{Solution to the Nonrelativistic Schr\"odinger Equation}
The foundation for the calculation, and the subsequent evaluation of
relativistic and QED corrections, is to find high-precision solutions
to the nonrelativistic Schr\"odinger equation for finite nuclear mass.
The past 20 years have seen important advances in developing
specialized techniques for doing this in the case of the three-body
problem (helium-like systems) \cite{Drake92,Drake93,%
Drake93Adv}, and more recently the four-body problem
\cite{Yan_Drake98,Yan_Drake02,Yan_Drake03,Puchalski2006,Puchalski2006b}. The
usual methods of theoretical atomic physics, such as the Hartree-Fock
approximation or configuration interaction methods, are not capable of yielding
results of spectroscopic accuracy, and so specialized techniques of the type
used here are needed.

For convenience, we begin by rescaling distances according to
$r\rightarrow (m/\mu)r$.  The advantage gained is that the Hamiltonian for a three-electron atomic system can then be written in
the form
\begin{eqnarray}
H &=& H_0+\lambda H_{\rm MP}\,,
\label{eq:aa1}
\end{eqnarray}
with
\begin{eqnarray}
H_0 &=& -\frac{1}{2}\sum_{i=1}^3\nabla_i^2-Z\sum_{i=1}^3{\frac{1}{r_i}} +
\sum_{i>j}^3 {\frac{1}{r_{ij}}}\,,
\label{eq:aa2}
\end{eqnarray}
and
\begin{eqnarray}
H_{\rm MP} &=& \sum_{i>j}^3\nabla_i\cdot\nabla_j\,,
\label{eq:aa3}
\end{eqnarray}
in units of $2R_M$, where the Rydberg constant for finite nuclear mass is defined by $R_M=
(1-\mu/M)R_{\infty}$, and $\lambda=-\mu/M$ can be treated as a
perturbation parameter. The Schr\"{o}dinger equation
\begin{eqnarray}
H\Psi &=& E\Psi
\label{eq:aa4}
\end{eqnarray}
was solved perturbatively by expanding $\Psi$ and $E$ according to
\begin{eqnarray}
\Psi &=& \Psi_0 +\lambda \Psi_1+\cdots\,,\\
E &=& \varepsilon_0 +\lambda \varepsilon_1 +\lambda^2 \varepsilon_2+
\cdots\,.
\label{eq:aa5}
\end{eqnarray}
Thus Eq.\ (\ref{eq:aa4}) becomes
\begin{eqnarray}
H_0 \Psi_0 &=& \varepsilon_0\Psi_0\,,\\
(\varepsilon_0-H_0) \Psi_1 &=& (H_{\rm MP}-\varepsilon_1)\Psi_0\,.
\label{eq:aa6}
\end{eqnarray}
$\varepsilon_1$ and $\varepsilon_2$ are
\begin{eqnarray}
\varepsilon_1 &=& \langle\Psi_0|H_{\rm MP}|\Psi_0\rangle\,,\\
\varepsilon_2 &=& \langle\Psi_0|H_{\rm MP}|\Psi_1\rangle -\varepsilon_1
\langle\Psi_0|\Psi_1\rangle\,.
\label{eq:aa7}
\end{eqnarray}
Both $\Psi_0$ and $\Psi_1$ were solved variationally in multiple basis
sets in Hylleraas coordinates containing terms of the form
\begin{eqnarray}
r_1^{j_1}\,r_2^{j_2}\,r_3^{j_3}\,r_{12}^{j_{12}}\,r_{23}^{j_{23}}
\,r_{31}^{j_{31}}\,e^{-\alpha_p r_1-\beta_p r_2-\gamma_p r_3}\, \times \nonumber \\
{\cal Y}
_{(\ell_1\ell_2)\ell_{12},\ell_3}^{LM}({\rm\bf r}_1,{\rm\bf r}_2, {\rm
\bf r}_3)\,\chi_1\,,
\label{eq:aa8}
\end{eqnarray}
where ${\cal Y}_{(\ell_1\ell_2)\ell_{12},\ell_3}^{LM}$ is a
vector-coupled product of spherical harmonics for the three electrons
to form a state of total angular momentum $L$, and $\chi_1$ is a spin
function with spin angular momentum $1/2$.

As described previously \cite{Yan1995,Yan1998}, the basis set is divided into
five sectors with different values of the scale factors $\alpha_p$,
$\beta_p$, and $\gamma_p$ in each sector, as labelled by the subscript
$p$.  The 15 independent scale factors are then optimized by a global
minimization of the energy.  Except for restrictions on which
correlation terms are included in each sector (see Ref.\ \cite{Yan1998}),
all terms from (\ref{eq:aa8}) are included such that
\begin{eqnarray}
j_1+j_2+j_3+j_{12}+j_{23}+j_{31} &\leq& \Omega\,,
\label{eq:aa9}
\end{eqnarray}
and the convergence of the eigenvalues is studied as $\Omega$ is
progressively increased. Further details may be found in Ref.\
\cite{Yan1998}. Since Eq.\ (\ref{eq:aa5}) is expressed in units of $(1+
\lambda)\,\, 2R_\infty$, the explicit mass-dependence of $E$ in units of $2R_\infty$ is
\begin{eqnarray}
E &=& \varepsilon_0 +\lambda (\varepsilon_0+\varepsilon_1) +\lambda^2 (
\varepsilon_1+\varepsilon_2)+O(\lambda^3).
\end{eqnarray}

\subsection{Relativistic and Relativistic Recoil Corrections}
The lowest-order relativistic corrections of $O(\alpha^2)$ and the
spin-dependent anomalous magnetic moment corrections of $O(\alpha^3)$
can be written in the form \cite{Stone1963,Drake92} (in atomic units)
\begin{eqnarray}
E_{\rm rel} &=&\langle\Psi |H_{\rm rel}| \Psi\rangle_J\,,
\label{eq:ax0}
\end{eqnarray}
where $\Psi$ is a nonrelativistic wave function and $H_{\rm rel}$ is
defined by
\begin{eqnarray}
\label{eq:ax1}
H_{\rm rel} &=& B_1+B_2+B_{3e}+B_{3z}+B_5 \\
&& -\pi\alpha^2\sum_{i>j}^3 \bigg(1+\frac{8}{3}{\bf s}_i \cdot{\bf s}_j\bigg) \delta({\bf r}_{ij})+ \nonumber \\
&& \frac{1}{2}Z\pi\alpha^2\sum_{i=1}^3 \delta({\bf r}_{i})
\mbox{}+\frac{m}{M}({\tilde \Delta}_{2}+{\tilde \Delta}_{3z}) \nonumber \\
&&+\gamma \,\bigg(2B_{3z}+\frac{4}{3}B_{3e} +\frac{2}{3}B_{3e}^{(1)}+2B_5\bigg) \nonumber \\
&& + \gamma\,\frac{m}{M}{\tilde \Delta}_{3z}\, \nonumber .
\end{eqnarray}
In (\ref{eq:ax1}),
\begin{eqnarray}
B_{1} = -\frac{\alpha^2}{8}(\nabla_1^4+\nabla_2^4+\nabla_3^4)\,,
\label{eq:ax2}
\end{eqnarray}
\begin{eqnarray}
B_{2} = \frac{\alpha^2}{2}\sum_{i>j}^3\bigg[\frac{1}{r_{ij}} \nabla_i
\cdot\nabla_j+\frac{1}{r^3_{ij}}{\bf r}_{ij}\cdot ({\bf r}_{ij}
\cdot\nabla_i)\nabla_j\bigg],
\label{eq:ax3}
\end{eqnarray}
\begin{eqnarray}
B_{3e} = \frac{\alpha^2}{2}\sum_{i\ne j}^3 \frac{1}{r_{ij}^3}{\rm\bf
r}_{ji}\times{\rm\bf p}_i\cdot ({\rm\bf s}_i+2{\rm\bf s}_j)\,,
\label{eq:ax4}
\end{eqnarray}
\begin{eqnarray}
B_{3z} = \frac{Z\alpha^2}{2}\sum_{i=1}^3 \frac{1}{r_i^3}{\rm\bf r}_{i}
\times{\rm\bf p}_i\cdot {\rm\bf s}_i\,,
\label{eq:ax5}
\end{eqnarray}
\begin{eqnarray}
B_{5} = \alpha^2\sum_{i>j}^3 \bigg[\frac{1}{r_{ij}^3}({\rm\bf s}_{i}
\cdot {\rm\bf s}_j) -\frac{3}{r_{ij}^5}({\rm\bf r}_{ij}\cdot {\rm\bf s}
_i) ({\rm\bf r}_{ij}\cdot {\rm\bf s}_j)\bigg],
\label{eq:ax6}
\end{eqnarray}
\begin{eqnarray}
\label{Delta_2}
{\tilde \Delta}_{2}= \frac{iZ\alpha^2}{2}\sum_{j=1}^3 \bigg[\frac{1}
{r_{j}} {\bf p}\cdot\nabla_j+\frac{1}{r^3_{j}}{\bf r}_{j}\cdot ({\bf r}
_j\cdot{\bf p})\nabla_j\bigg],
\label{eq:ax7}
\end{eqnarray}
\begin{eqnarray}
\label{Delta3Z}
{\tilde \Delta}_{3Z} = Z\alpha^2\sum_{i=1}^3 \frac{1}{r_i^3}{\rm\bf r}
_{i}\times{\rm\bf p}\cdot {\rm\bf s}_i\,,
\label{eq:ax8}
\end{eqnarray}
\begin{eqnarray}
B_{3e}^{(1)} = \frac{\alpha^2}{2}\sum_{i\ne j}^3 \frac{1}{r_{ij}^3}{
\rm\bf r}_{ji}\times{\rm\bf p}_i\cdot ({\rm\bf s}_i-{\rm\bf s}_j)\,,
\label{eq:ax9}
\end{eqnarray}
with ${\rm\bf p}={\rm\bf p}_1+{\rm\bf p}_2+{\rm\bf p}_3$, and $\gamma$
is
\begin{eqnarray}
\gamma = \frac{\alpha}{2\pi}+(-0.328\,47)\bigg(\frac{\alpha}{\pi}
\bigg)^2 +\cdots\,.
\label{eq:ax10}
\end{eqnarray}
The operator $-\pi\alpha^2\sum_{i>j}(1+\frac{8}{3}
{\bf s}_i \cdot{\bf s}_j) \delta({\bf r}_{ij})$ can be replaced by
$\pi\alpha^2\sum_{i>j}\delta({\bf r}_{ij})$ and the expectation value
of the spin-spin term $B_5$ vanishes. The terms proportional to $m/M$
are the nuclear relativistic recoil corrections and the terms
proportional to $\gamma$ are the anomalous magnetic moment corrections.

The perturbing effect of mass polarization on the expectation values of
Breit operators can be obtained using
\begin{eqnarray}
\Psi &=& \Psi_0 + \lambda\, (\Psi_1-\langle\Psi_1|\Psi_0\rangle \Psi_0)+
\cdots\,,
\label{eq:aa15}
\end{eqnarray}
where the extra term $-\langle\Psi_1|\Psi_0\rangle\Psi_0$ is added to
$\Psi_1$ so that the first two terms of the right hand side are
orthogonal to each other \cite{Schiff1968}. Thus, for a Breit operator $A$,
one has
\begin{eqnarray}
\langle\Psi|A|\Psi\rangle &=&a_0+\lambda a_1+\cdots\,,
\label{eq:aa16}
\end{eqnarray}
where
\begin{eqnarray}
a_0 &=& \langle\Psi_0|A|\Psi_0\rangle\,,
\label{eq:aa17}
\end{eqnarray}
and
\begin{eqnarray}
a_1 &=& 2\langle\Psi_0|A|\Psi_1\rangle -2\langle\Psi_0|\Psi_1\rangle
\langle\Psi_0|A|\Psi_0\rangle\,.
\label{eq:aa18}
\end{eqnarray}
Furthermore, due to the use of $\mu$-scaled atomic units in Eq.\
(\ref{eq:aa1}), the units of $\langle\Psi|A|\Psi\rangle$ in Eq.\
(\ref{eq:aa16}) are $(\mu/m)^n\, 2R_\infty$, where $-n$ is the degree
of homogeneity of operator $A$ in the three-electron coordinate space such
that
\begin{eqnarray}
A(\beta{\bf r}_1,\beta{\bf r}_2,\beta{\bf r}_3) &=& \beta^{-n}A({\bf r}
_1,{\bf r}_2,{\bf r}_3)\,.
\label{eq:aa19}
\end{eqnarray}
Using
\begin{eqnarray}
\bigg(\frac{\mu}{m}\bigg)^n &=&(1+\lambda)^n \approx 1+n\lambda\,,
\label{eq:aa20}
\end{eqnarray}
one has the explicit mass-dependent formula
\begin{eqnarray}
\langle\Psi|A|\Psi\rangle &=& a_0 +\lambda\,(na_0+a_1)+O(\lambda^2)\,,
\label{eq:aa21}
\end{eqnarray}
in units of $2R_\infty$.

\subsection{QED Corrections}
Until recently, the QED contributions of lowest order $\alpha^3$\,Ry
were the major source of uncertainty in calculations of atomic energy
levels and the isotope shift for atoms more complicated than hydrogen.
However, tremendous progress has been made in recent years.  Complete
results to lowest order $\alpha^3$\,Ry are now readily obtainable, and
higher order corrections can be estimated in a screened hydrogenic
approximation. For a many-electron atom, it is convenient to express
the total QED shift in the form
\begin{equation}
E_{\rm QED} = E_{\rm L,1} + E_{\rm M,1} + E_{\rm R,1} + E_{\rm L,2}~,
\end{equation}
where $E_{\rm L,1}$ is the mass-independent part of the
electron-nucleus Lamb shift (the Kabir-Salpeter term \cite{Kabir1957}),
$E_{\rm M,1}$ contains mass scaling and mass polarization corrections,
$E_{\rm R,1}$ contains recoil corrections (including radiative recoil),
and $E_{\rm L,2}$ is the electron-electron term originally obtained by
Araki \cite{Araki1957} and Sucher \cite{Sucher1958}. We now discuss each of
these in turn.

The term $E_{\rm L,1}$ closely resembles the corresponding hydrogenic
Lamb shift \cite{Eides2001}, except that an overall multiplying
factor of $\langle\delta({\bf r})\rangle = Z^3\delta_{L,0}/(\pi n^3)$
for the hydrogenic case is replaced by the correct expectation value
$\langle\sum_{j=1}^N\delta({\bf r}_j)\rangle$ for the multi-electron
case, summed over the $N$ electrons. The residual state dependence due
to other terms such as the Bethe logarithm discussed below is then
relatively weak.

Following the notation of Eq.\ (\ref{eq:aa1}) for the mass polarization
corrections, the main electron-nucleus part for infinite nuclear mass
is (in atomic units throughout)
\begin{eqnarray}
\label{E_L}
E_{\rm L,1} &=& \frac{4Z\alpha^3\langle\delta({\bf r}_i)\rangle^{(0)}}
{3}\left\{\ln(Z\alpha)^{-2} - \beta(n\,^2\!L) + \frac{19}{30}\right.
\nonumber\\
&&\left.\mbox{} + (3\pi \alpha Z)0.765\,405\,577\right.\nonumber\\
&&\mbox{}+ \frac{\alpha}{\pi}\left[0.404\,17 - (3\alpha Z/4)21.556\,85
\right] \nonumber\\
&&\mbox{}+ (Z\alpha)^2\left[- \frac34\ln^2(Z\alpha)^{-2} + C_{61} (1s^x
\,nL)\ln(Z\alpha)^{-2} \right.\nonumber\\
&&\left.\left.\mbox{} + C_{60}(1s^x\,nL)\right]\right\}\,,
\end{eqnarray}
the mass scaling and mass polarization corrections are
\begin{eqnarray}
\label{E_M}
E_{\rm M,1} &=& \frac{\mu\langle\delta({\bf r}_i)\rangle^{(1)}}{M
\langle \delta({\bf r}_i)\rangle^{(0)}}E_{\rm L,1}\nonumber\\
&&\mbox{} + \frac{4Z\alpha^3\mu\langle \delta({\bf r}_i)\rangle^{(0)}}
{3M}\left[1 - \Delta\beta_{\rm MP}(n\,^2 \!L)\right],
\end{eqnarray}
and the recoil corrections (including radiative recoil) are given by
\begin{eqnarray}
E_{\rm R,1} &=& \frac{4Z^2\alpha^3\mu\langle\delta({\bf r}_i)
\rangle^{(0)}} {3M}\left[ \frac{1}{4}\ln(Z\alpha)^{-2} - 2\beta(n\,^2
\!L) - \frac{1}{12}\right.\nonumber\\
&&\mbox{}\left. - \frac{7}{4}a(1s^x\,nL) - \frac{3}{4}(\pi\alpha)
1.36449 \right.\nonumber\\
&&\mbox{}\left.+ \frac{3}{4} \pi Z \alpha D_{50}(n\,^2\!L) + \frac{1}{2}
\alpha^2Z\ln^2(Z\alpha)^{-2}\right] \,.
\label{E_R}
\end{eqnarray}
These equations involve contributions to the hydrogenic Lamb shift obtained by
many authors, as summarized by Eides {\it et al.}\ \cite{Eides2001}. The
quantity $\beta(1s^x\,nL) = \ln(k_0/Z^2R_\infty)$ is the two- or three-electron
Bethe logarithm, and the two terms $1 - \Delta\beta_{\rm MP}(1s^x\,nL)$ in Eq.\
(\ref{E_M}) account for the mass scaling and mass polarization corrections to
$\beta(1s^x\,nL)$ respectively.  These terms are further discussed below.  The
term $a(1s^x\,nL)$ is a well-known part of the hydrogenic Lamb shift.  Its
many-electron generalization is given by
\cite{Pachucki_98,Yan_Drake02,Pachucki_Sapirstein_00}
\begin{equation}
a(1s^x\,nL) = \frac{2Q_1^{(0)}}{\textstyle\sum_{j=1}^N\langle\delta({
\bf r}_j)\rangle} - 3,
\end{equation}
where
\begin{equation}
\label{Q1}
Q_1^{(0)} = \frac{1}{4\pi}\lim_{\epsilon\rightarrow 0}\sum_{j} \langle
r_{j}^{-3}(\epsilon)+ 4 \pi (\gamma +\ln \epsilon) \delta({\rm\bf r}_{j}
)\rangle~.
\end{equation}
$\gamma$ is Euler's constant and $\epsilon$ is the radius of a sphere
about $r_{j} = 0$ that is excluded from the integration.

The orders of magnitude for the other state-dependent coefficients
$C_{61} (1s^x\,nL)$, $C_{60}(1s^x\,nL)$, and $D_{50}(1s^x\,nL)$ are all
estimated from the generic formula
\begin{equation}
\label{X}
X(1s^x\,nL) = \frac{x\tilde{X}(1s) + \tilde{X}(nL)/n^3} {x +
\delta_{L,0}/n^3}~,
\end{equation}
where $\tilde{X}(nL)$ is the corresponding one-electron coefficient,
evaluated directly with the full nuclear charge $Z$ for $L = 0$ and
with a fully screened nuclear charge $Z - x$ for the outer $nL$
electron for $L > 0$ \cite{Drake_Martin_1998}. The numerical values can be
immediately calculated from the hydrogenic values, as discussed by
Drake and Martin \cite{Drake_Martin_1998} for helium, and Yan and Drake
\cite{Yan_Drake02} for lithium.  Their contribution to the transition
energy is taken to be the QED uncertainty.

The electron-electron QED shift $E_{\rm L,2}$ can similarly be
separated into mass-independent and mass-dependent parts according to
\begin{equation}
E_{\rm L,2} = E_{\rm L,2}^{(0)} + \frac{\mu}{M}E_{\rm L,2}^{(1)} +
\cdots,
\end{equation}
where
\begin{equation}
E_{\rm L,2}^{(0)} = \alpha^3\left(\frac{14}{3}\ln\alpha +\frac{164}{15}
\right) \sum_{i>j}\langle\delta({\bf r}_{ij})\rangle^{(0)} - \frac{14}
{3}\alpha^3 Q^{(0)},
\end{equation}
and the mass scaling and mass polarization corrections are
\begin{eqnarray}
E_{\rm L,2}^{(1)} &=& -3E_{\rm L,2}^{(0)} + \alpha^3\left(\frac{14}{3}
\ln\alpha + \frac{164}{15}\right) \sum_{i>j}\langle\delta({\bf r}_{ij})
\rangle^{(1)} \nonumber\\
&&\mbox{}-\frac{14}{3}\alpha^3 \left(Q^{(1)} +\sum_{i>j}\langle\delta({
\bf r}_{ij})\rangle^{(0)}\right)~.
\end{eqnarray}
Following our notation, the $Q^{(0)}$-term for infinite mass is given
by
\begin{equation}
\label{Q0}
Q^{(0)} = \frac{1}{4\pi}\lim_{\epsilon\rightarrow 0}\sum_{i>j} \langle
r_{ij}^{-3}(\epsilon)+ 4 \pi (\gamma +\ln \epsilon) \delta({\rm\bf r}
_{ij})\rangle.
\end{equation}
The $Q^{(1)}$ term is the correction due to the mass polarization
correction to the wave function. As a word of explanation, the infinitesimal limiting quantity $\epsilon$ has dimensions of distance, and so it generates an additional finite mass correction when distances are rescaled for the finite mass case according to $\epsilon \rightarrow (\mu/m)\epsilon$.

The principal computational challenge is the calculation of the Bethe logarithm $\beta(1s^2\,nL)$ in $E_{\rm L,1}$, originating from the emission and re-absorption of a virtual photon, and the finite mass correction $\Delta\beta_{\rm MP}(1s^2\,nL)$ due to mass polarization. The Bethe logarithm is the logarithmic remainder after mass renormalization and is defined by
\begin{eqnarray}
\label{beta}
\beta(1s^2\,nL) = \frac{\cal N}{\cal D} = \\
\frac{\displaystyle\sum_i |
\langle\Psi_0|{\bf p}_1 + {\bf p}_2|i\rangle|^2 (E_i - E_0) \ln|E_i -
E_0|} {\displaystyle\sum_i |\langle\Psi_0|{\bf p}_1 + {\bf p}_2|i
\rangle|^2 (E_i - E_0)}~. \nonumber
\end{eqnarray}

The accurate calculation of $\beta(1s^2\,nL)$ has been a long-standing
problem in atomic physics. This has been solved by use
of a discrete variational representation of the continuum in terms of
pseudostates \cite{Drake_Goldman_1999}.  The key idea is to define a
variational basis set containing a huge range of distance scales
through multiple sets of exponential scale factors $\alpha$ and $\beta$
that themselves span many orders of magnitude. The Bethe logarithm
comes almost entirely from virtual excitations of the inner $1s$
electron to $p$-states lying high in the photoionization continuum, and
so the basis set must be extended to very short distances for this
particle. The outer electrons are to a good approximation just
spectators to these virtual excitations.

Table \ref{tab:Li_beta} compares the Bethe logarithms for the two lowest
$s$-states of lithium with those for the Li-like ions Li$^+(1s^2\;^1{\rm
S})$ and Li$^{++}(1s\;^2{\rm S})$.  The comparison emphasizes that the
Bethe logarithm is determined almost entirely by the hydrogenic value
for the $1s$ electron and is rather independent of the state of
excitation of the outer electrons or the degree of ionization.  In
order to make the connection with the hydrogenic Bethe logarithm more
obvious, the quantity tabulated is $\ln(k_0/Z^2Ry)$.  The effect of
dividing by a factor of $Z^2$ is to reduce all the Bethe logarithms to
approximately the same number $\beta(1s) = 2.984\,128\,556$ for the
ground state of hydrogen.

\begin{table}[t]
\caption{\label{tab:Li_beta}
Comparison of Bethe logarithms for lithium and its ions.}
\begin{center}
\begin{tabular}{l r@{}l r@{}l r@{}l r@{}l }
\hline\noalign{\smallskip}
Atom &\multicolumn{2}{c}{Li($1s^22s$)}
&\multicolumn{2}{c}{Li($1s^23s$)}
&\multicolumn{2}{c}{Li$^+(1s^2)$}
&\multicolumn{2}{c}{Li$^{++}(1s$)}\\
\noalign{\smallskip}
\hline
\noalign{\smallskip}
$\ln(k_0/Z^2Ry)$ &    2&.981\,06(1)  & 2&.982\,36(6) & 2&.982\,624 &
2&.984\,128\\
\hline
\end{tabular}
\end{center}
\end{table}

\subsection{Nuclear Polarizability}

The interaction of the  nucleus with an electromagnetic field
can be described by the Hamiltonian
\begin{equation}
H_{\rm int} = q\,A^0 - {\bf d}\cdot {\bf E}  - \mbox{\boldmath{$\mu$}}\cdot{\bf B}
-\frac{q}{6}\,\langle r^2\rangle \,\nabla\cdot {\bf E}  , \label{01}
\end{equation}
which is valid as long as the characteristic momentum of the electromagnetic field
is smaller than the inverse of the nuclear size. Otherwise, one has to use
a complete description in terms of form factors and structure functions.
Under this assumption, the dominant term for
the nuclear excitation is the electric dipole interaction. This is the main approximation of
this approach,  which may not always be valid.
It was checked however that higher order polarizabilities are quite small
(below 1~kHz) for deuterium \cite{Leidemann_Rosenfelder_95,Friar1997}, and this is similar for
He, Li and Be isotopes. Within this low electromagnetic momentum approximation, the nuclear
polarizability correction to the energy
is given by \cite{Puchalski2006} (in units $\hbar = c = 1, e^2 = 4\,\pi\,\alpha$)
\begin{equation}
E_{\rm pol} = -m\,\alpha^4\,\Bigl\langle\sum_a\delta^3(r_a)\Bigr\rangle\;
(m^3\,\tilde\alpha_{\rm pol}), \label{02}
\end{equation}
where $m$ is the electron mass and the expectation value of the Dirac $\delta$
is taken with the electron wave function in atomic units.
For hydrogenic systems it is equal to $Z^3/\pi$. In the equation
above,  $\tilde\alpha_{\rm pol}$  is a {\em weighted} electric polarizability of the nucleus,
which is given by the double integral
\begin{eqnarray}
\tilde\alpha_{\rm pol} &=& \frac{16\,\alpha}{3}\,\int_{E_T}^\infty dE\,
\frac{1}{e^2}\,|\langle\phi_N|{\bf d}|E\rangle|^2\,\int_0^\infty\,\frac{d
w}{w}\, \frac{E}{E^2+w^2} \nonumber \\ &&\hspace*{-9ex}
\times\frac{1}{(\kappa+\kappa^\star)}\,\biggl[1+
\frac{1}{(\kappa+1)(\kappa^\star+1)}\,\biggl(\frac{1}{\kappa+1}+\frac{1}{\kappa^\star+1}
\biggr)\biggr]~,\label{03}
\end{eqnarray}
where $ \kappa = \sqrt{1+2\,i\,m/w}$ and
$E_T$ is the excitation energy for the nuclear breakup threshold.

The kets $|\phi_N\rangle$ and $| E \rangle$ denote
the ground state of the nucleus and a dipole excited state with excitation energy $E$, respectively.
The square of the dipole moment is related to the so called $B(E1)$ function
by the relation
\begin{equation}
|\langle\phi_N|{\bf d}|E\rangle|^2 = \frac{4\,\pi}{3}\,\frac{dB(E1)}{dE} \label{04}
\end{equation}
in units $e^2$ fm$^2$ MeV$^{-1}$, which explains the presence of $e^2$ in the
denominator in Eq.~(\ref{03}).

If $E_T$ is much larger than the electron mass $m$, one can perform a
small electron mass expansion and obtain a simplified formula \cite{Pachucki1993,Pachucki1994}:
\begin{eqnarray}
\tilde \alpha_{\rm pol} &=& \frac{19}{6}\,\alpha_{\rm E} + 5\,\alpha_{\rm Elog} \label{06}
\end{eqnarray}
with the static electric dipole polarizability
\begin{eqnarray}
\alpha_{\rm E} &=&
\frac{2\,\alpha}{3}\,\frac{1}{e^2}\,\biggl\langle\phi_N\biggl|
\,{\bf d}\,\frac{1}{H_N-E_N}\,{\bf d}\;\biggr|\phi_N\biggr\rangle \nonumber \\
&=&\frac{8\,\pi\,\alpha}{9}\,\int_{E_T} \frac{dE}{E}  \,\frac{1}{e^2}\frac{d B(E1)}{dE}
\end{eqnarray}
and the logarithmically modified polarizability
\begin {eqnarray}
\alpha_{\rm Elog} &=&
\frac{2\,\alpha}{3}\,\frac{1}{e^2}\,\biggl\langle\phi_N\biggl| \,{\bf
d}\,\frac{1}{H_N-E_N}\, \nonumber\\
& & \hspace*{+8ex} \times \ln\biggl(\frac{2(H_N-E_N)}{m}\biggr)\,{\bf d}
\;\biggr|\phi_N\biggr\rangle \nonumber \\
&=&\frac{8\,\pi\,\alpha}{9}\,\int_{E_T} \frac{dE}{E}  \,\frac{1}{e^2}\frac{d
  B(E1)}{dE}\,\ln\biggl(\frac{2\,E}{m}\biggr).
\end{eqnarray}
This approximation is for example valid for $^3$He  and $^4$He isotopes, but
not for $^{11}$Li or $^{11}$Be.

\begin{figure}[bt]%
\includegraphics[width=\linewidth]{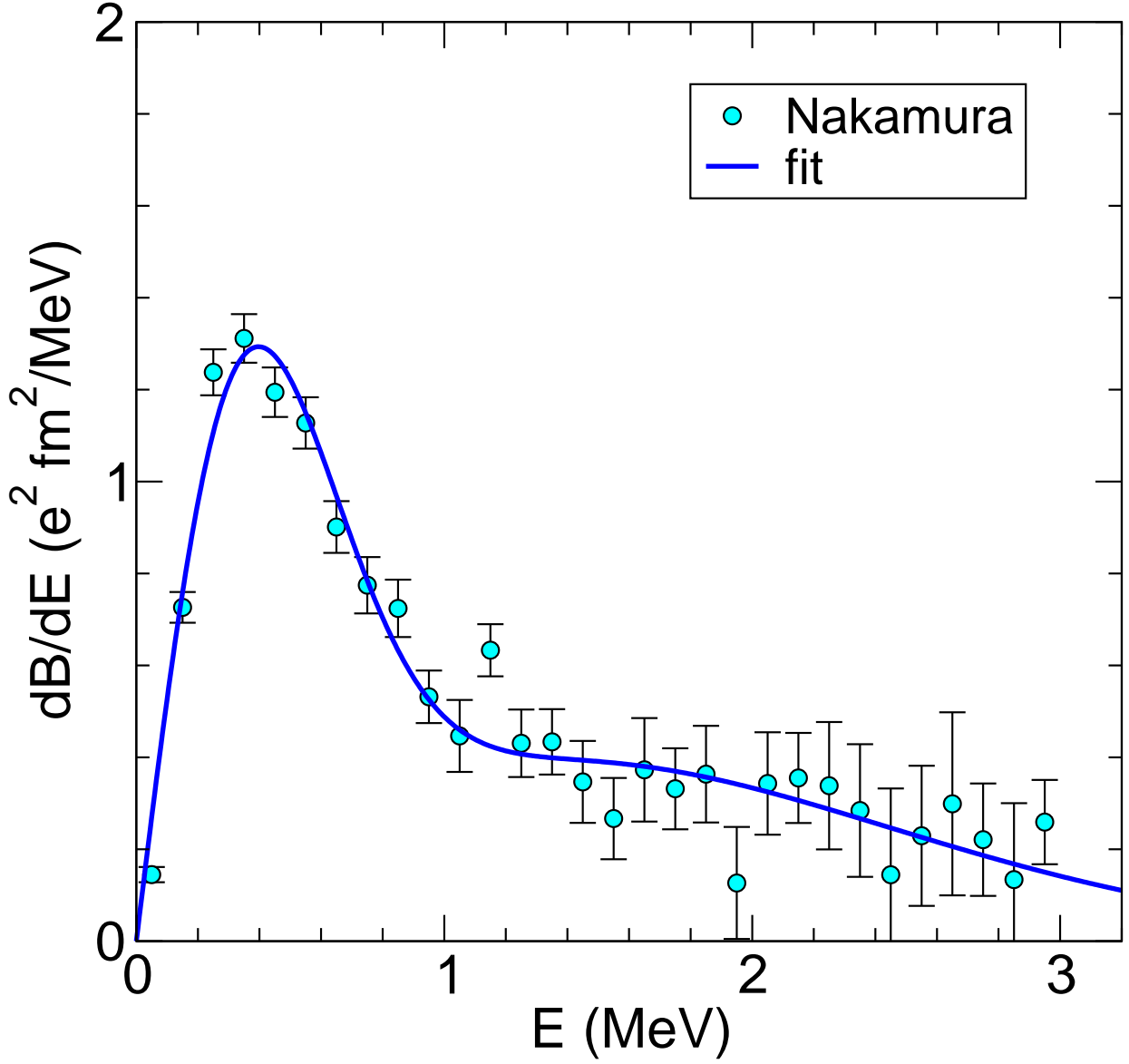}%
\caption{\label{fig:E1Strength} (Color online) Electric dipole line strength by Nakamura {\em
et al}. \cite{Nakamura2006} adapted to the new value of $E_T=369.15(65)$~keV
from \cite{Smith2008}.}
\end{figure}
In the opposite situation, i.e., when $m$ is much larger than $E_T$, we can use the nonrelativistic limit, and the polarizability correction adopts
the form (with $m$ being the reduced mass here)
\begin{eqnarray}
E_{\rm pol} &=& -m\,\alpha^4\,\left\langle\sum_a\delta^3(r_a)\right\rangle\;
\frac{32\,\pi\,\alpha\,m^2}{9}\, \nonumber \\
& & \times \, \int_{E_T} dE\, \frac{1}{e^2}\frac{d B(E1)}{dE}\,
\sqrt{\frac{m}{2\,E}}~. \label{07}
\end{eqnarray}
This approximation is justified for muonic helium atoms or ions, because the
muon mass ($\sim$106 MeV) is much larger than the threshold energy
$E_T=369.15(65)$~keV \cite{Smith2008}. This formula requires, however, a few
significant corrections, namely Coulomb distortion and form factor corrections.
They were obtained by Friar in \cite{Friar1977} for the calculation of the
polarizability correction in $\mu\,^4$He. This nonrelativistic approximation,
however, is not valid for electronic atoms since the typical nuclear excitation
energy in light nuclei is larger than the electron mass. This indicates a
possible limitation in the comparison of charge radii obtained from electronic
and muonic atoms, as the nuclear structure effects are much different.

The $B(E1)$ function is directly related to the photoabsorption cross Section
at photon energies $E$
\begin{equation}
\sigma(E) = \frac{16 \, \pi^3}{9} \, \alpha E \, \biggl( \frac{1}{e^2} \, \frac{d B(E1)}{dE}\biggr)~,
\end{equation}
which allows us to obtain the $B(E1)$ function from experimental data presented
in \cite{Nakamura2006} and shown in Fig.~\ref{fig:E1Strength}. With the
two-neutron separation energy $E_T = 369.15(65)$~keV from \cite{Smith2008} one
obtains
\begin{equation}
\tilde \alpha_{\rm pol} = 60.9(6.1)\,{\rm fm}^{-3} = 1.06(0.11)\,10^{-6}\,\lambdabar_C^{3},
\end{equation}
and a polarizability correction to the $2s \; ^2{\rm S}_{1/2} \rightarrow 3s \;
^2{\rm S}_{1/2}$ transition frequency of $\nu_{\rm pol} = 39(4)\,{\rm kHz}$. The
polarizability correction for the lighter lithium isotopes is expected to be
negligible due to much larger separation energies.

\subsection{Extraction of Nuclear Charge Radii}
\label{IS_theory}

\begin{table*} 
\caption{Contributions to the mass shift $\delta \nu_{\rm MS,Theory}^{6,A}$ of $^{A}$Li ($A=$ 7, 8, 9, 11)
relative to $^6$Li in the $2s\;^2{\rm S}_{1/2} \rightarrow 3s\;^2{\rm S}_{1/2}$ transition. The mass-dependent terms are calculated using the masses $M$ listed in the first row. The mass of the reference isotope $^6$Li is
6.015122794(16) amu \cite{Audi2003}. The nuclear polarizability contribution $\nu_{\rm pol}$ is included for $^{11}$Li and expected to be negligible for the other isotopes. The unit of the electronic factor $C_{6,A}$ is MHz/fm$^2$. All other values are in MHz.  Both sets of theoretical results are given in cases where they differ (see text).}
\label{Li_isotope}
\begin{center}
\begin{tabular}{l r@{}l r@{}l r@{}l r@{}l}
\hline
Term   & \multicolumn{2}{c}{$^{7}$Li} & \multicolumn{2}{c}{$^{8}$Li}& \multicolumn{2}{c}{$^{9}$Li} & \multicolumn{2}{c}{$^{11}$Li} \\
\hline %
$M$ (amu)  &  7.016003&4256(45)$^{\rm a}$     %
           &  8.022486&24(12)$^{\rm b}$     %
           &  9.026790&20(21)$^{\rm b}$     %
           & 11.043723&61(69)$^{\rm b}$    \\
$\mu/M$             & 11\,454&.655\,2(2)$^{\rm c}$ & 20\,090&.837\,3(9)$^{\rm c}$ & 26\,788&.479\,2(13)$^{\rm c}$&  36\,559&.175\,4(27)$^{\rm c}$\\
$(\mu/M)^2$         &     --1&.794\,0              &     --2&.964\,4              &     --3&.764\,2              &     --4&.761\,9                 \\
$\alpha^2\mu/M$     &       0&.017\,2$^{\rm d}$    &       0&.030\,2$^{\rm d}$    &       0&.040\,2$^{\rm d}$    &        0&.055\,0$^{\rm d}$      \\
                    &       0&.016\,8(1)$^{\rm e}$ &       0&.029\,5(2)$^{\rm e}$ &       0&.039\,3(3)$^{\rm e}$ &        0&.053\,7(4)$^{\rm e}$   \\
$\alpha^3\mu/M$     &     --0&.048\,5(6)           &     --0&.085\,1(11)          &     --0&.113\,5(15)          &      --0&.154\,8(21)            \\
$\alpha^4\mu/M$     &     --0&.009\,2(23)$^{\rm d}$&     --0&.016\,1(40)$^{\rm d}$&     --0&.021\,5(63)$^{\rm d}$&      --0&.029\,4(73)$^{\rm d}$  \\
                    &     --0&.008\,4(28)$^{\rm e}$&     --0&.014\,7(41)$^{\rm e}$&     --0&.019\,6(66)$^{\rm e}$&      --0&.026\,8(90)$^{\rm e}$  \\
$\nu_{\rm pol}$     &                             &&                             &&                             &&        0&.039(4)                \\
Total               & 11\,452&.820\,7(24)$^{\rm d}$& 20\,087&.801\,9(42)$^{\rm d}$& 26\,784&.620\,2(66)$^{\rm d}$&  36\,554&.323(9)$^{\rm d}$      \\
                    & 11\,452&.821\,1(28)$^{\rm e}$& 20\,087&.802\,6(50)$^{\rm e}$& 26\,784&.621\,3(67)$^{\rm e}$&  36\,554&.325(9)$^{\rm e}$      \\
$C_{6,A}$    &     --1&.571\,9(16)$^{\rm f}$&     --1&.571\,9(16)$^{\rm f}$&     --1&.572\,0(16)$^{\rm f}$&      --1&.570\,3(16)$^{\rm f}$\\
\hline
\end{tabular}
\end{center}
\begin{flushleft}
Ref. $^{\rm a}$ \cite{Nagy2006}, $^{\rm b}$ \cite{Smith2008}.\\
$^{\rm c}$ Uncertainties for this line are dominated by the nuclear mass
uncertainty.\\
$^{\rm d}$ Calculation by Puchalski and Pachucki (this work).\\
$^{\rm e}$ Calculation by Yan and Drake (this work).\\
$^{\rm f}$ A 25$\%$ error is assumed for the relativistic correction due to
the estimation of the relativistic correction to the wave function at the
origin on the basis of a known result for hydrogenic systems.
\end{flushleft}
\end{table*}

The various contributions to the isotope shift as they have been discussed so
far are listed in Table \ref{Li_isotope} for all lithium isotopes relative to
$^6$Li. The mass values that were used for the calculations are
also included for reference.  In addition, there is a significant electronic binding
energy correction $\Delta M= -E_{\rm binding}/c^2$. The terms are classified
according to their dependence on $\mu/M$ and the fine structure constant $\alpha$.
As an important check, most of the results have been calculated independently
by Puchalski and Pachucki (P\&P) in Poland \cite{Puchalski2008},
and by Yan and Drake (Y\&D) in Canada \cite{Yan2008}.
Two exceptions are the Bethe logarithm part of the radiative recoil correction,
which have been calculated only by Y\&D \cite{Yan2008}, and the nuclear polarizability
correction, which has been calculated only by P\&P \cite{Puchalski2006}.
In some cases, the two sets of results are slightly different due to different
methods of calculation, as noted in the table.
In these cases, both sets of results are given for comparison.
The differences are not large enough to affect the determination of the nuclear charge radius from experimental results, but they indicate the areas where further work is desirable. 

The various contributions are as follows: The term labelled $\mu/M$ contains the sum of the reduced mass
scaling of the nonrelativistic ionization energy, and the first-order mass
polarization correction.  For example, the mass-scaling term is
$2R_\infty[E_{\rm NR}({\rm Li}^+,\infty) - E_{\rm NR}({\rm Li},\infty)]
[(\mu/M)_{^{A}{\rm Li}} - (\mu/M)_{^6{\rm Li}}]$, where $E_{\rm NR}({\rm
Li},\infty)$ is the nonrelativistic energy for infinite nuclear mass (in atomic
units), and $R_\infty$ is the corresponding Rydberg constant.

The term of order $(\mu/M)^2$ comes from second-order mass polarization.  
The relativistic recoil terms of order $\alpha^2\mu/M$ results from mass scaling, mass polarization, and
the Stone terms as expressed by Eqs.\ (\ref{Delta_2}) and (\ref{Delta3Z}).
The two sets of results here do not quite agree because the Y\&D results
include partial contributions from the next-higher-order terms of order
$\alpha^2(\mu/M)^2$.  The difference between the two calculations is an indication of the uncertainty due
to the partial neglection of these higher-order terms.
The numbers in this row are anomalously small because of nearly complete numerical
cancelation between the mass polarization and mass scaling plus Stone contributions.
For example, for the case of $^{11}$Li, the two parts are  -17.9198(4) MHz
and +17.9735 MHz, respectively, resulting in a final recoil term of only 0.0537(4) MHz. Because
of this cancelation, the percentage uncertainty is relatively large.

The radiative recoil terms of order $\alpha^3\mu/M$ similarly come from a
combination of mass scaling, mass polarization, and higher-order recoil
corrections, as discussed in \cite{Yan_Drake98,Yan_Drake02,Yan_Drake03,Yan2008,Pachucki_Sapirstein_03,Puchalski2006b,Puchalski2006}.
The most difficult part of the calculation is the Bethe logarithm contribution
obtained in Ref.\ \cite{Yan2008}.

The sum of all these terms as given in the penultimate row in Table
\ref{Li_isotope} is the mass-dependent part of the isotope shift $\delta
\nu_{\rm MS,Theory}^{6,A}$ between $^6$Li and $^A$Li. It includes the nuclear
polarizability correction and the adjusted relativistic recoil term as
calculated for the first time in \cite{Puchalski2006}. The difference between
this value and the measured isotope shift must arise from the finite nuclear
size effect. To extract nuclear charge radii information from this value, the
electronic factor $C_{A,A^\prime}$ as defined in Eq.~(\ref{eq:IsotopeShift}) is required.

The contribution of the finite nuclear size effect to the total transition
energy can be written as
\begin{equation}
\Delta E_{\rm nuc} = C \cdot \bar{r}_{\rm c}^2,
\end{equation}
with the mean square nuclear charge radius $\bar{r}_{\rm c}^2 = \langle r_{\rm c}^2\rangle$.
The electronic factor $C$ can be expanded into a power series of $\alpha$ and $(\mu
/ M)$
\begin{eqnarray}
C & = & \frac{R_\infty m^2 c^3 \alpha^2}{\hbar^2}  \times \nonumber \\
& & \left ( C^{(4,0)} + \frac{\mu}{M}\,C^{(4,1)} + \alpha^2 C^{(6,0)}  + \ldots  \right )~,
\end{eqnarray}
where the $C^{(m,n)}$ are dimensionless coefficients that refer to corrections
on the order of $\alpha^m (\mu/M)^n$. The leading-order term is proportional to
the nonrelativistic wave function at the origin
\begin{eqnarray}
 C^{(4,0)} & = & \frac {2 \pi Z}{3} \Delta \left | \Psi(0) \right |^2,
\end{eqnarray}
and $C^{(6,0)}$ comes from the relativistic correction to the wave function at
the origin \cite{Friar1979}
\begin{eqnarray}
 C^{(6,0)} & \approx & -Z^2 \ln(Z\alpha m \bar{r}_{\rm c})  C^{(4,0)}.
\end{eqnarray}
Contrary to these terms, $C^{(4,1)}$ is mass-dependent. It consists of a term
originating from the mass scaling and one from the mass polarization
\cite{Puchalski2010}. For example, the contributions of these terms to the $C$-coefficients
of the transition frequency for $^{6,11}$Li are listed in
Table~\ref{tab:CFiniteSizeCorrection}. As compared with the leading term
$C^{(4,0)}$, the $C^{(6,0)}$ and $C^{(4,1)}$ corrections are of the order
$10^{-3}$ to $10^{-4}$. Thus, they are significant in the systematic studies of
the isotope shift at the current level of theoretical uncertainties in
$\delta \nu_{\rm MS, Theory}^{A,A^\prime}$.

\begin{table}[t]
\renewcommand{\arraystretch}{1.0}
\caption{Finite size corrections $C^{(m,n)}$ of the constant $C$ in MHz/fm$^{2}$ for the $2s \rightarrow 3s$ transition in
$^{11}$Li and $^{6}$Li.} \label{tab:CFiniteSizeCorrection}
\begin{tabular}{l r@{.} l r@{.} l}
\hline
Coefficient & \multicolumn{2}{l}{$^{11}$Li} & \multicolumn{2}{l}{$^{6}$Li} \\
\hline
$C^{(4,0)}$  &  --1&566501 & --1&566501 \\
$C^{(6,0)}$  &  --0&006675 & --0&006641 \\
$C^{(4,1)}$  &    0&000231 &   0&000424 \\[0.5ex]
$C$          &  --1&572945 & --1&572718 \\
\hline
\end{tabular}
\end{table}

Since the correction term $C^{(6,0)}$ depends on the nuclear size given by $\bar{r}_{\rm
c}$, we have to use initial approximate values $r_{A0}$ for the charge radii,
{\it e.g.}, those which are obtained in \cite{Puchalski2006}. The influence of
this approximation is negligible. With these values we can obtain the
$C_{A,A^\prime}$ coefficient for the isotopes $A$ and $A^\prime$ as
\begin{eqnarray}
 C_{A,A^\prime} & = & \frac{r_{A^\prime 0}^2 C_{A^\prime} - r_{A0}^2 C_{A} }%
                       {r_{A^\prime 0}^2 - r_{A 0}^2}.
 \label{eq:c_coef}
\end{eqnarray}
The obtained numerical results for constants $C_{A,A^\prime}$ for the relevant
isotope shifts using this relation are included in Table~\ref{Li_isotope}.

Concerning the accuracy of the calculated mass shift contributions, the
uncertainty is largely dominated by the nuclear mass uncertainty for $^{11}$Li,
as listed in the table. An exception is the relativistic recoil term of
order $\alpha^2\mu/M$. Because of the almost complete numerical cancelation of individual contributions that were already mentioned,
the percentage uncertainty is correspondingly large.

For convenience, the theoretical result relating the measured isotope shift
$\delta\nu$ to the change in the mean square nuclear charge radii between $^6$Li and
$^{11}$Li is
\begin{eqnarray}
\delta\nu^{6,11}_{2s \rightarrow 3s} &=& 36\,554.324(9)
\nonumber\\
&&\mbox{}+ 1.570\left[\bar{r}_{\rm c}^2(^6{\rm Li}) - \bar{r}_{\rm c}^2(^{11}{\rm
Li})\right] \label{2s3s},
\end{eqnarray}
where frequencies are given in MHz and radii in fm.
The results presented in Table~\ref{Li_isotope} are the basis for the
extraction of the change in the mean square charge radius for all lithium isotopes by
comparison with the experimental values, according to
\begin{eqnarray}
\left [ \bar{r}_{\rm c}^2(^A{\rm Li}) - \bar{r}_{\rm c}^2(^{6}{\rm Li})\right] =
\frac{\delta\nu^{6,A}_{\rm exp} - \delta\nu^{6,A}_{\rm MS,
Theory}}{C_{6,A}}.
\label{eq:delta_rc}
\end{eqnarray}

\section{Experimental Setup}

\begin{figure}[b]
\includegraphics[width=\columnwidth,angle=0]{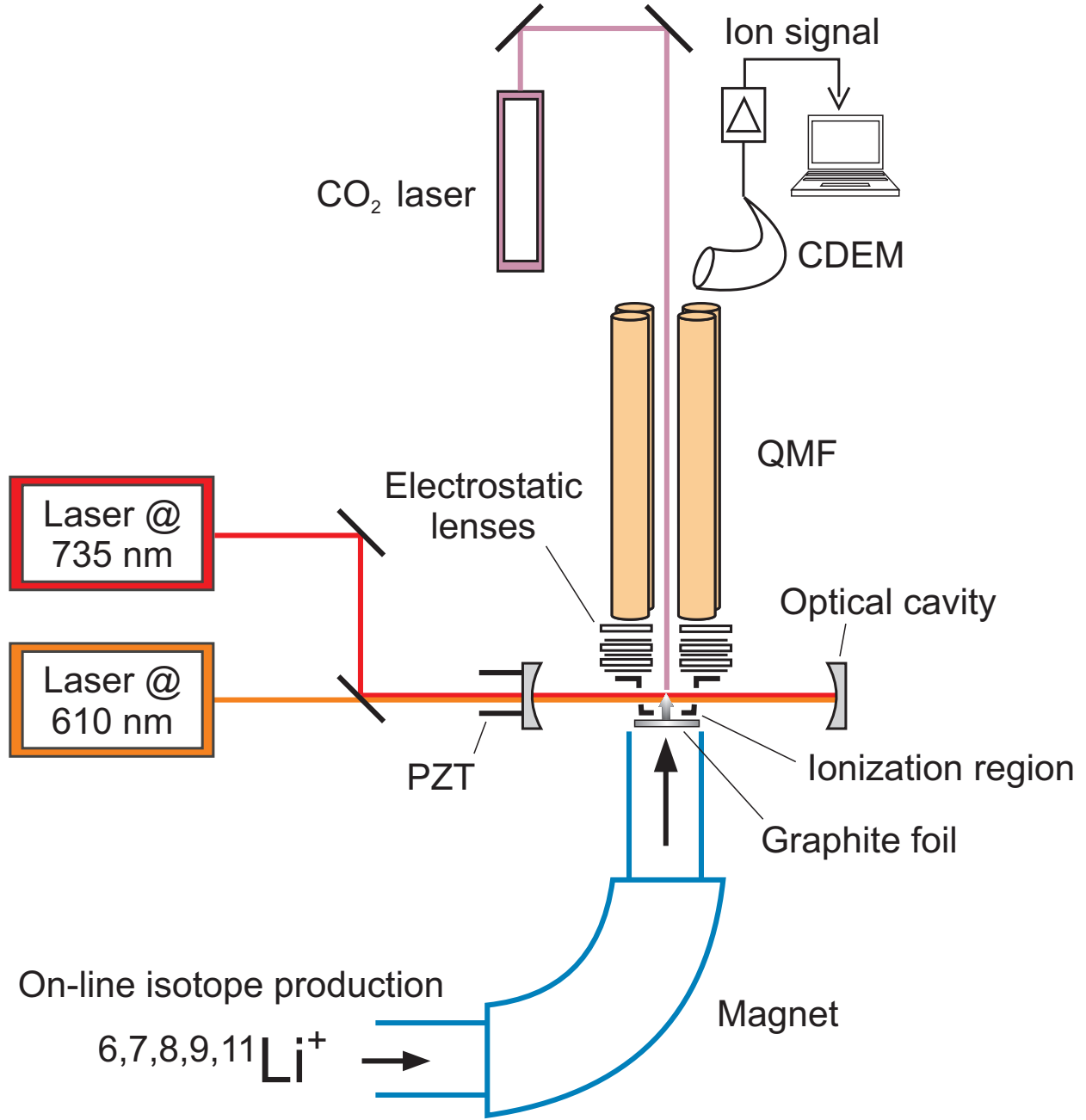}
{\caption{(Color online) Experimental setup to measure the ${2s\;^2{\rm
S}_{1/2}}\rightarrow 3s\:{^2{\rm S}_{1/2}}$ electronic transition in
lithium. CDEM: Continuous dynode electron multiplier, QMF: Quadrupole mass filter, PZT: Piezoelectric transducer.}\label{Fig:exp_arr_li11}}
\end{figure}

Figure \ref{Fig:exp_arr_li11} shows an overall schematic of the
experimental apparatus used for laser spectroscopy of the  
$2s\;^2{\rm S}_{1/2} \rightarrow 3s\;^2{\rm S}_{1/2}$ two-photon
transition in lithium. Ions of both stable and short-lived lithium isotopes, produced at an accelerator and accelerated to 40 keV, are mass separated and transported to the
experimental area. Here they are stopped and neutralized in a thin
carbon foil catcher.
This foil is heated with a CO$_{2}$ laser beam to about
1700-1800\,$^{\circ}$C , so that the implanted lithium atoms diffuse quickly to
the surface. Atoms released in the forward direction drift into the
ionization region in front of a quadrupole mass filter (QMF). Here, the
lithium atoms are resonantly ionized with laser light at 735 nm and
610~nm according to the three-step four-photon resonance ionization
scheme
\begin{eqnarray}
2s\;^2{\rm S}_{1/2}\overset{2\text{x}735~\text{nm}}{\longrightarrow\longrightarrow}3s\;^2{\rm S}_{\text
1/2}\xrightarrow[\tau\approx{\rm 30ns}]{\rm decay}2p\;^2{\rm P}_{3/2,1/2} \\
\xrightarrow{\rm 610~nm}3d\;^2{\rm D}_{3/2,5/2}
\xrightarrow{\rm 610~nm,735~nm}\text{Li}^{+}~,\nonumber
\end{eqnarray}
with a two-photon transition followed by spontaneous decay of the $3s$ state with a lifetime of $\tau\approx 30$ ns and subsequent resonance ionization as discussed in more detail in Section \ref{sec:ExcIon}. For brevity, if the meaning is clear, the laser-driven resonance transitions are abbreviated as the $2s \rightarrow 3s$ and $2p \rightarrow 3d$
transitions, respectively.  The photo-produced ions are then mass analyzed with the QMF and
detected with a continuous dynode electron multiplier (CDEM)
detector. The isotope shift in the $2s \rightarrow 3s$ transition is measured by tuning the 735 nm
light across the lithium two-photon resonances. The individual parts
of the system will be described in detail in the following Sections.

\subsection{Production of Radioactive Lithium Isotopes}

Radioactive lithium isotopes were produced at the on-line mass separator\footnote{The mass separator was shutdown in early 2004.} at GSI Darmstadt in 2003 \cite{Ewald04} and in a second experiment
at the ISAC mass separator facility at TRIUMF in September and October 2004 \cite{Sanchez06} (see Table \ref{tab:yield_li}). The experimental setups were almost identical for the two experimental runs. If changes were made, the version used for the TRIUMF experiment is presented in this publication. 

At GSI, $^{8,9}$Li were produced by directing a $^{12}$C beam from the
UNILAC with an energy of 11.4~MeV/u onto a 100~mg/cm$^2$ tungsten target. Fast reaction products
entering the hot ion source of the mass separator through a tungsten
window were stopped in a sintered graphite catcher. Atoms
diffusing out of the catcher were surface-ionized and extracted
through a small hole. Ion yields towards the experiment are listed in Table \ref{tab:yield_li}.
\begin{figure}[t]
\includegraphics[width=\columnwidth,angle=0]{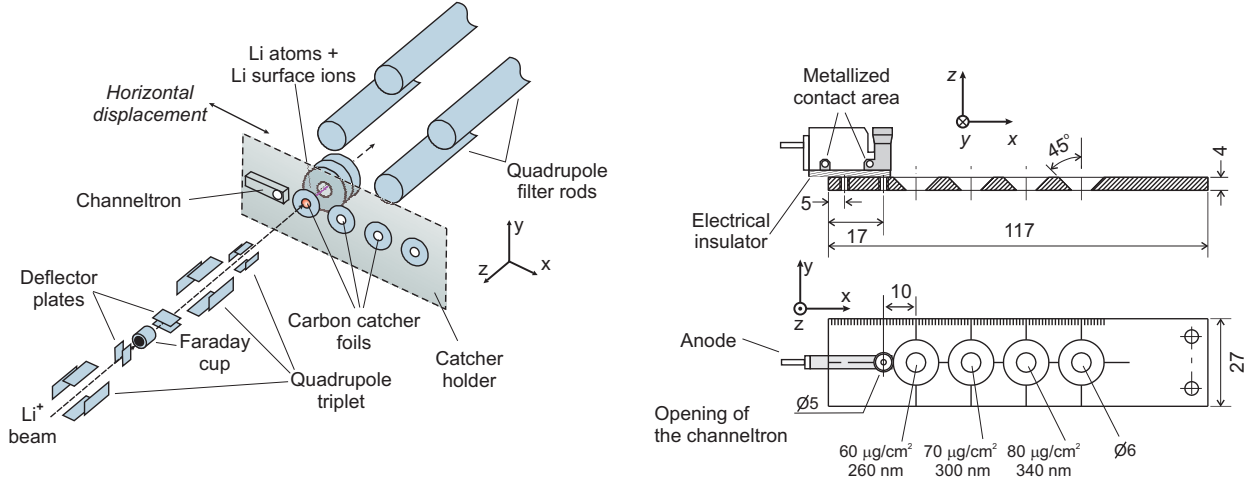}
{\caption{(Color online) Schematics of the ion optical setup (left) in front of the catcher holder which holds a channeltron-type continuous dynode multiplier detector for aligning the $^{11}$Li ion beam and several carbon foils of different thickness. Details of the catcher holder are shown on the right. Dimensions are given in mm. The holder, fixed to a linear feedthrough, can be moved in the $x$-direction horizontally and vertically in the $y$-direction.}\label{fig:IonOpticsOverview}}
\end{figure}

At TRIUMF, $^{8,9,11}$Li were produced with a 500~MeV
primary proton beam of 40~$\mu A$ extracted from the H$^-$ cyclotron. A stack
of tantalum foils was used as target in order to allow fast release of the short-lived
lithium isotopes that are produced by target fragmentation. The reaction products were surface ionized and extracted from the ion
source with a beam energy of 40~keV. Mass separation was obtained with a 60$^\circ$
preseparator magnet followed by a 120$^\circ$ main separator \cite{Dombky2000}. A fast
switch (kicker) installed behind the main mass separator was used to turn the ion beam on and off. 

\begin{table}
\caption{\label{tab:yield_li}Half-lifes and typical ion yields in ions/s for the
radioactive lithium ions at GSI and ISAC-TRIUMF.}
  \begin{ruledtabular}
    \begin{tabular}{llll}
      Beamtime               & \mbox{$^8$Li}     & \mbox{$^9$Li}     & \mbox{$^{11}$Li}\\
      \hline
      Half-life (ms)         & 838(6)            & 178.3(4)          & 8.59(14)\\
      Yield (GSI 2003)       &   3.6$\times10^5$ &   1.8$\times10^5$ & \mbox{-}\\
      Yield (TRIUMF 09/2004) &  10$^8$           &  10$^6$           & \mbox{-}\\
      Yield (TRIUMF 10/2004) &   8$\times10^8$   &   9$\times10^7$   & 3.5$\times10^4$\\
    \end{tabular}
  \end{ruledtabular}
\end{table}

The mass separated ion beam was transported into the ISAC low energy
experimental area, where the ToPLiS experiment was installed. The
existing low energy beamline was extended to allow installation of
deflector plates and a quadrupole doublet, as depicted in Fig.~\ref{fig:IonOpticsOverview}, for precise shaping and
steering of the ion beam. A similar ion optics was existing at the GSI mass separator. To detect the small number of
$^{11}$Li ions delivered to the experiment, a channeltron-type
detector was installed on a linear feedthrough that also carries the
carbon catcher foils and could be moved into the beam in front of
the QMF.

\subsection{Neutralization and Atomic Beam Generation}
After production and mass separation, the 40 keV ion beam had to be converted into a thermal beam of neutral atoms. This process must be efficient and considerably faster than the 8.6~ms half-life of the $^{11}$Li ions. Therefore, the ion beam was directed onto a thin carbon foil (Fig.~\ref{fig:IonOpticsOverview}). In this `catcher'
foil the ions were stopped and neutralized. ``Stopping and Range
of Ions in Matter'' (SRIM) calculations \cite{Ziegler2003} were used to estimate the thickness
of foil that would stop the ions shortly before they reached the
back side of the foil. This enabled rapid release of the neutralized
atoms in the preferred direction towards the ionization region (Fig.~\ref{Fig:exp_arr_li11}). Catcher foils with
thickness around 300 nm were prepared in the GSI Target Laboratory
and glued onto a stainless steal holder, providing three positions
equipped with catcher foils of different thickness (ranging from
about 60 to 80~$\mu$g/cm$^2$). The target holder was mounted on a
linear feedthrough that moves the holder along the $x$-position so
that foils of different thickness could be located in front of the quadrupole mass
filter. The holder was also movable in $y$ and $z$
directions to optimize the vertical position of the catcher foil and
the distance to the QMF ion optics.
\begin{figure}
\includegraphics[width=\columnwidth,angle=0]{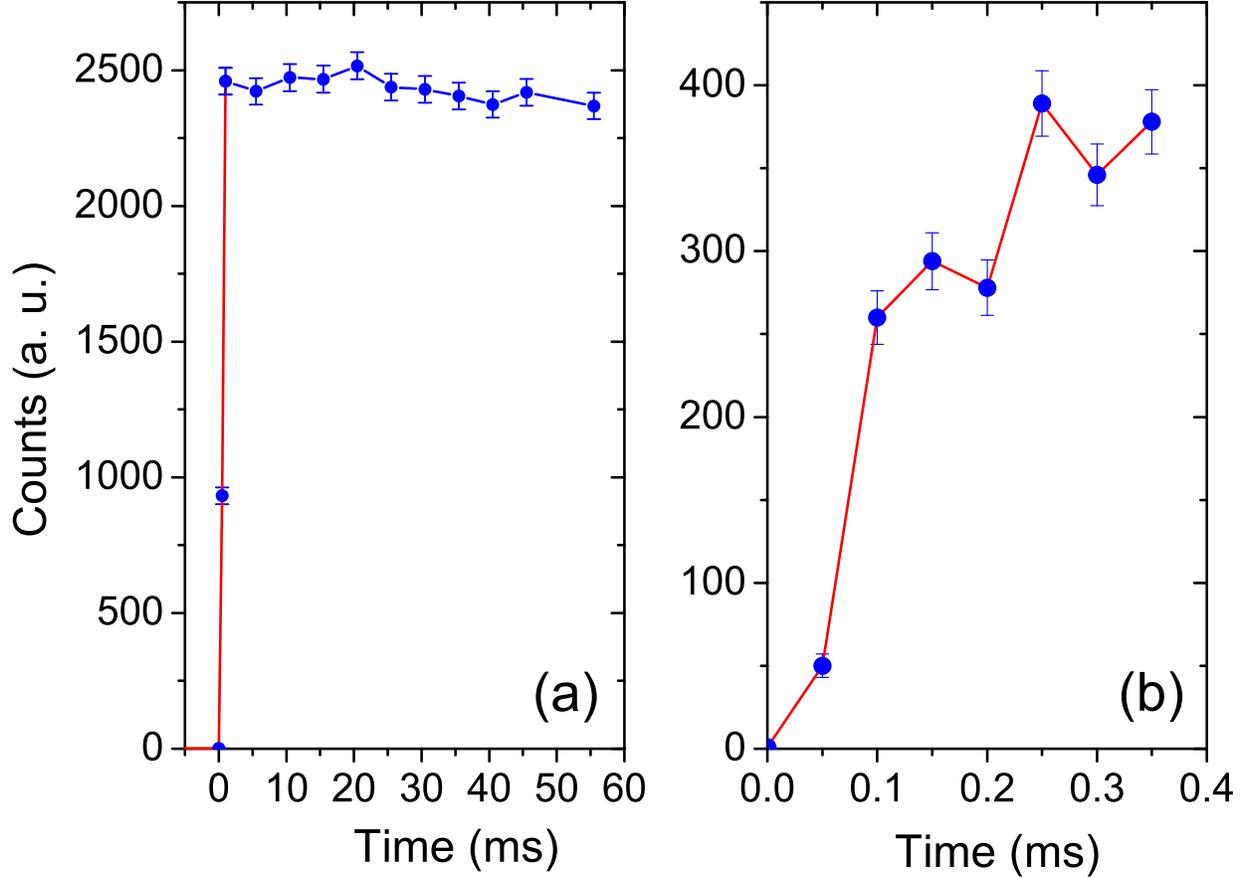}
{\caption{(Color online) Time-resolved surface-ion release recorded for $^6$Li to
determine the release time at a catcher temperature of about 1700-1800$^\circ$C. 
The $t=0$ position is approximately the time when the ion beam was turned on. Time resolution was 0.5~ms in (a) but after the first three data points only every tenth channel is shown as an individual data point for clarity. In (b) a resolution of 50~${\rm \mu}$s was used.}\label{Fig:spc_dwe_tim_li11}}
\end{figure}
As mentioned before, fast release is essential for measurements on $^{11}$Li. In order to achieve this, a
2 mm diameter spot on the catcher foil was heated to about
1800$^{\circ}\mathrm{C}$ with a CO$_{2}$ laser beam that was coupled
into the system along the axis of the QMF (see below). As shown in Fig.~\ref{Fig:spc_dwe_tim_li11}, a release of the implanted atoms within 0.5~ms was
observed under these conditions. The temperature of the catcher foil was still sufficiently low that only a very small fraction of about $10^{-4}$ of the incoming ions were released as surface ions.
The release time of the lithium atoms from the graphite catcher foil
was measured by observing the increase of surface ions as a function
of time after turning the ion beam on. For this purpose the fast
kicker at the mass separator was used, the electrostatic ion optics
of the QMF were set for surface ion detection and the output of the
CDEM detector was fed into a fast timing amplifier for linear pulse
amplification. The amplified signal was recorded with a multichannel
scaler. The measurement was performed with a beam of stable
$^{6}$Li and the observed signal with a time resolution of 1 ms and 0.05 ms is shown in Fig.~\ref{Fig:spc_dwe_tim_li11}(a) and \ref{Fig:spc_dwe_tim_li11}(b), respectively. The increase in ion beam intensity from a practically background-free
baseline is very fast and is not resolved with the time
resolution of 1~ms. The graph with the higher time resolution (Fig.~\ref{Fig:spc_dwe_tim_li11}(b)) shows two contributions: The ion signal rises to about 70\% of
the final level within 100~${\rm \mu}$s and is then followed by a slightly
slower increase to the full intensity which is reached after about
300~${\rm \mu}$s. A large contribution of ions that have sufficient energy
for simply penetrating the foil was never observed. Hence, it is assumed that even the flank was originating from ions that were first stopped in the catcher foil and released again. However, even if the 300~${\rm \mu}$s time scale would be the relevant one for the release time of neutral atoms, this amounts to only 3\% of the $^{11}$Li half-life and is clearly sufficient to have a good release efficiency. The total conversion efficiency of the foils was estimated to be about 50 \% and the transport efficiency into the laser beam to 20\%.

\subsection{Excitation and Ionization Scheme }
\label{sec:ExcIon}

The laser excitation and ionization scheme has to provide both, high resolution and high efficiency in order to detect the signal for ion yields of only a few thousand ions/s with an accuracy of about 0.1 MHz.

\begin{figure}
\includegraphics[width=\columnwidth,angle=0]{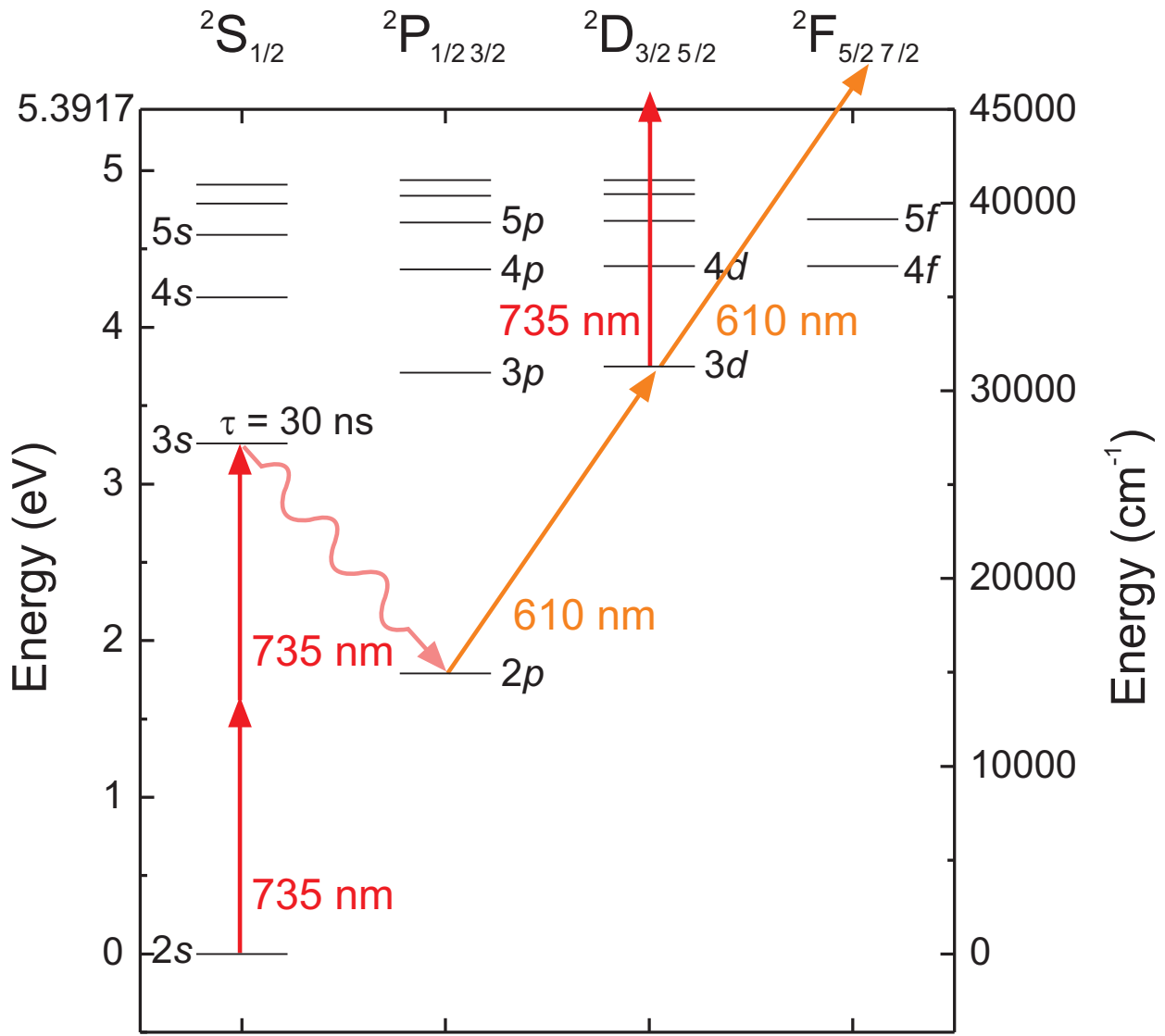}
\caption{(Color online) Resonance ionization scheme for
lithium as used in the reported experiments.}\label{Fig:ionization_scheme}
\end{figure}

Figure\;\ref{Fig:ionization_scheme} shows the level scheme for neutral lithium with
the excitation path chosen for resonance ionization: Lithium
atoms in the $2s\;^2{\rm S}_{1/2}$ ground level are excited via the
Doppler-free two-photon transition at 735~nm to the $3s\;^2{\rm S}_{1/2}$ level. This transition offers a narrow resonance and
leads to efficient excitation since all velocity classes can be excited simultaneously. However, relatively large intensities are required for saturation.
The two-photon excitation is followed by a spontaneous decay into
the $2p\;^2{\rm P}_J$ levels. Atoms in the
$2p\;^2{\rm P}_J$ levels are subsequently excited into the $3d\;^2{\rm D}_{J'}$ levels with resonant laser light at 610 nm and then photoionized by absorption of a photon at either 735 nm or 610 nm. In this way the states used for measuring the isotope shift are decoupled from the states used for ionization and detection which results in a strong reduction of the ac Stark
broadening and ac Stark shifts \cite{Schmitt00}. 
Compared with fluorescence detection, this multi-step excitation and
ionization followed by mass-selective ion detection has the advantage that the $2s \to 3s$ resonance can be
detected with a very high efficiency and an extremely high signal-to-background ratio.

\subsection{Mass Spectrometer}

To achieve the additional background suppression, mass selective ion detection is used. Quadrupole mass filters (QMF) provide ideal conditions to mass separate the ions with thermal energies that are produced in the laser ionization process described above \cite{Bushaw89, Blaum00}.

Therefore, the laser beams have to be overlapped with the thermal atom cloud in a region from which the created ions can be extracted into the QMF. Moreover, surface ions that are produced on the hot catcher foil must be efficiently suppressed, because every surface ion of the isotope under investigation entering the laser beam cannot be distinguished from a photo ion and will produce background events.

\begin{figure}[b!]
\includegraphics[width=\columnwidth,angle=0]{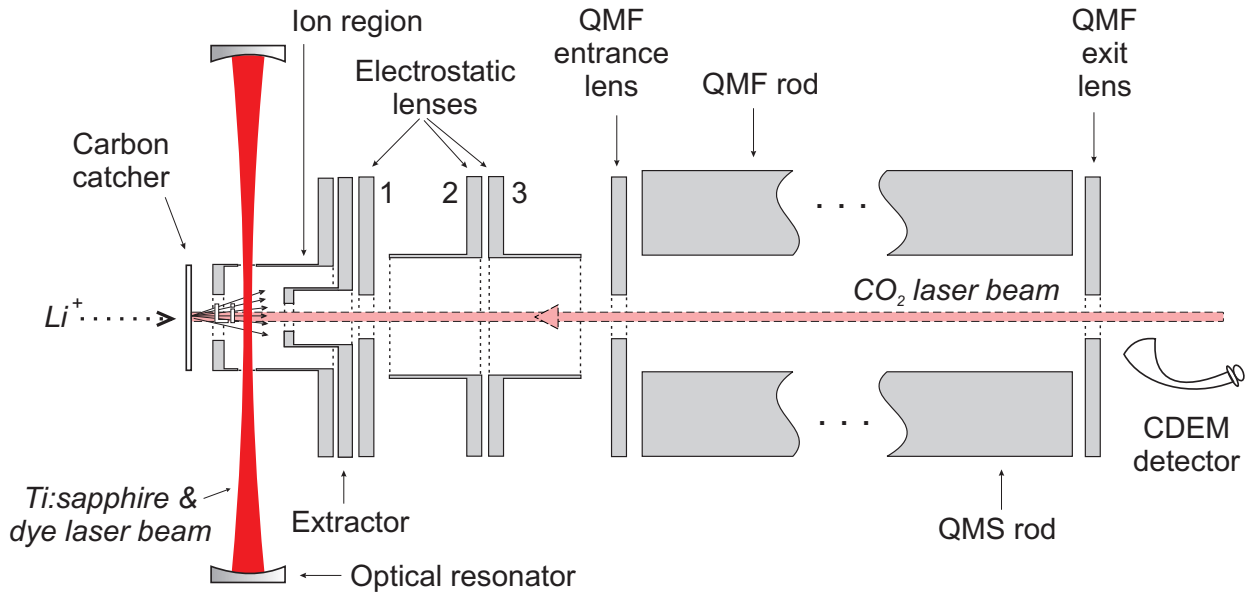}
{\caption{(Color online) Schematics of the optical resonator, the ion optics, the quadrupole mass filter (QMF) and the continuous dynode electron multiplier (CDEM) detector.}
\label{Fig:qms_ionoptics}}
\end{figure}

The QMF used for the reported experiment was a commercial instrument from
ABB Extrel (ABB Extrel, Pittsburgh, USA, Model No.\ 150 QC) with
9.39~mm radius rods of 21~cm length and a free-field radius of $r_0
= 8.33$~mm. The system is driven at a frequency of 2.9 MHz and can
be used for ions up to $A/q \leq 64$. This model provides
transmission close to 100~\% and an excellent neighboring mass
suppression as has been demonstrated in simulations
\cite{Blaum98,Blaum00}, ultratrace analysis applications
\cite{Wendt99,Bushaw00}, and measurements discussed below. Ions transmitted through the QMF were
detected with an off-axis continuous dynode electron multiplier (CDEM) detector. The entrance opening of the channeltron was biased with about -2000~V such that the positive ions were accelerated in the CDEM.

For the measurements reported here, the ionization region of the QMF
was modified as shown in Fig.\,\ref{Fig:qms_ionoptics}. The original axial
electron impact ion source of the EXTREL device was removed and the electrodes before the quadrupole structure were replaced by specifically designed electrodes allowing free access for the laser beams. The distance between the catcher foil and the ionization region was 2 - 3~mm to have an efficient
transfer of the released atoms into the ionization region. The QMF
ion optics could be operated to accept either laser-created ions
from inside the ion region or surface ions created on the hot carbon
catcher foil. In the first mode, the ion region (Fig.~\ref{Fig:qms_ionoptics}) was held at 3.9~V
relative to the grounded catcher foil. This repels catcher surface
ions, while neutral atoms can enter the laser ionization region. The
potential difference between the catcher foil and the ionization
region was small enough to ensure that electrons emitted from the hot
catcher surface and accelerated into the ion region, do not gain
sufficient energy to ionize lithium atoms by electron impact
ionization (Li ionization potential: 5.39 eV). Photo ions were then
extracted by the negative extractor voltage and focussed into the
QMF rod structure using the remaining electrostatic lenses. In the
second mode, surface ions produced at the hot catcher foil were accelerated into the QMF with a
negative voltage at the ion region and the extractor was operated as
another lens for adapting the beam properties to the QMS acceptance.
Ion optical settings of all lenses for both detection modes are
summarized in Table\,\ref{Table:qms_set}.
\begin{table*}
\begin{center}
{\caption{Settings for the extractor, the electrostatic lenses and the quadrupole mass filter (see Fig.~\ref{Fig:qms_ionoptics}) for the
detection of either surface ions or photo ions produced by resonance ionization. All values are in V. Pole bias is an additional common DC voltage that is applied to all four QMF rods to change the kinetic energy of the ions entering the rod system.}\label{Table:qms_set}} \vspace{2mm}
\begin{scriptsize}
\begin{tabular*}{1\textwidth}{@{\extracolsep{\fill}}@{\hspace{2mm}} l | c c c c c c c c}
\hline\hline
 & Ion Region & Extractor & Lens 1 & Lens 2 & Lens 3 & Entrance Lens & pole bias & exit lens \\
\hline
Laser Ions   & +3.9 & ~~-9 & -43 & ~~+8 & -43 & +2  & 0     & ~-2\\
\hline
Surface Ions & ~-4.9 & -165 & -60 & -310 & -60 & ~-7 & ~-1.4 & +9\\
\hline\hline
\end{tabular*}
\end{scriptsize}
\end{center}
\end{table*}
Ions transmitted through the rod system were focussed with an exit
lens and detected with the CDEM detector model DeTech 5402AH-021,
which was chosen for its very low dark count rate of typically 5 to
20~mHz.

\begin{figure}
\includegraphics[width=\columnwidth,angle=0]{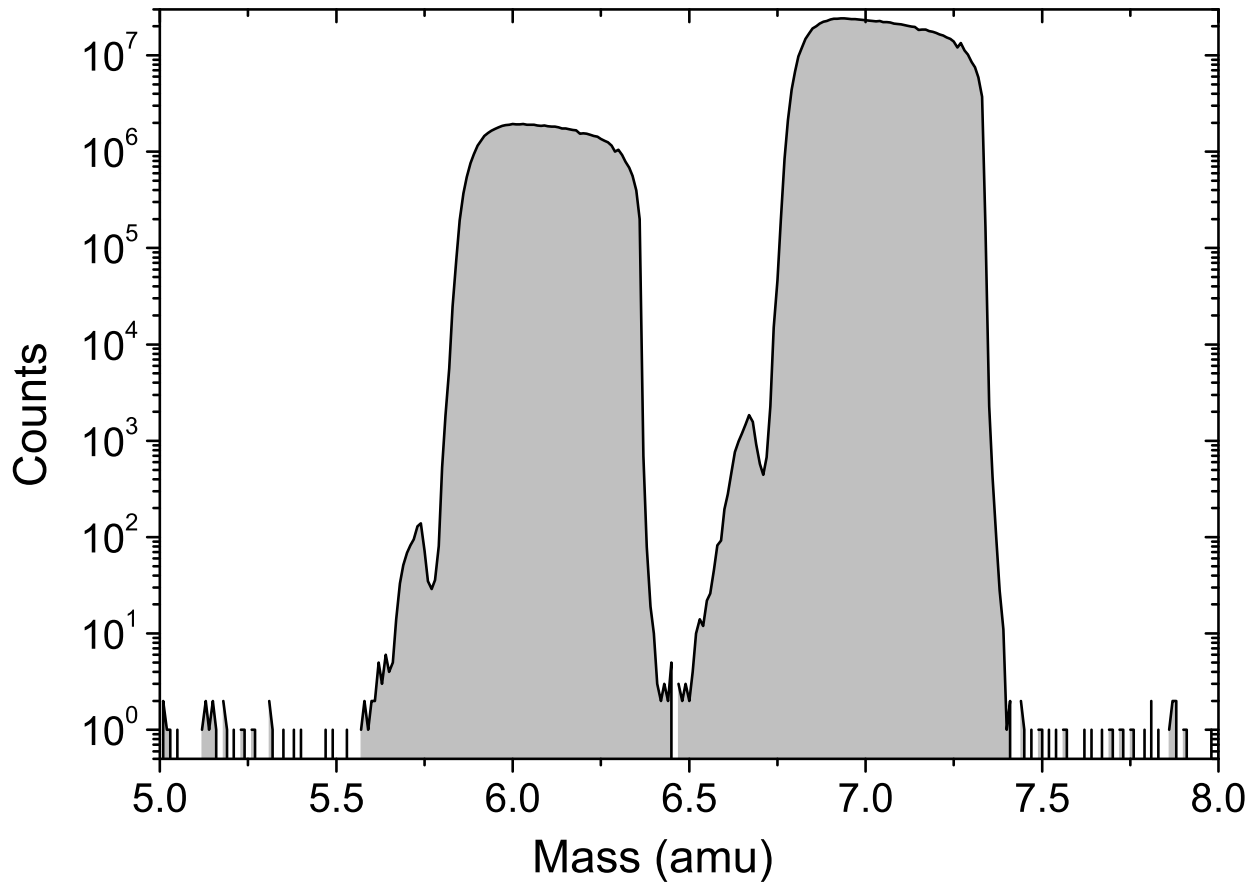}
\caption{(Color online) Mass spectra of the stable lithium isotopes $^6$Li and $^7$Li.} \label{Fig:Li6_Li7_mas_pro}
\end{figure}

The lithium isotopes were produced with rates that differ by orders of magnitude (see Table~\ref{tab:yield_li}). For additional suppression of neighboring isotopes, the QMF was
tested with ions on the stable isotopes $^{6,7}{\rm Li}$ produced by
surface ionization on the hot carbon catcher foil. Ion optical
parameters of the QMF extraction and transport optics were optimized in order to obtain a flat top profile with
steep flanks and maximum transmission. An example for the peak
profile of surface ions is shown in Fig.~\ref{Fig:Li6_Li7_mas_pro}.
Typically, a suppression of neighboring masses of $>10^{8}$ was
achieved. The dark count rate of the CDEM detector was about 10~mHz.

The entire QMF system consisting of ion source, quadrupole mass
filter and ion counting was computer controlled using the commercial
Extrel Merlin data acquisition and control electronics. The system
was installed in a vacuum chamber and pumped by a turbomolecular pump,
which provided residual gas pressures in the range of
$10^{-7}$~mbar, even when the graphite catcher was heated.

\subsection{Laser System and Enhancement Cavity}
The excitation and ionization scheme discussed in Section~\ref{sec:ExcIon} requires laser systems that provide the high power required to saturate the two-photon transition and to efficiently drive the ionization steps and at the same time a sufficiently small bandwidth, high stability and precise frequency control to reach the required accuracy for the isotope shift measurements.

For the two-photon transition it is important to have two exactly counterpropagating laser beams in the laser ionization region. It is also favourable to have these beams well balanced in power to avoid an asymmetry in the background signal due to Doppler-broadened two-photon excitation as discussed below. The laser system that fulfilled all these requirements was composed of two argon ion laser pumped ring lasers, a titanium:sapphire (Ti:sapphire) laser and a dye laser, combined with a Fabry-Perot cavity around the laser interaction region to enforce the counterpropagating beams and the high intensities that are required for the two-photon transition. Because both laser beams must interact with the atoms simultaneously, this solution required a cavity that is kept in resonance with the two laser beams having strongly different wavelenghts. High-accuracy frequency determinations were achieved by referencing the Ti:sapphire laser to an iodine-stabilized diode laser.

\subsubsection{Iodine-Locked Reference Laser}
\label{sec:IodineLock} 
\begin{figure}
\includegraphics[width=\columnwidth,angle=0]{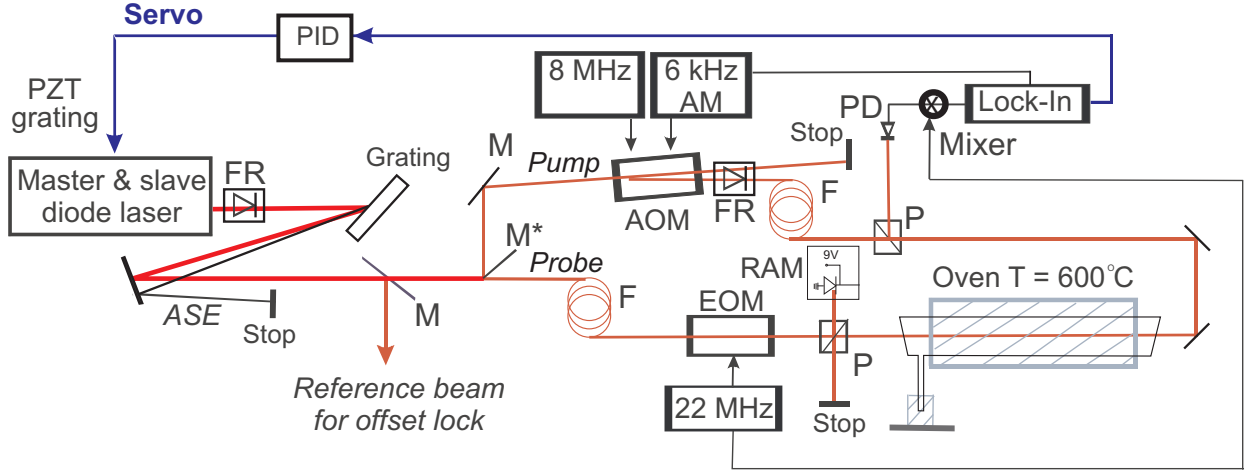}
{\caption{(Color online) Diode laser system locked to an iodine line for frequency reference. ASE: Amplified spontaneous
emission, AOM: Acousto-optical modulator, EOM: Electro-optical modulator, F: Fiber, FR: Faraday rotator, M: Mirror, P: Polarizer, PD: Photodiode, PID: Proportional-integral-differential regulator, PZT: Piezoelectric transducer for tuning the grating of the master laser, RAM: Residual amplitude modulation.}
\label{dio_las_li11}}
\end{figure}

A stable reference frequency for the isotope
shift measurement was realized by an amplified diode laser system
locked to a hyperfine component in the molecular spectra of
$^{127}$I$_2$. To obtain a detectable beat frequency between the
Ti:sapphire laser and the reference diode laser while the Ti:sapphire laser frequency was tuned
across the $2s\rightarrow 3s$ transition in lithium, the iodine transition had to be within
approximately 50~GHz of the two-photon resonance frequency. Hence,
the $X^{\,1}\Sigma_{g}^{+}\rightarrow BO_{u}^{+}\,\,R$(114)~11-2
transition in iodine was chosen. The a$_1$ hyperfine component has a
predicted resonance frequency of $407\,815\,138\,(30)$~MHz according
to the `iodine spec 4' program \cite{Knoeckel04}. Recent
measurements accurately determined the frequency as
$407\,815\,137.15\,(30)$~MHz \cite{Reinhardt2007}. This is very
close to the lithium two-photon transitions with the isotopes
$^{6,7,8}$Li having resonance frequencies below and $^{9,11}$Li above the iodine transition frequency. The
largest separation is for $^6$Li and is approximately 11.5~GHz.
Stabilization of the diode laser to the iodine transition was
achieved using frequency modulation (FM) saturation spectroscopy
\cite{Hall1981}, with the experimental arrangement shown in
Fig.~\ref{dio_las_li11}.
The light produced by the master laser (Toptica, Model PDL 100) was
amplified in a broad-area diode laser (BAL 740-100-1, Sacher) and
separated from the weak amplified spontaneous emission (ASE) of the
amplifier with a diffraction grating. The spatially-extended amplified beam was split at the
edge of a mirror into a pump and a probe beam with intensity ratio
of about 2:1. The pump beam was frequency-shifted (8 MHz) and
amplitude-modulated (AM) at 6.6~kHz with an acousto-optical modulator
(AOM) and then sent through a single-mode fiber (Newport S-FS-C) to
obtain a good TEM$_{00}$ mode structure. The probe beam was first
sent through a mode-cleaning fiber and then frequency-modulated
(22 MHz) with an electro-optical modulator (EOM) for side-band
generation. The signal in FM saturation spectroscopy after simultaneous interaction of the counterpropagating probe and pump beam with the iodine vapour is carried by the amount of amplitude modulation of the probe beam at the EOM frequency. Hence, residual amplitude modulation (RAM) of the phase-modulated probe beam introduced by a non-ideal matching of the laser beam polarization to the EOM would have caused an offset signal in the detection that shifts the
locking point away from the resonance center. Therefore the RAM was
actively suppressed by separating a small part of the laser beam
after the EOM and detecting intensity fluctuations at the EOM
frequency on a fast photodiode. A feedback loop that regulated a
high-voltage DC offset on the EOM was used to remove the spurious RAM
as described in \cite{Wong1985}.

The probe and pump beams were superimposed in a counterpropagating
geometry with perpendicular polarization in an iodine vapor cell
(pump beam: 5~mW, probe beam: 3~mW, beam diameters 1~mm). The beams
were combined and separated with Rochon polarizers. After passage
through the iodine cell and separation from the pump beam, the probe
beam was directed onto a photodiode. The photodiode signal was
amplified and demodulated with a mixer at the 22~MHz EOM frequency. The
mixer intermediate frequency (IF) output was then fed into a lock-in
amplifier for phase-sensitive detection to extract the signal at the
6.6 kHz frequency that was applied to the AOM driver to amplitude-modulate the pump beam. 

\begin{figure}
\includegraphics[width=\columnwidth,angle=0]{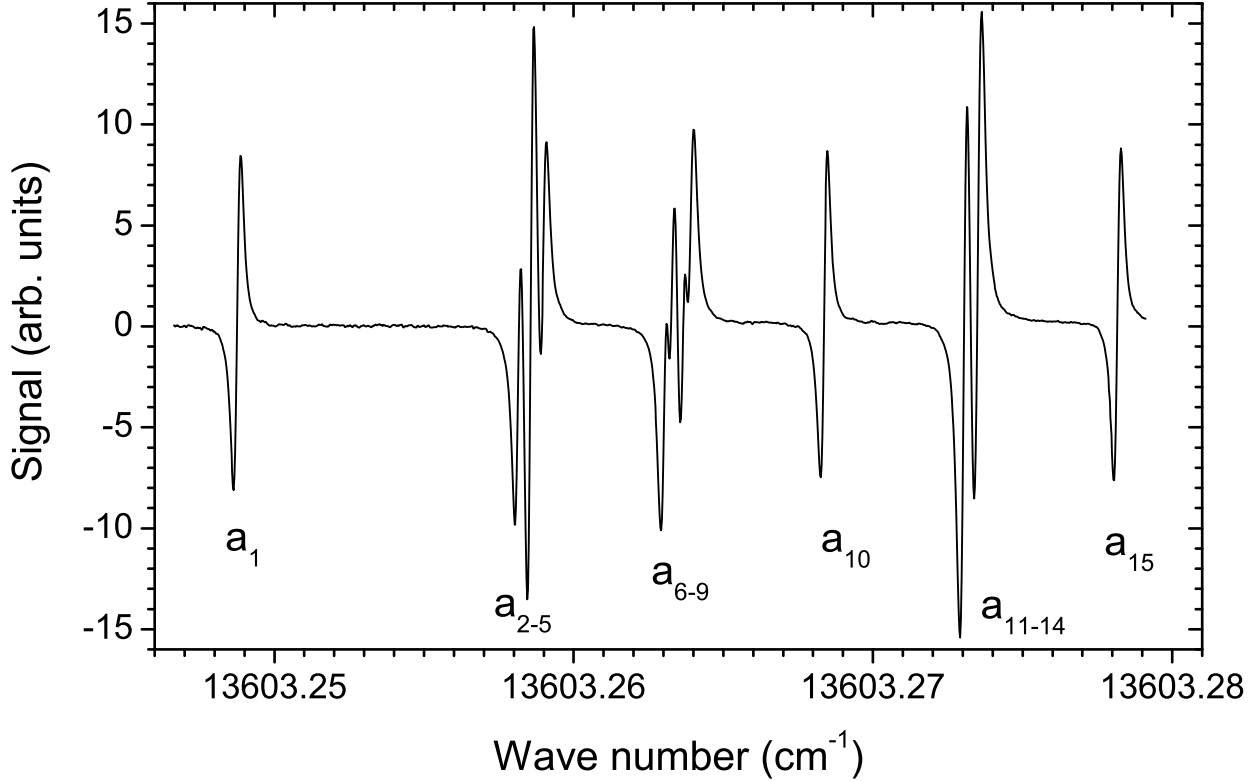}
{\caption{(Color online) Hyperfine structure spectrum of the $X^{\,1}\Sigma_{g}^{+}\rightarrow BO_{u}^{+}\,\,R$(114)~11-2 reference iodine line measured with the set-up shown in Fig.~\ref{dio_las_li11}.} \label{iod_lin_li11}}
\end{figure}

A typical spectrum of the iodine
transition obtained by scanning the unlocked diode laser across the
hyperfine resonances is shown in Fig.~\ref{iod_lin_li11}.
To lock the laser to one of the isolated hyperfine components, it was
coarsly tuned to the frequency of the I$_2$ a$_1$ transition guided
by a wavemeter. The output of the lock-in amplifier was used as
input signal for a PID regulator to produce a servo signal for the
diode laser stabilization. The a$_1$ hyperfine component of the
transition was chosen as the reference transition since it is
clearly separated from the other hyperfine components and could be
easily identified to facilitate reliable relocking during the
measurements.

To populate sufficiently the $\nu =11$ vibrational level of the
I$_2$ electronic ground state, the iodine cell is heated in an oven to
600$^{\circ}$C. The iodine reservoir is kept outside the oven in a
cold finger at a fixed temperature of $29^{\circ}$C to control the
vapor pressure inside the cell. Pressure broadening of the lines was
only observed at cold-finger temperatures well above $30^{\circ}$C,
the pressure-dependent shift of a few kHz/Pa \cite{Reinhardt2007} is
not relevant for the accuracies targeted in the measurements
reported here.

\subsubsection{The Titanium:Sapphire Laser}

\begin{figure}[tb]
\includegraphics[width=\columnwidth,angle=0]{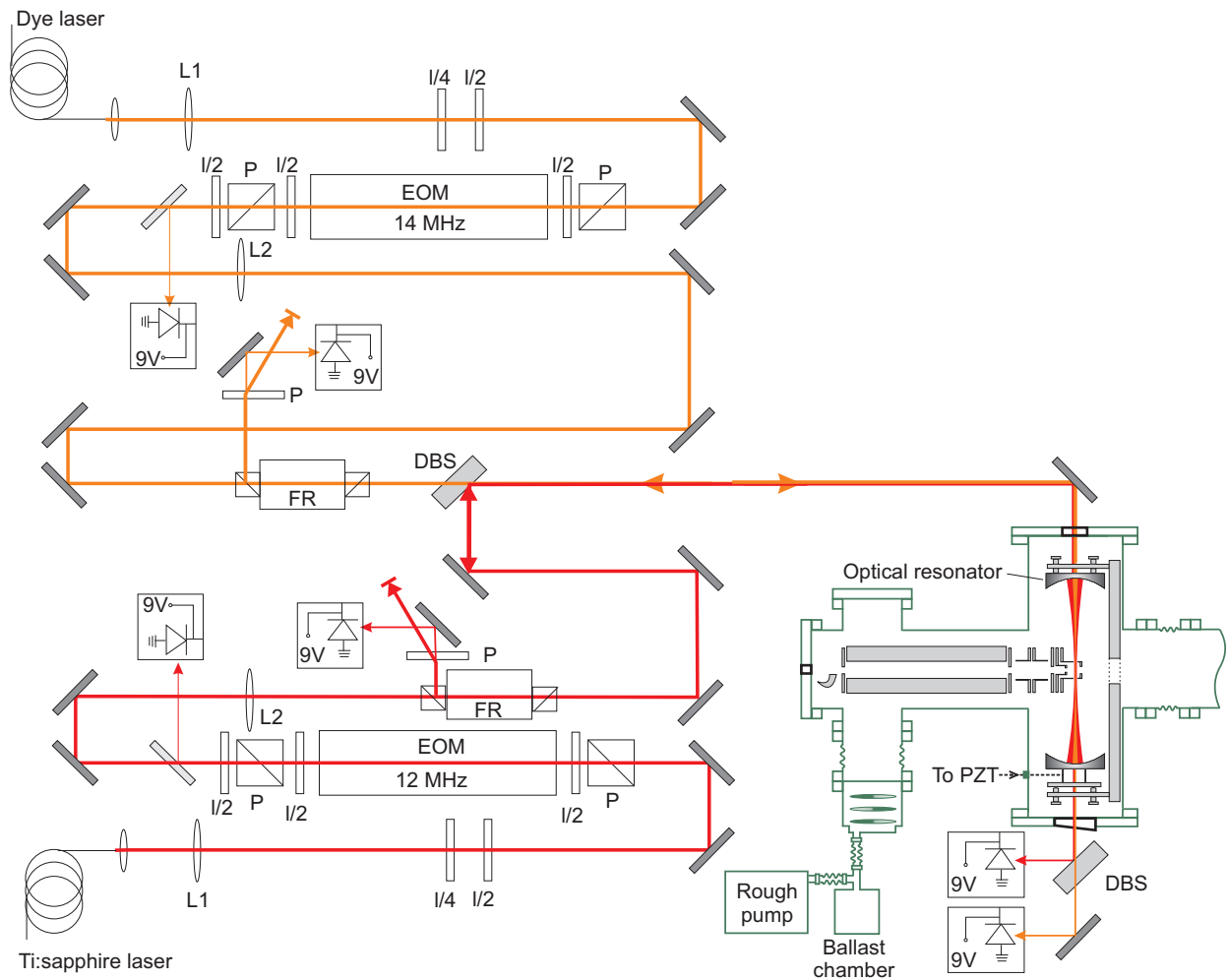}
{\caption{(Color online) Optical setup of the Pound-Drever-Hall stabilisation of
the dye laser (coming from top) and the Ti:sapphire laser (coming from the bottom) and injection of the laser beams into the experimental setup for resonance ionization spectroscopy of lithium isotopes. DBS: Dichroic beamsplitter mirrors, EOM: Electro-optical modulators for side-band generation, FR: Faraday rotators for separation of the reflected laser beams and detection with fast photodiodes, L: Mode-matching lenses, $\lambda /4$ and $\lambda /2$: Waveplates for
polarisation adaption, P: Polarizers, PZT: Piezoelectric transducer.}
\label{fig:CavityOptics}}
\end{figure}

The 735 nm light for the lithium $2s \to 3s$ two-photon transition was produced by a Coherent
899-21 titanium:sapphire (Ti:sapphire) ring laser pumped with 15 W from an
argon ion laser (multi-line visible). The Ti:sapphire laser provided up to
1~W of single-frequency output with a typical linewidth of
approximately 1~MHz. Short-term frequency fluctuations were
suppressed by locking the laser to the external Fabry-Perot cavity
that is part of the Coherent 899-21. Medium ($\geq 30$~ms) and long-term 
stabilization was achieved with a frequency-offset lock of the
Ti:sapphire laser relative to the iodine stabilized diode laser. About
1~mW of light from the reference diode laser was superimposed with a few mW
from the Ti:sapphire laser by coupling each into the two arms of a fiber-optical 
beamsplitter. One of the output arms with the mixed laser
beams was connected to the fiber-optical input of a fast photodiode
(New Focus, Model 1434, 25~GHz), while the output of the second arm
was collimated and mode matched into a 300~MHz Fabry-Perot
interferometer for spectral analysis of the diode and the Ti:sapphire laser beams. The radio frequency (RF) output signal of the fast
photodiode, {\em i.e.,} the beat signal between the Ti:sapphire laser and
the diode laser, was amplified with an ultra-wideband amplifier
(MITEQ Model JS4-001020000-30-5A) and then divided with a two-way
splitter. One part was guided to a microwave frequency counter
(Hewlett Packard, Model 5350B), while the second part was used for
frequency offset locking of the Ti:sapphire laser. To obtain the
respective servo signal, the beat frequency was mixed with the RF
output of a synthesized sweeper (Hewlett Packard, Model 83752A) and
a frequency difference of exactly 160 MHz between the sweeper and
the beat frequency was maintained. This was accomplished with a
frequency discriminator (MiTeq, Model FD-2PZ-160/10PC) that provided
an output voltage proportional to the deviation of the input
frequency from 160 MHz. Its output was fed into a PID regulator and
the correction signal applied to the external scan input of the
Coherent Ti:sapphire laser. The bandwith of the feedback loop was
approximately 30~Hz. By changing the frequency of the synthesized
sweeper, the Ti:sapphire laser could be set and stabilized to any arbitrary
frequency around the iodine reference line within the bandwidth of
the fast photodiode ($\pm$25~GHz). An upper limit for the laser linewidth could be
obtained from the frequency spectrum of the beat signal. At GSI it
was usually on the order of 1 MHz \cite{Ewald04}, while at TRIUMF a
slightly larger linewidth of 2 to 3~MHz was observed. This increase
was mainly due to the frequency jitter of the diode laser, caused by
the acoustic noise of a CAMAC crate located nearby. However, with
about 1~s integration time, the average frequency measured by the
frequency counter was stable to within a few 10~kHz.

The Ti:sapphire laser light was transported from the laser
laboratory to the experimental hall with a 25 meter long photonic-crystal fiber (Crystal Fibers, LMA-020) that can transmit high cw
powers without significant losses and without indication of
nonlinear Brioullin scattering due to its large mode area
\cite{Russel2003}. Typical transmission through the fiber was about
80\% for 1~W input Ti:sapphire laser power.

\subsubsection{Light Enhancement Cavity}
High intensities are required to approach saturation for the $2s \to 3s$ two-photon transition and hence efficient
detection. One approach that is often used in Doppler-free two-photon spectroscopy is strong focussing of two
counterpropagating laser beams to reach saturation intensity.
However, this has the disadvantage of poor spatial overlap between
the two laser beams and the atomic beam released from the graphite
catcher. Hence, we developed an optical enhancement cavity to both
ensure collinearity of the counterpropagating beams and to obtain
higher intensities. A critical point is that both laser beams, one for 
the two-photon resonant excitation and the other for the $2p \to 3d$ excitation, {\it i.e.,}
the Ti:sapphire and the dye laser beams, must be coupled to the cavity
simultaneously. In designing this cavity one has also to consider that the spatial profile of the resonator mode should provide a relatively large focus to ensure sufficient spatial overlap with the atomic beam.

Figure~\ref{fig:CavityOptics} shows the cavity together with the optical setup for stabilization. The symmetrical cavity has a length of approximately 30~cm and
mirror curvature radius of 50~cm. The diameter of the TEM$_{00}$
mode in the focus is therefore relatively large and is approximately
500~${\rm \mu}$m. The cavity is placed completely inside the vacuum
chamber with two mirror holders mounted on a vertically oriented
baseplate. The input coupling mirror above the ionization region of
the QMF has a transmittance of 98\% while the high reflector mounted below
has reflectivities of $R>99.93\%$ for both wavelengths (735~nm,
610~nm). The high reflector is fixed to a piezoelectric transducer
(PZT) for fine tuning of the cavity length.

The cavity length is actively stabilized to the Ti:sapphire laser
frequency using Pound-Drever-Hall locking
\cite{Drever1983,Black2001}. Collimating optics for the fiber,
lenses for spatial mode-matching and sideband generation with an EOM
required for the locking scheme is mounted on a breadboard on top of
the vacuum chamber. The Ti:sapphire light is first focused through an EOM
operated at 12~MHz. Similar to the diode laser modulation discussed
in Section \ref{sec:IodineLock}, an active feedback loop suppresses
residual amplitude modulation. After passing through an initial
polarizer and Faraday rotator, the main Ti:sapphire beam is superimposed
on the dye laser beam (see below) using a dichroic beamsplitter
that is highly reflective for 735~nm at $45^\circ$ incidence and
antireflection coated for $610$~nm. Behind the dichroic
beamsplitter, a broadband mirror directs the combined laser beams
vertically through an antireflection-coated viewport into the vacuum
chamber and to the enhancement cavity. Light reflected by the input
coupler of the cavity is separated at the input polarizer after
returning through the Faraday rotator and detected with a fast
Si-PIN photodiode. Depending on the resonance condition of the
cavity, the two sidebands created by the EOM modulation exhibit
different phase shifts and a dispersionlike signal is obtained after
demodulation of the photodiode signal at the EOM frequency. This
signal is used to generate a PID-regulated servo signal (0-500 V),
which is applied to the PZT to change the cavity length. This
stabilizes the cavity length to the Ti:sapphire frequency and tracks it
while tuning across the resonance transitions of the different
lithium isotopes. The effective tuning range of the cavity for PZT
voltage variations of 500~V is about 600~MHz and thus sufficient to
cover the hyperfine structure of each lithium isotope (typically 300
MHz) without relocking the servo loop.

Such an optical resonator is extremely sensitive to vibration because the distance of the mirror positions must be stabilized within a small fraction of the laser wavelength, {\it i.e.} less than 10 nm.
To decouple the vacuum chamber with the optical cavity from
vibrations caused by vacuum pumps and other mechanical devices, the
two cross pieces housing the experiment (Fig.~\ref{fig:CavityOptics}) are connected to the beam
line only through a flexible metal bellow, while the roughing pump
is isolated from the chamber turbo pump by flexible tube couplings
to and from a massive  ballast chamber mounted on the floor.

Fluctuating light intensity inside the cavity has a strong influence
on the observed lineshape for two reasons: The quadratic dependence of
the excitation efficiency increases the sensitivity of the signal intensity to vibrationally induced power changes. Even more important, changing intensities cause fluctuating ac Stark shifts of the transition
frequencies. These are discussed in more detail below. Therefore it is
important to monitor the light power contained within the resonant
cavity. To do so, the light transmitted through the high reflector,
which is about 0.05\% and 0.07\% of the power inside the
cavity for 735~nm and 610~nm, respectively, leaves the vacuum
chamber through a second viewport. A second dichroic beamsplitter
separates the two wavelengths and send them to two
photodiodes where their intensity is recorded. From this averaged signal a power enhancement of about
80 to 100 was determined for the cavity.

\subsubsection{Dye Laser}
Laser light at 610~nm is required to drive the $2p\;^2{\rm P}_J \to 3d\;^2{\rm D}_{J^\prime}$ transitions in lithium. This was produced by a Coherent 699-21 dye
laser operated with a Rhodamine 6G solution in ethylene glycol and
pumped with 6~W of multiline visible output of a second argon ion
laser.  The laser light was transported from
the laser laboratory to the experimental hall with a second
large-mode-area photonic-crystal fiber (LMA-020). In this case,
transmission efficiencies up to about 70\% for 600 mW input dye
laser power were achieved. This light had also to be resonantly
coupled into the optical cavity together with the Ti:sapphire laser
light. Hence, the dye laser was locked to a longitudinal cavity mode that was as close as possible to the $2p\;^2{\rm P}\to 3d\;^2{\rm D}$ transition frequency. This was achieved
using a second Pound-Drever-Hall servo loop depicted in
Fig.~\ref{fig:CavityOptics}, but the servo signal was applied to the
external-scan input of the Coherent 699 dye laser controller (rather
than to the PZT controlling the length of the enhancement cavity). This approach does not allow tuning
the dye laser exactly to the $2p\to 3d$ resonance
transitions, but the high laser intensities caused strong power
broadening of the allowed dipole transition, as will be discussed in
the next Section. With typical power broadened linewidths of 7~GHz
(FWHM) and a free spectral range of the optical cavity of 500 MHz it
was always possible to find a locking point that provided maximum
excitation efficiency and kept the dye laser frequency in a region of constant excitation efficiency.

\subsection{Data Acquisition}
\label{sec:dat_acq_li11}

Data acquisition (DAQ) of the experiment was based on the
Multi-Branch System (MBS) developed at GSI. A CAMAC-GPIB controller allowed
communication with the RF synthesizer and the microwave counter. Digital-to-analog, analog-to-digital converters and scalers were directly implimented in the CAMAC crate. The MBS system worked stand-alone. It read all data from a CAMAC crate and wrote it on a local tape drive. No other computers were necessary to take data and to store them. For analysis, the data of each scan was transferred via a TCP/IP connection into the data analysis package Origin 7.0, running on a PC. Routines written with C++ in the Origin environment were finally used to analyze the data. Nonlinear least-square fits of lineshapes were performed with a Levenberg-Marquardt algorithm adapted from Numerical Recipes \cite{Press1996}.

\section{Results and Discussion}
\label{sec:Results}
We present and sumarize results and data from the preparatory experimental phase at GSI and TRIUMF and the four beamtimes that were performed at these facilities: an on-line beamtime at GSI in December 2003, an off-line beamtime at TRIUMF in June 2004, and two on-line beamtimes at TRIUMF in September and October 2004.

\subsection{Lineshapes}

Determination of the isotope shift on the level of 100~kHz or better
requires a detailed understanding and description of the resonance
lineshape and all factors affecting it. Hence, this Section
starts with an analysis of the observed lineshapes and the influence of
the ac Stark effect in the $2s\;^2{\rm S}_{1/2} \rightarrow 3s\;^2{\rm S}_{1/2}$ as well as in the $2p\;^2{\rm P}_J \rightarrow 3d\;^2{\rm D}_{J'}$ transitions. Afterwards, the measurements of the hyperfine structure (HFS) and isotope shift (IS) of the stable and short-lived isotopes are discussed.

\subsubsection{$2s\rightarrow3s$ Two-Photon Transition}
Resonance profiles of the $2s \rightarrow 3s$ transition were
recorded in the following way: First, the Ti:sapphire laser was tuned to a
frequency below the $2s \rightarrow 3s$ resonance of the respective
isotope and frequency-offset locked to the iodine-stabilized diode
laser. Then, the enhancement cavity was locked to the Ti:sapphire laser
and finally the dye laser was locked to the longitudinal mode of the
cavity closest to the $2p \rightarrow 3d$ transition frequency of
the respective isotope. The Ti:sapphire laser was scanned by slowly
varying the RF frequency for the frequency-offset lock in steps of
1~MHz. Maximum scan ranges were about 600~MHz, limited by the
voltage range of the high-voltage power supply for the piezoelectric
transducer (PZT) at the enhancement cavity. Since the HFS of all lithium isotopes in the two-photon
transition is smaller than this scanning range the complete resonance structure can be covered with a single scan.

\begin{figure}[ht]
\includegraphics[width=\columnwidth, clip=]{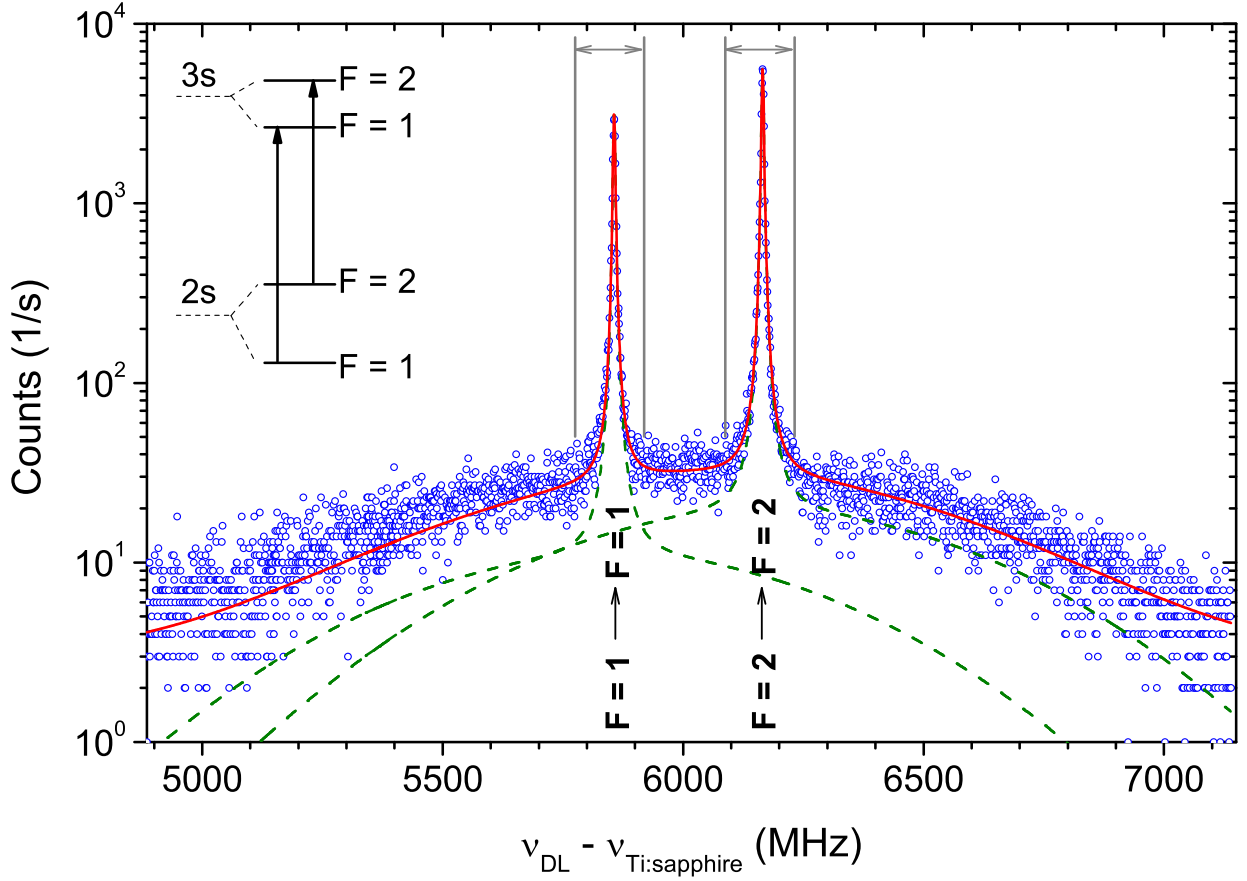}
\caption{\label{fig:LiLineProfiles} (Color online) Wide-range line profile
of the $^{7}$Li $2s \to 3s$ two-photon transition with
narrow Doppler-free hyperfine transitions and the Doppler-broadened
background as observed off-line at GSI. On the $x$-axis, the beat frequency between the iodine-stabilized diode laser and the Ti:sapphire laser is given. The solid (red) line is a fit of two Voigt profiles including a wide Doppler background pedestal. The dashed (green) line shows the individual contributions of the two hyperfine components. The regions marked by vertical lines are the scanning regions in $^7$Li on-line experiments. The inset shows schematically the hyperfine splitting in the $2s$ and $3s$ states of the $I=3/2$ isotopes $^{7,9,11}$Li and the allowed ($\Delta F=0$) two-photon transitions.}
\end{figure}

Figure \ref{fig:LiLineProfiles} shows an overall resonance profile
obtained for $^{7}$Li. The observed count rate at the
channeltron-type detector is plotted as a function of the measured beat
frequency between the iodine-locked diode laser and the Ti:sapphire
laser. Here, a large tuning range was used to record also the
far-reaching wings of the resonance profile. Therefore the complete
scanning range of about 2~GHz was divided into scans of
approximately 400~MHz range in order to stay inside the tuning
range of the cavity PZT and to limit the
frequency variation of the dye laser (see below). After each 400 MHz scan,
the cavity and the dye lasers were relocked to keep the dye laser frequency as
close to the $2p \to 3d$ resonance frequency as possible.

The profile exhibits two narrow Doppler-free components that are
labeled with their $F$ quantum numbers. According to the
selection rules for a two-photon transition between $s$
states, only hyperfine transitions with $\Delta F = 0$ are allowed.
The wide Doppler-broadened background with a relative amplitude of
about 1\% of the peak intensity and a Gaussian width of approximately 1.7~GHz is caused by
the absorption of two photons from copropagating beams
\cite{Grynberg77}. To account for this process, each of the two
peaks is fitted by applying a Levenberg-Marquardt minimization
procedure with a Voigt plus a background Gaussian function
represented by the dashed (green) lines. Both HFS components are constrained to have identical width parameters for the Voigt as well as the Gaussian lineshape of the background. The Voigt profile shows a
Lorentzian linewidth $\Gamma_{\rm L}$ of 4.5~MHz which is already
slightly larger than the natural linewidth of 2.6~MHz, attributed to
saturation broadening, and a Gaussian linewidth $\Gamma_{\rm G}$ of
1~MHz, fitting well to the observed laser linewidth. Here, all width parameters refer to the Ti:sapphire laser frequency scale and have to be doubled to obtain the value for the $2s\to 3s$ two-photon transition. The solid (red)
line is the overall fitting function which shows an excellent
agreement with the experimental data points over more than 2 GHz
frequency range and about three orders of magnitude in signal intensity.

\begin{figure*}
\includegraphics[width=\textwidth, clip=]{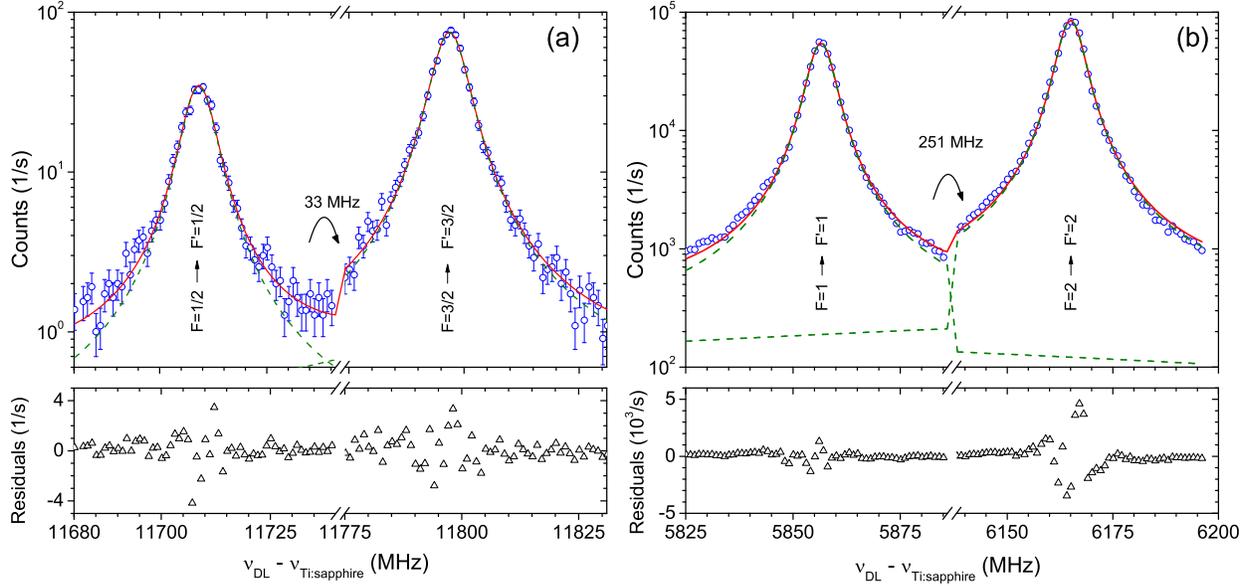}
\caption{\label{fig:Li6Li7SkipSpectra} (Color online) $2s \to 3s$ two-photon resonances for (a) $^{6}$Li obtained at GSI and for (b) $^{7}$Li recorded at TRIUMF. The $x$-axis break indicates the intermediate region that was skipped. The solid (red) line is the common fit function while the dashed (green) lines are the contributions of the individual hyperfine components.
The residuals (differences between the experimental points and the common fit function) are plotted at the bottom.}
\end{figure*}

For precise frequency determination, only the regions around the two
Doppler-free peaks are important and scans of the radioactive
isotopes were thus usually performed by scanning about $\pm30$ or
$\pm15~$MHz around the resonance centers, as indicated in Fig.~\ref{fig:LiLineProfiles}, skipping the intermediate
part in fast steps without data taking. Such spectra of $^6$Li and
$^7$Li, taken at the GSI on-line mass separator and the Off-Line
Isotope Separator (OLIS) at TRIUMF, respectively, are depicted in
Fig.~\ref{fig:Li6Li7SkipSpectra}. The incoming ion yield was about
$3\cdot 10^{5}$ ions/s for $^{6}{\rm Li}$ and $2\cdot 10^{8}$ ions/s for $^{7}{\rm Li}$, while approximately 80 and 10$^5$ ions/s were obtained in resonance on the strongest hyperfine transition. This corresponds to overall
efficiencies of $\approx 2 \cdot 10^{-4}$ at GSI and $5\cdot
10^{-4}$ at TRIUMF. The observed efficiencies agree quite well with
those calculated and estimated during the design of the setup as listed in Table~\ref{tab:DetecEff}.

\begin{table*}
\caption{\label{tab:DetecEff}Estimated partial efficiencies and
expected overall efficiency for the detection of lithium ions compared with that experimentally observed.}
  \begin{ruledtabular}
    \begin{tabular}{ll}
      Release efficiency of the catcher foil                          & 50 \% \\
      Overlap between laser beams and atomic beam (diam. 0.5 mm)      & 20 \% \\
      Excitation efficiency for the $2s \to 3s$ two-photon transition & 25 \% \\
      Ionization efficiency via $2p\to 3d \to {\rm continuum}$        &  3 \% \\
      Signal reduction by HFS splitting                               & 62 \% \\
      Transmission of the quadrupole mass filter                      & 90 \% \\
      Quantum efficiency of the detector                              & 80 \% \\
      \hline
      Expected overall efficiency on resonance                        & $2.5 \cdot 10^{-4}$\\
      \hline
      Experimental overall efficiency at GSI                          & $2 \cdot 10^{-4}$\\
      Experimental overall efficiency at TRIUMF                       & $5 \cdot 10^{-4}$
    \end{tabular}
  \end{ruledtabular}
\end{table*}

Before fitting the recorded spectra, the observed number of ion events ($N_{\rm Raw}$) was normalized for each channel with the Ti:sapphire laser power ($P_{\,\rm Ti:sapphire}$) that was recorded with the photodiode located behind the enhancement cavity (see Fig.~\ref{fig:CavityOptics}) while the atoms were irradiated with the constant laser frequency:
\begin{equation}
{N}_{\rm Norm} = {N}_{\rm Raw} \cdot \left(\frac{P_{\,\rm
Ti:sapphire}}{\langle P_{\,\rm Ti:sapphire} \rangle} \right)^{-2}.
\label{eq:nor_tis_li11}
\end{equation}
Here $\langle P_{\,\rm Ti:sapphire} \rangle$ is the average Ti:sapphire laser
power while recording the complete spectrum. The uncertainty of $N_{\rm
Norm}$ was calculated from the statistical uncertainty and from the
laser power fluctuations using Gaussian error propagation. To
confirm the normalization function, a $\chi^2$ optimization was
performed using different normalization exponents and checking for
the lowest $\chi^2$ value in the subsequent lineshape fitting. The
optimum was found to scatter between 1.7 and 2.8, and
hence the theoretically expected factor of 2 seems to be well
justified.

The red fit curves are Voigt profiles
including a Doppler background with a width fixed to 1.7~GHz as obtained from
Fig.~\ref{fig:LiLineProfiles}. The Lorentzian and Gaussian
linewidths of the narrow Voigt profile as well as the relative intensity of the Doppler background are constrained to be equal
for both hyperfine structure components. The width of the Gaussian pedestal was varied within a reasonable range ($\approx$ 1 -
3~GHz) and does not show considerable influence on the fitted peak
centers. The fitting curves of the individual peaks are indicated by the green dashed lines. This
procedure was used for all isotopes.

In the lower part of Fig.~\ref{fig:Li6Li7SkipSpectra}, the residuals
of the fit are plotted. A slight systematic asymmetry is visible in the $^{7}$Li spectrum while it is much less pronounced in the residuals of the $^6$Li resonance fit. In later experiments performed off-line at GSI, a much stronger and clearly visible asymmetry of the peaks was observed which is discussed in detail in a recent publication
\cite{Sanchez2009}. There it is shown that the asymmetry is caused
by the Gaussian profile of the laser beam. Atoms passing through the laser beam in the interaction region experience an
intensity-dependent ac Stark shift as will be discussed in the next
section. Therefore, the exact atomic resonance frequency depends on the
position within the laser beam where the excitation and ionization occur. This line profile distortion is similar for all
isotopes under identical experimental conditions. Therefore, the line shape
calculations described in \cite{Sanchez2009} show that the
influence of this asymmetry on the extracted isotope shift is very small. However, its contribution is significant if the total transition frequency is to be extracted. Then a total correction by about 160~kHz is required \cite{Sanchez2009}.

To calculate the isotope shift, the resonance positions of the
individual hyperfine components obtained from the fit must be
converted into center of gravity (c.g.) frequencies. Since the nuclear
quadrupole moment does not affect the $J=1/2$ states in the
transition, only the magnetic hyperfine interaction must be considered. The energy shift of the hyperfine state with angular
momentum $F$ relative to the $J$-level energy is in first order
given by
\begin{equation}
E_{\mathrm{HFS}} = \frac{A}{2}C_{F} = \frac{A}{2} \left[ F(F+1)-J(J+1)-I(I+1)\right]
\end{equation}
with the Casimir factor $C_{F}$ and the magnetic dipole hyperfine
constant $A$. In first-order perturbation theory, the hyperfine
structure cg coincides with the unperturbed $J$-level energy and can
be calculated from the two hyperfine resonances in the  $2s \to 3s$ transition of lithium according to
\begin{equation}
\nu_{\mathrm{cg}}=\frac{C_{F}~\nu_{F^{\prime}}-C_{F^{\prime}}~\nu_{F}}
{C_{F}-C_{F^{\prime}}}, \label{eq:Licg}
\end{equation}
where $\nu_{F}$ is the transition frequency of the $F
\to F$ transition. For $^{7,9,11}$Li with nuclear spin
$I=3/2$, this leads for example to
$\nu_{\rm cg}=\frac{5}{8}\nu_{2}+\frac{3}{8}\nu_{1}$.

\begin{figure}[b]
\includegraphics[width=0.7\columnwidth,angle=0]{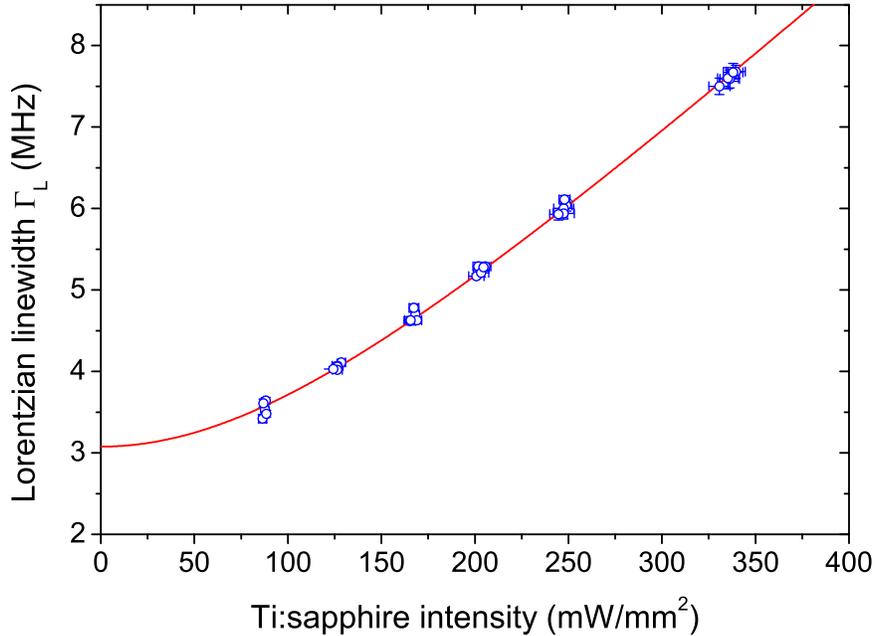}
\caption{\label{fig:power_broad} (Color online) Lorentzian linewidth of $^6$Li resonances as a
function of the Ti:sapphire laser intensity as obtained in off-line experiments at TRIUMF. The solid line is a fit by $\Gamma_{\rm L} = \Gamma_{0}\sqrt{1+(I/I_{\rm Sat})^2}$, describing the saturation for a two-photon transition. The Ti:sapphire laser intensity is calculated based on the laser power behind the high reflector of the enhancement cavity as described in the text. Please note that the linewidth refers to the Ti:sapphire laser frequency and must be doubled for the full $2s \to 3s$ transition.}
\end{figure}

Figure \ref{fig:power_broad} shows the Lorentzian linewidth of the two-photon transition as a function
of the Ti:sapphire laser intensity inside the cavity $\langle I_{\rm
Ti:sapphire} \rangle$. The
intensity was calculated from the power transmitted through the high
reflector of the enhancement cavity with a measured cavity mirror
transmission of 0.05(1)\% and a calculated cavity mode diameter of 0.46
mm.
The solid red line is a fit of the function for the power-broadened
linewidth of a two-photon transition
\begin{equation}
\Gamma_{\rm L}=\Gamma_{\rm 0}\sqrt{1+\left(\langle I_{\,\rm Ti:sapphire}\rangle/
I_{\rm Sat}\right)^2}
\end{equation}
to the data points, where $\Gamma_{\rm 0}=3.08(2)$~MHz is the
extrapolated natural linewidth for vanishing intensity of the Ti:sapphire laser and ${I}_{\rm Sat}=148(2)$ W/mm$^2$ is the saturation intensity. These results are in good agreement with values obtained in a previous
beamtime at GSI, where $\Gamma_{\rm 0}=3.2(1)$~MHz and ${I}_{\rm
Sat}=167(6)$~W/mm$^2$ were obtained. The uncertainty of the absolute
value of ${I}_{\rm Sat}$ is solely the fitting uncertainty and includes neither the uncertainty in the
mirror transmission and the transmission through the entrance
window\footnote{The window transmission shows an etalon effect
discussed below and changes slightly as a function of the laser wavelength.}
nor the uncertainty of the effective diameter of the laser focus. Hence
an additional uncertainty of about 20\% should be
added if these values are to be compared with calculations or other measurements. The small deviation of the fit value $\Gamma_0$ from the natural linewidth $\Gamma_{\rm 0\,nat}=2.65$ MHz is either
caused by laser intensity fluctuations inside the cavity, resulting in
varying ac Stark shifts for the atoms of the ensemble, or an artefact
from the fit program.

\subsubsection{$2p \rightarrow 3d$ Resonance}
\begin{figure*}
\includegraphics[width=\textwidth,angle=0]{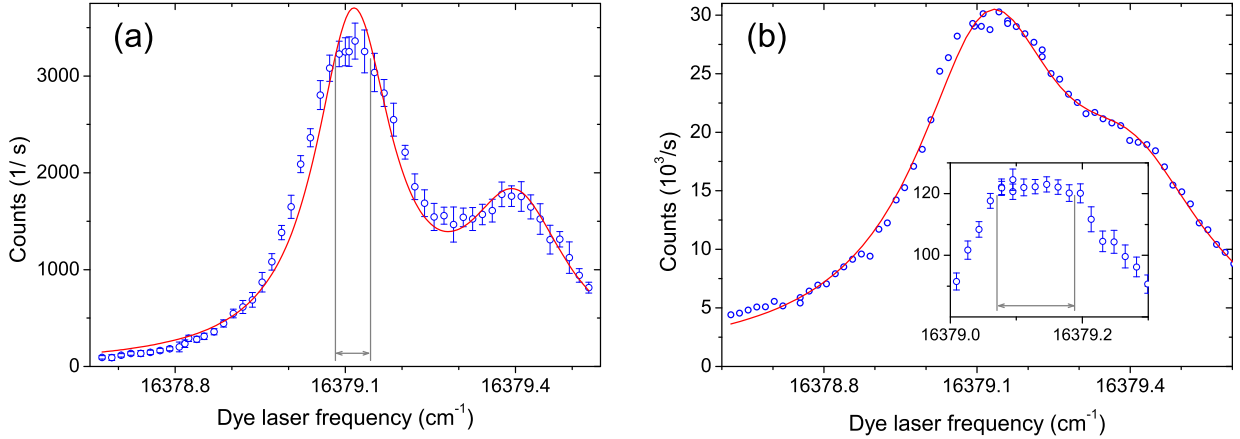}
\caption{\label{fig:li6_pd_tra_jun} (Color online) Strongly saturated resonance profiles of the $2p\to 3d$ transition in $^7$Li as obtained with 20 mW laser power entering the resonator at GSI (a) and with 80~mW at TRIUMF (b). Frequencies were measured with a Fizeau-type interferometer and have uncertainties of about $3\cdot 10^{-3}$ cm$^{-1}$. Circles indicate the experimentally obtained signal and the solid line is a fit of two Lorentzian line profiles with identical widths to the data. This curve is only to guide the eye since the strong saturation and the unresolved $^2$D$_{3/2,5/2}$ levels lead to a distorted line shape. The flat-top regions used for spectroscopy in the $2s\to 3s$ transition are indicated by vertical lines. The inset in (b) is the result of another scan with higher statistics around the flat-top region.}
\end{figure*}

The lineshape of the two-photon transition is distorted if the
ionization efficiency along the $2p \to 3d  \to {\rm Li}^+$ ionization path changes during the scan. This could
happen because the change in the Ti:sapphire laser frequency during a scan of the $2s\to 3s$ transition
induces also a change of the dye laser frequency. This cannot be
avoided since the dye laser is locked to the enhancement cavity and
the resonator length must be changed when scanning the Ti:sapphire laser
frequency. Hence, the dye laser frequency is changed typically by a few hundred~MHz during the scan. Helpful is
here the strong power broadening of the $2p \to 3d$
transition. The high intensity of the 610~nm light in the cavity
focus considerably broadens the linewidth of the $2p \to 3d$ transition. This broadening was studied to find
maximum excitation efficiency and to ensure a constant ionization
efficiency along the whole scan. The results for different laser powers as they were available at TRIUMF and GSI are shown in Fig.~\ref{fig:li6_pd_tra_jun}. The main reason for the different power levels is the fiber transport from the laser laboratory to the experimental hall. While a 50 m long standard single-mode fiber was used for the dye laser at GSI, a large mode area (LMA) photonic crystal fiber was applied at TRIUMF with a transport distance of 25 m. Hence 20 mW of dye laser light were coupled to the enhancement cavity at GSI while 80 mW were available at TRIUMF.

The resonance profiles in Fig.~\ref{fig:li6_pd_tra_jun} were obtained in the following way:
First the QMF was set to detect photo ions of the respective isotope and the Ti:sapphire laser frequency
was fixed at the resonance of the strongest hyperfine transition and operated at about 25\% of the maximum
achievable power. Then, the frequency of the dye laser was set to a
value clearly below resonance and changed manually until a
longitudinal mode of the enhancement cavity was reached. There, the dye laser
frequency was locked to the cavity and resonant laser ions were detected
for a period of 3~s before the dye laser was taken out of
lock and its frequency changed until the next longitudinal cavity mode was reached. The free spectral range of
the cavity was approximately 500~MHz which determined the step size.
The intensity of the incoming Li$^{+}$ ion beam as well as of the
Ti:sapphire laser was sufficiently constant during the measurements and a
normalization of the count rate therefore not required.

In Fig.\,\ref{fig:li6_pd_tra_jun} the detected ion count rates (circles) are plotted as functions of the dye laser frequency
that was determined with a Fizeau-type wavemeter. The solid line is a fit of two Lorentzian profiles with equal widths to the data and serves only to guide the eye. The profile (a) obtained at GSI with lower dye laser power shows a flat-top region with an extension of about 1.5 GHz. That is sufficient to ensure constant ionization efficiency since the scan width of the Ti:sapphire laser was less than 500~MHz for all isotopes. In (b), recorded with the higher dye laser power that was available at TRIUMF, the flat region extends to approximately 0.1 cm$^{-1}$ (3~GHz) width, ideal for looking the dye laser during a scan. The region of the first resonance agrees roughly with the reported transition frequencies of $16\,379.0661$~cm$^{-1}$ and $16\,379.1021$~cm$^{-1}$ \cite{Radziemski95} for the $2p\;^2{\rm P}_{3/2}\to 3d\;^2{\rm D}_{3/2}$ and $2p\;^2{\rm P}_{3/2}\to 3d\;^2{\rm D}_{5/2}$ resonances in $^7$Li, respectively. Obviously, the dye laser power at TRIUMF was sufficient to even excite the tails of the $2p\;^2{\rm P}_{1/2} \to 3d\;^2{\rm D}_{3/2}$ transition at $16\,379.4014$~cm$^{-1}$ \cite{Radziemski95}. Contrary, in the measurements at GSI (Fig.~\,\ref{fig:li6_pd_tra_jun}(a)), where less dye laser power was available, the $2p\;^2{\rm P}_{1/2} \to 3d\;^2{\rm D}_{3/2}$ transition is still clearly resolved. Thus, we can expect the overall efficiency at TRIUMF to be about 30\% larger than that obtained at GSI since the dye laser will also excite and ionize the fraction of atoms decaying from the $3s\;^2{\rm S}_{1/2}$ to the $2p\;^2{\rm P}_{1/2}$ state. This is in accordance with the higher efficiencies obtained at TRIUMF as reported above.

The frequency of the $2p\;^2{\rm P}_{3/2} \to 3d\;^2{\rm D}_{5/2}$ resonances for the unstable isotopes were
calculated from the isotope shift formula
\begin{equation}
  \delta \nu^{AA^\prime}= K_{\rm MS} \cdot \frac{M_A M_{A^\prime}}{M_A-M_{A^\prime}},
\end{equation}
with $K_{\rm MS}=117.812$~GHz$\cdot$amu being the mass shift coefficient obtained from the known $^{6,7}$Li isotope shift data \cite{Radziemski95} and
neglecting all field shift contributions. The calculated resonance frequencies that were used for setting the dye laser frequency are listed in Table~\ref{tab:MS2P3D}.

\begin{table*}
\caption{\label{tab:MS2P3D} Mass shift
$\delta\nu^{6,A}_{2p-3d,{\rm MS}}$ and resonance frequencies
$\nu_{2p-3d}$ of the $2p\;^2{\rm P}_{3/2} \to 3d\;^2{\rm D}_{5/2}$ transition in lithium isotopes based on the $^{6,7}$Li
transition frequencies reported in \cite{Radziemski95}. }
  \begin{ruledtabular}
    \begin{tabular}{cccccc}
    Isotope  & $^6$Li & $^7$Li & $^8$Li & $^9$Li & $^{11}$Li\\
    \hline
    $\delta\nu^{6,A}_{2p-3d,{\rm MS}}$ (MHz)& 0 &  2\,794 & 4\,901 & 6\,535 & 8\,918\\
    $\nu_{2p-3d}$ (cm$^{-1}$)& 16\,379.0089 & 16\,379.1021 & 16\,379.1724 & 16\,379.2269 & 16\,379.3064
    \end{tabular}
    \end{ruledtabular}
\end{table*}

\subsection{Ac Stark Shift}

Since atoms experience relatively strong fields when crossing the
laser beam focus in the resonator, the atomic level energies
will be altered by ac Stark shifts. To correct for these shifts,
spectra were recorded at different light powers.
\begin{figure*}
  \includegraphics[width=\textwidth,angle=0]{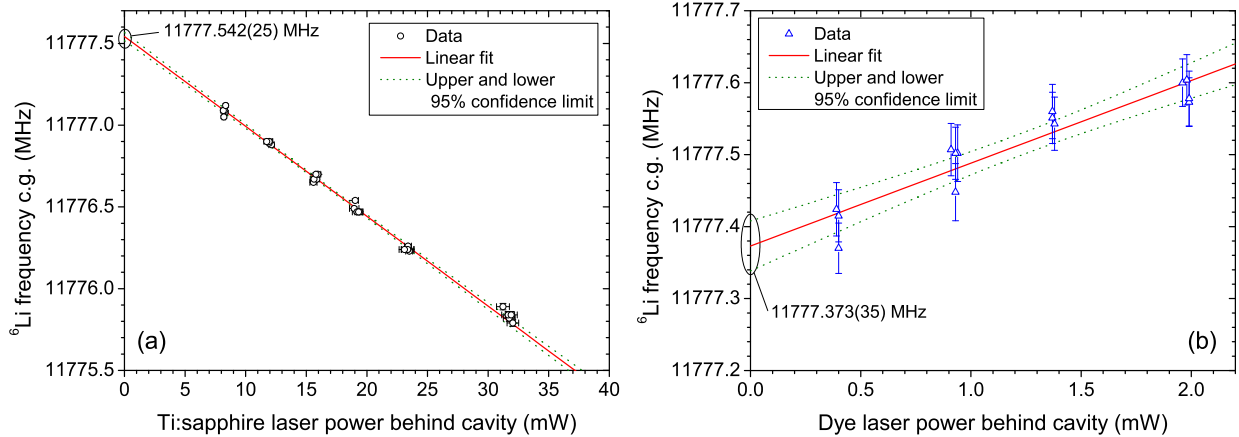}
  \caption{\label{fig:ac_stark_6li}
  (Color online) Ac Stark shift in the $2s \to 3s$ transition of $^6$Li caused by (a) the Ti:sapphire laser light at 735~nm and (b) the dye laser light at 610~nm. In (a) the power of the dye laser was fixed to 50 mW in front of the cavity and in (b) the power of the Ti:sapphire laser was fixed to $330$ mW during the measurements. The solid lines are linear regresion lines while the dashed lines indicate the 95\% confidence bands of the fit.}
\end{figure*}
To distinguish
between the effects caused by the Ti:sapphire and the dye laser light, the power of one of the
lasers was varied while keeping the other laser intensity constant
during the measurements. The observed cg frequencies were plotted
against the power level of the respective laser detected on the
photodiodes behind the cavity high-reflector. These values were not converted into intensities at the laser focus since this conversion includes a large uncertainty due to the mirror and window transmission functions and the effective laser diameter inside the focus. Figure~\ref{fig:ac_stark_6li}(a) shows the $^6$Li cg frequency as a
function of the Ti:sapphire (a) and the dye laser power (b). In both cases
a linear relation is observed as is expected for an off-resonance ac Stark shift. Hence, the linear function
\begin{equation}
\nu_{\rm cg} = b_{0} + b_{1} \langle P \rangle~ \label{eq:lin_acs_li11}
\end{equation}
is fitted to the data points from which the slopes
\begin{eqnarray}
b_{1}^{{\rm Ti:sapphire}} &=& -0.0549(14) \; {\rm MHz/mW \hspace{1ex} and} \label{eq:b_TiSa}\\
b_{1}^{{\rm Dye}} &=& 0.115(14) \; {\rm MHz/mW} \label{eq:b_dye}
\end{eqnarray}
are obtained. In both fits of Fig.~\ref{fig:ac_stark_6li}, the 95\% confidence bands are indicated.
The dye and the Ti:sapphire laser intensities influence the level energies
in opposite directions, resulting in the different signs of the
slope. Even though the sensitivity on the dye laser intensity is
stronger, the absolute shift that it causes is much smaller due to
the relatively low power levels compared with the Ti:sapphire laser.
Varying from highest to lowest applied power, the Ti:sapphire laser light shifts
the $2s \to 3s$ frequency by about 1.5~MHz, the dye laser light only
by 150~kHz. Hence, it was concluded that it is sufficient to keep the
dye laser power as constant as possible during all measurements
whereas such power series measurements were performed with the Ti:sapphire laser for all isotopes excluding $^{11}$Li. For the latter isotope the production rates were not high enough to produce sufficient statistics at
low laser power. A remaining small influence of the ac Stark shift, caused by power
variations of the dye laser while taking the Ti:sapphire laser power-series, was included in the
uncertainty of the isotope shifts as discussed below.

\subsection{Hyperfine Structure and Isotope Shifts}
A large number of measurements on the stable isotopes was performed
during the preparation of the experiments and the four beamtimes
that were performed at GSI and TRIUMF. Spectra
like those shown in Fig.~\ref{fig:Li6Li7SkipSpectra} were
analyzed by fitting the lineshapes as discussed above and
determining the c.g. frequency $\nu_{\rm cg}$ and the
hyperfine splitting $\Delta \nu^{\rm HFS}_A$ from the peak
positions. Spectra of the radioactive isotopes will be presented in
the following Sections.

\subsubsection{Hyperfine Structure}
The hyperfine structure splitting in an atomic transition is determined in first-order perturbation theory by the following nuclear ground state properties: spin $I$, magnetic moment $\mu$, and the spectroscopic electric quadrupole moment $Q_s$. Values of those are listed for all lithium isotopes in Table~\ref{tab:hfsli}.
\begin{table*}
 \caption{\label{tab:hfsli}
Nuclear spins and parity $I^\pi$, experimental dipole $\mu_I$ and spectroscopic quadrupole
moments $Q_s$ of the investigated lithium isotopes from \cite{Borremans05,Neugart08}. $A_{2s}$ values are taken from
\cite{Arimondo1977} for the stable isotopes and are calculated based on the ratio of the magnetic moments for the short-lived
ones. }
  \begin{ruledtabular}
    \begin{tabular}{llllll}
      Property  & \mbox{$^6$Li} & \mbox{$^7$Li} & \mbox{$^8$Li} & \mbox{$^9$Li} & \mbox{$^{11}$Li}\\
      \hline
      $I^\pi$       &  $1^+$        & $3/2^-$       & $2^+$         & $3/2^-$       & $3/2^-$     \\
      $\mu_{I}~(\mu_{N})$
                &  0.8220473(6) & 3.2564268(17) & 1.653560(18)  & 3.43678(6)    & 3.6712(3)       \\
      $Q_s$ (mb)  & -0.806(6)     & -40.0(3)      & +31.4(2)      & -30.6(2)      & (-)33.3(5)      \\
      $A_{2s}$ (MHz)
                & 152.136840(3) & 401.752044(3) & 153.017(3)    & 424.031(7)    & 452.954(37)
  \end{tabular}
  \end{ruledtabular}
\end{table*}
Additionally, the magnetic hyperfine constant $A_{2s}$ in the $2s\;^2{\rm S}_{1/2}$ electronic ground state, which is directly proportional to the magnetic moment, is included in the table. For the stable isotopes they are well known from atomic
beam magnetic resonance measurements \cite{Arimondo1977} and for the short-lived isotopes they are calculated based on the ratio of the magnetic moments  taken from \cite{Borremans05,Neugart08}.

The spectra of the two-photon transitions recorded here, provides only information about the hyperfine splitting $\Delta \nu^{\rm HFS}$ between the $F=I+1/2 \to F=I+1/2$ and the $F=I-1/2 \to F=I-1/2$ transitions. Separate determinations of the hyperfine splitting in the ground and the excited state are therefore not possible. However, using the information listed in Table~\ref{tab:hfsli}, the hyperfine splitting in the excited  $3s\;^2{\rm S}_{1/2}$ level can be extracted according to
\begin{equation}
A_{3s} (^A{\rm Li}) = A_{2s}(^A{\rm Li}) - \frac{\Delta \nu^{\rm HFS}_A}{I+J}.
\label{eq:A3s}
\end{equation}
and the results will be used to check for an indication of hyperfine anomaly which might be expected for such an extended nucleus like $^{11}$Li.

\begin{figure}[btp]
    \includegraphics[width=\columnwidth,angle=0]{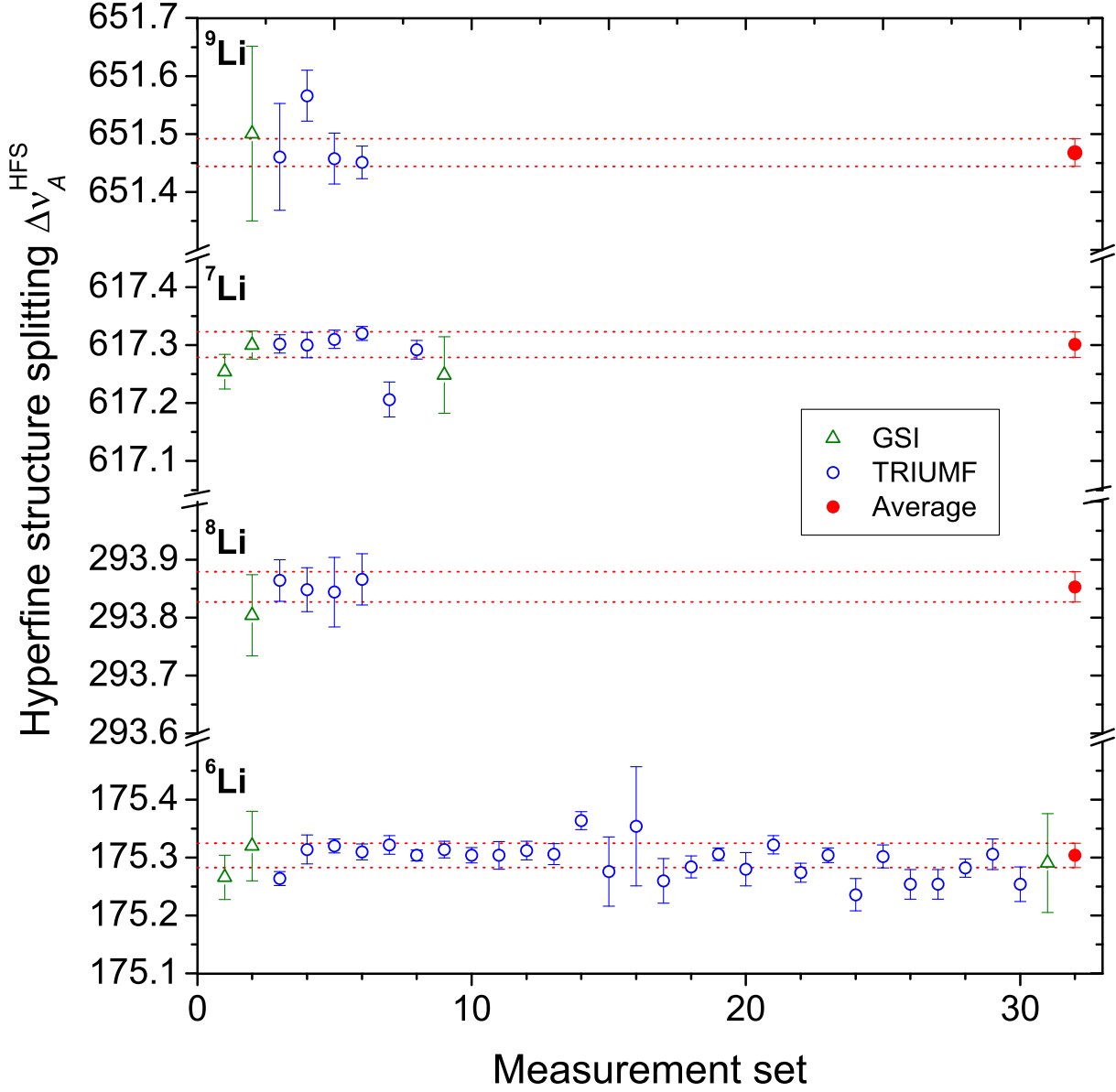}
  \caption{\label{fig:hfs_statistics}
(Color online) Results of all hyperfine splitting measurements in the $2s\;^2{\rm S}_{1/2}\to 3s\;^2{\rm S}_{1/2}$ transition of Li isotopes as obtained during the off-line and on-line beamtimes at GSI and TRIUMF. For $^{6,7}$Li, additional off-line measurements were performed at GSI after the TRIUMF measurements using a frequency comb for laser frequency stabilization \cite{Sanchez2009}. These results are shown as the rightmost GSI datapoint and are in good agreement with previous data.} Plotted are the measured two-photon frequency intervals $\Delta\nu_{A}^{\rm HFS}$ (twice the Ti:sapphire laser frequency difference) between $F\to F$ transitions for $F=I-1/2$ and $F=I+1/2$. Each data point (circle) is the average of 10-20 measurements. The
rightmost data point (full circle) represents the weighted average. For their uncertainties see caption of Table~\ref{Table:hfs_results}.
\end{figure}

Results for $\Delta \nu^{\rm HFS}$ are plotted in Fig.~\ref{fig:hfs_statistics}. Here, each data point is the average
of typically 10 to 20 measurements with the standard error of the mean (SEM) as statistical uncertainty.  The rightmost data
point is the overall weighted average, where the standard deviation of all results is represented with the dashed lines and used as the final uncertainty.
Additionally, $\chi^2$-fitting of a constant to the data points was performed. It resulted in reduced-$\chi^2$ values between 0.5 and 2.0 for all isotopes, indicating that the SEM of such a set of measurements is a realistic estimation of the corresponding statistical uncertainty. Hence, for $^{11}$Li, for which in total 24
spectra were obtained (see below), the corresponding SEM is used as uncertainty of the weighted average of all individual measurements.
The average values and uncertainties obtained in this way are listed in the first line of Table~\ref{Table:hfs_results}.

The $A$ factors of the $3s$ state were calculated using Eq.~(\ref{eq:A3s}) and are listed in in the last line of Table~\ref{Table:hfs_results}. The current result for $A_{3s} (^7{\rm Li})$ disagrees with the previous experimental determination by Stark spectroscopy \cite{Stevens1995}, but is in excellent agreement with a previous laser spectroscopic measurement on an atomic beam \cite{Bushaw2003} and with theoretical values, obtained by both Hylleraas variational \cite{Yan1996} and multiconfiguration Hartree-Fock methods \cite{Godefroid2001}. After the measurements at TRIUMF, frequency-comb based measurements of the $2s \to 3s$ transition were performed at GSI using the same apparatus but a considerably changed laser system, where the Ti:sapphire laser was locked to a frequency comb \cite{Sanchez2009}. The $A_{3s}$ factors obtained from these measurements are in excellent agreement with the prior ones and are included in Fig.~\ref{fig:hfs_statistics}. Also, the ratio $A_{3s}
(^7 {\rm Li}) / A_{3s} (^6 {\rm Li}) = 2.6407(10)$ is in good
agreement with the $2s$ ground state ratio of $2.6407282(1)$ from Ref.\
\cite{Beckmann1974}, indicating no hyperfine anomaly at the level of
precision available in these experiments. The $A_{3s}$ factor of the short-lived isotopes $^{8,9}$Li as obtained from our measurements are also listed.

\begin{table*}
{\caption{\label{Table:hfs_results} $2s \to 3s$ hyperfine splitting $\Delta \nu^{\rm HFS}$ (twice the difference in Ti:sapphire laser resonance frequencies between the $F=I+1/2 \to F=I+1/2$ and the $F=I-1/2 \to F=I-1/2$ transitions) of the lithium isotopes and calculated $A_{3s}$ constants according to Eq.~(\ref{eq:A3s}). The $A_{2s}$ constants listed in Table~\ref{tab:hfsli} are used. Uncertainties of the average are the geometric sum of the statistical uncertainty and a 10~kHz systematic contribution caused by a possible unresolved Zeeman effect in the stray magnetic field at the beamline as discussed in Section~\ref{sec:systematics}. Theoretical values $\Delta \nu^{\rm HFS}_{\rm Theo}$ for the hyperfine splitting are obtained from a calculation of the Fermi contact term \cite{Yan1996} and experimental nuclear magnetic
moments. They include a finite nuclear size (Zemach) correction with exponential ($e^{-\Lambda r}$) and Gaussian (Gauss) electromagnetic distributions
\cite{Yan1996} as discussed in the text.} }
  \begin{ruledtabular}
    \begin{tabular}{llllll}
                       & \mbox{$^6$Li} & \mbox{$^7$Li} & \mbox{$^8$Li} & \mbox{$^9$Li} & \mbox{$^{11}$Li}\\
      \hline
      $\Delta\nu^{\rm HFS}_{\rm Exp}$(Average)
      & 175.304(21)   & 617.301(22)   & 293.853(34)   & 651.468(29)   & 696.09(10)\\
      \hline
      $\Delta\nu^{\rm HFS}_{\rm Exp}$(PNNL) \cite{Bushaw2003}
      & 175.311(24)   & 617.291(22)   & \mbox{-}      & \mbox{-}      & \mbox{-}\\
      $\Delta \nu^{\rm HFS}_{\rm Theo}$ ($R_c \propto e^{-\Lambda r}$)
      & 175.295(17)   & 617.264(60)   & 293.851(29)   & 651.470(65)   & 695.88(9)\\
      $\Delta \nu^{\rm HFS}_{\rm Theo}$ ($R_c \propto$ Gauss)
      & 175.305(17)   & 617.289(60)   & 293.867(23)   & 651.502(64)   & 695.92(9)\\
        \hline
      $A_{3s}$   &   35.267(14)  &  93.103(11)   &   35.476(14)  & 98.297(16)    & 104.91(6)\\
    \end{tabular}
  \end{ruledtabular}
\end{table*}

According to \cite{Yan1996}, the magnetic dipole coupling constant in a $^2{\rm S}_{1/2}$ state can be written as
\begin{equation}
    A_{1/2}({\rm exp}) = 95.410\,67(7) \frac{g_e \mu_I f_{\rm exp}}{3I}
\end{equation}
in MHz. The experimental Fermi contact term $f_{\rm exp}$ can now be extracted from the known magnetic hyperfine constant $A_{2s}$. According to theory $f_{\rm exp}$ is connected to the uncorrected Fermi contact term $f_c$ by the factorization:
\begin{equation}
    f_{\rm exp}= \frac{2(1+a_e)}{g_e} C_{\rm rel} C_{M} C_{R} C_{\rm QED} f_c.
\end{equation}
Here, $f_c$ is the fermi contact term as obtained from nonrelativistic Hylleraas variational calculations for a point dipole, $a_e$ denotes the anomaly of the electron magnetic moment, $C_{\rm rel}$ is a relativistic correction factor, $C_{M}$ the
finite mass correction, $C_{R}$ the finite size correction and
$C_{\rm QED}$ are QED corrections other than the $a_e$. The finite size effect originates from the
nuclear moment distribution in the nucleus and cannot be evaluated
independently of a nuclear model. For a one-electron ion, the
correction is given by the Zemach correction
\begin{equation}
  C_R = 1-2Z \langle R_{\rm em} \rangle /a_0
\end{equation}
with the Bohr radius $a_0$ and the average electromagnetic charge
radius $\langle R_{\rm em} \rangle$ for the nucleus obtained by
folding the charge and magnetization distributions with sizes $R_c$
and $R_m$, respectively. The result for $\langle R_{\rm em} \rangle$
depends somewhat on the model chosen for the distributions. To
calculate the size of possible effects, two models are applied
that are discussed also in \cite{Yan1996}. In both cases it is
assumed that $R_c=R_m$ and either an exponential distribution
$e^{-\Lambda r}$ or a Gaussian distribution is
used. This results in different prefactors and therefore different values for $\langle R_{\rm em}
\rangle$
\begin{equation}
    \langle R_{\rm em} \rangle_{\rm exp}   = \frac{35}{16\sqrt{3}} R_c
    ~~{\rm and}~~
    \langle R_{\rm em} \rangle_{\rm Gauss} = \frac{4}{\sqrt{3\pi}} R_c.
\end{equation}
The charge radii obtained from our measurement
and discussed in the next Chapter as well as the magnetic moments from literature
(Table~\ref{tab:hfsli}) allows us to calculate the Zemach contributions \cite{Yan1996} for all isotopes and the hyperfine
splitting $\Delta\nu_{\rm Theo}^{\rm HFS}(2s\to 3s)$. The results are included in Table~\ref{Table:hfs_results}. The influence of the two models is small and no model can be favored within the accuracy of our
results. Even in the case of the halo nucleus $^{11}$Li where the assumption $R_c = R_m$ is for sure not fulfilled, the values for the theoretical and experimental HFS splittings lie almost all in the range of the $1\sigma$ uncertainties.

\subsubsection{Isotope Shift of $^{6,7}$Li}

\begin{figure}
    \includegraphics[width=\columnwidth,angle=0]{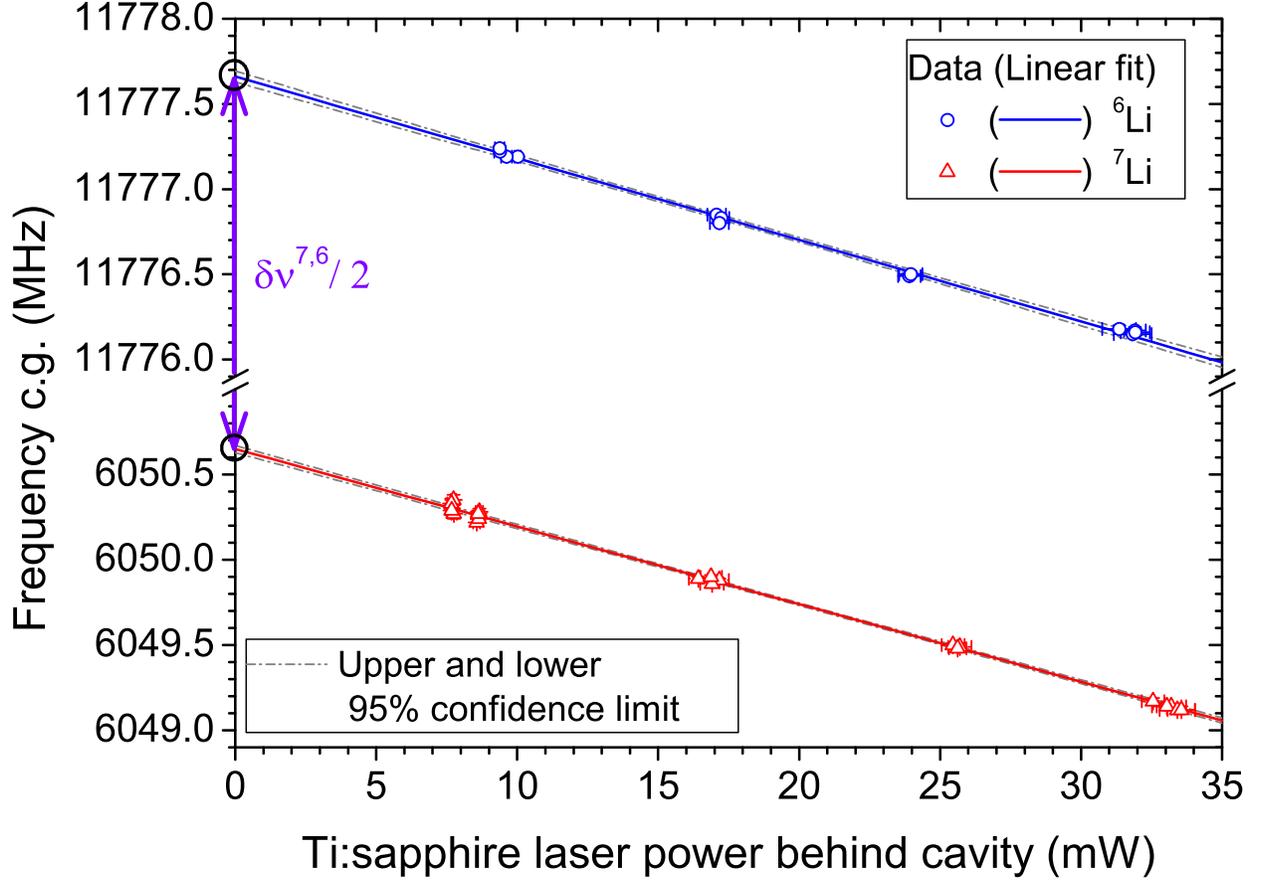}
    \caption{\label{fig:ac_stark_6li_7Li.eps} (Color online) Determination of the $\delta \nu_{\rm IS}^{6,7}$ isotope shift from the ac Stark shift measurement. The frequency offset $\left|\nu_{\rm Ti:sapphire} - \nu_{\rm DL}\right|$ between the Ti:sapphire and the iodine-locked diode laser at the center-of-gravities of the $^6$Li (upper line) and $^7$Li (lower line) hyperfine structures is plotted as a function of the Ti:sapphire laser power. The $x$-axis is the average Ti:sapphire laser power observed on a photodiode behind the enhancement cavity while recording the resonance. This power is proportional to the light intensity at the waist of the resonator mode where the atoms cross the laser beams. It is varied between the maximum obtainable value and 25\% of the maximum value. The dye laser power was kept constant at the maximum obtainable power during all measurements. A linear shift of the resonance positions is observed which is caused by the ac Stark shift effect. The solid lines are linear fits to the data points and the dashed lines indicate the 95\% confidence bands. The isotope shift $\delta \nu ^{6,7}$ in this two-photon transition was calculated by taking twice the distance between the intercepts with the $y$-axis as indicated by the arrow.}
\end{figure}

The isotope shift $\delta\nu_{\rm IS}^{6,7}$ between the two stable
isotopes was extracted from measurements of the dependence of $\delta\nu_{\rm IS}$ on the Ti:sapphire laser power and by
calculating the ac Stark shift regression lines for both isotopes as discussed above. A
typical result is shown in Fig.~\ref{fig:ac_stark_6li_7Li.eps}. Note that the $y$-axis frequency scale in the ac Stark shift plots is always given in terms of the Ti:sapphire laser frequency. Hence, the isotope shift in the $2s \to 3s$ two-photon transition is twice the
distance between the $y$-axis intercepts of the two ac Stark shift regression lines, as indicated in the figure.
At this point, the laser power is zero and the atomic level energies
are unperturbed by the laser light. For both regression lines the
95\% (2$\sigma$) confidence interval (CFI) was also calculated and is plotted in
the graph (dashed lines). The statistical uncertainty (1$\sigma$) of the extracted isotope shift was determined from the fitting result ($1\sigma$ confidence limit) and an additional uncertainty of 10~kHz was added to account for
dye laser power fluctuations during the measurements of the Ti:sapphire-induced ac Stark shift. This value was obtained using the $b_1^{\rm Dye}$ coefficient according to Eq.~(\ref{eq:b_dye}) and assuming dye laser power fluctuations up to $1/3$ of the total power as a conservative estimate. Such fluctuations might tilt the ac Stark shift line systematically in a single set. However, they will fluctuate statistically between different sets and the further statistical treatment is expected to be justified.

\begin{figure}
    \includegraphics[width=\columnwidth,angle=0]{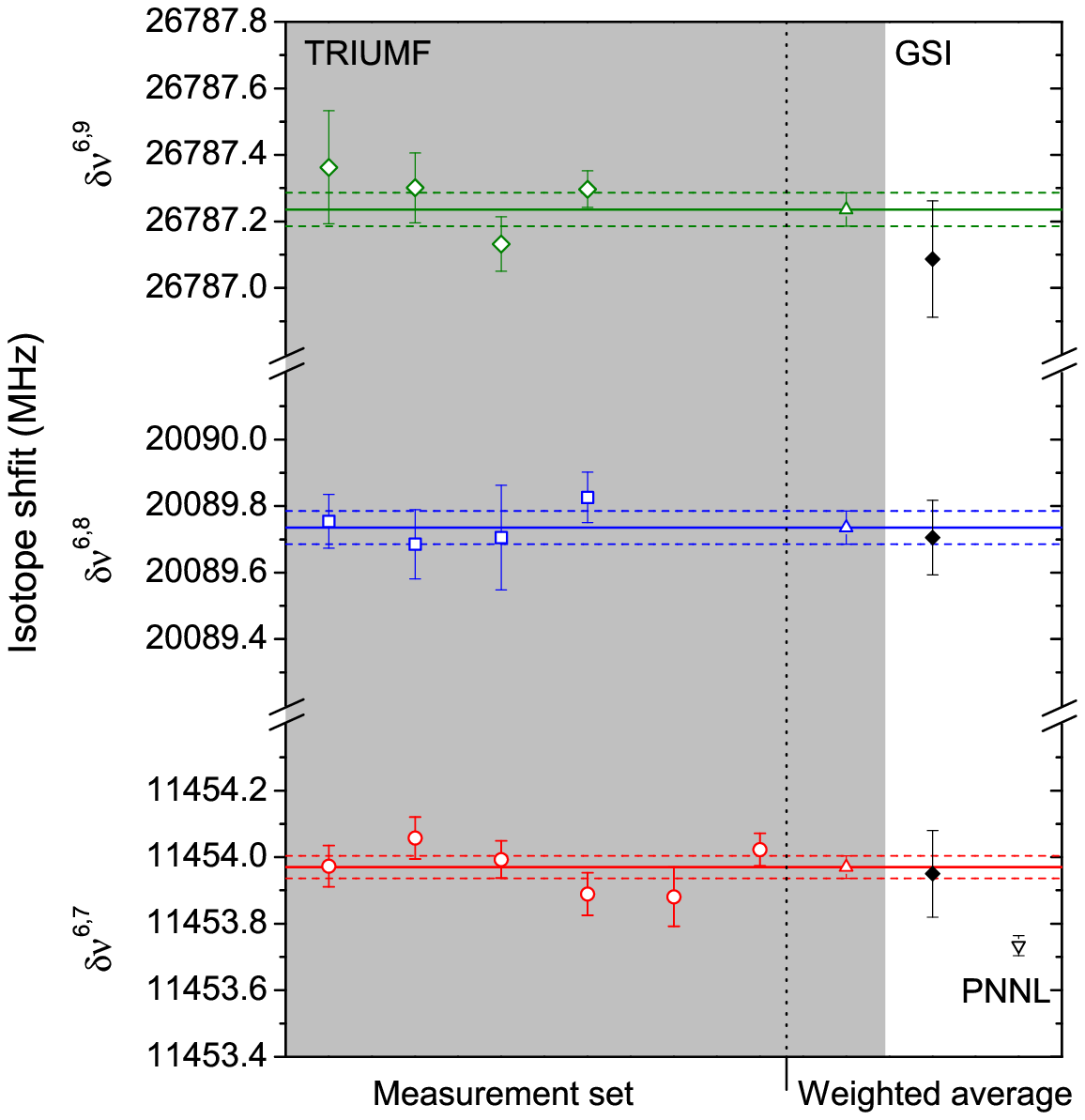}
    \caption{\label{fig:IS_Statistics_7-8-9}
    (Color online) Results of isotope shift measurements in the $2s \to 3s$ transition for $^{7}$Li, $^{8}$Li and $^{9}$Li at TRIUMF (left) as listed in Table~\ref{Table:isotope_shifts}. The average of all TRIUMF measurements ($\vartriangle$) is plotted on the right and compared to the results of the GSI beamtime \cite{Ewald04} for $^{7,8,9}$Li ($\blacklozenge$) and a measurement of the stable isotopes ($\triangledown$) at the Pacific Northwest National Laboratory (PNNL) with a different technique \cite{Bushaw2003}. Uncertainties include the statistical and the systematic uncertainty as listed in Table~\ref{Table:isotope_shifts} and discussed in the text.}
\end{figure}

Results of all isotope shift measurements performed at GSI and at TRIUMF are summarized in Table~\ref{Table:isotope_shifts} and plotted in Fig.~\ref{fig:IS_Statistics_7-8-9}. Each entry corresponds to an extrapolation from an ac Stark shift measurement as it is shown in Fig.~\ref{fig:ac_stark_6li_7Li.eps}.
The isotope shifts obtained in the individual sets at TRIUMF were averaged, weighted with the uncertainty, and the weighted error is taken as the final statistical uncertainty. Fitting a constant to the data points, as shown in Fig.~\ref{fig:IS_Statistics_7-8-9} (bottom graph for $^{6,7}$Li), results in a reduced-$\chi^2$ between 0.5 and 1 for all isotopes. This $\chi^2_{\rm red}$ was not used to scale the final uncertainty. Excellent agreement between the TRIUMF average and the previous measurements at GSI is found for the $^{6,7}$Li isotope shift. The combined values from all ToPLiS experiments, including those at GSI, are listed in the table and marked ``Average".

Since the dye laser power fluctuates statistically, its influence on the line position was added to the statistical uncertainty. Other effects may contribute with systematic shifts in a preferred direction. The estimated size of these systematic uncertainties is given in the caption of Table~\ref{Table:isotope_shifts}. It includes contributions from the small lineshape asymmetry mentioned before (10~kHz) and possible contributions from unresolved Zeeman shifts caused by stray magnetic fields (10~Hz) which will be discussed in Section~\ref{sec:systematics}.

The isotope-dependent velocities cause a differential second-order Doppler shift (SODS) between the isotopes $^6$Li and $^A$Li with mean square velocities $\langle \upsilon_6^2 \rangle$ and $\langle \upsilon_A^2 \rangle$ according to
\begin{equation}
\delta \nu_{\rm SODS}^{6,A} = - \frac{\nu_0}{2}\frac{\langle \upsilon_A^2 \rangle - \langle \upsilon_6^2 \rangle }{c^2}.
\label{eq:SecondOrderDoppler}
\end{equation}
Here, we used the value $\langle \upsilon_A^2 \rangle = 3/2 \cdot 2k_B T / M_A$ for the most probable velocity of the atoms in the ``quasi''-collimated atomic beam that enters the laser focus. A catcher temperature of $T \approx 2000$~K is estimated and  $\nu_0 = 815.61$~THz is the $2s \to 3s$ transition frequency ($2 \cdot \nu_{\rm TiSa}$) \cite{Sanchez2009}. Due to the uncertainty of the absolute temperature and the insufficient knowledge of the exact velocity distribution, an uncertainty of about 30\% is assumed for this correction. Absolute shifts are on the order of 30-60~kHz and the differential values are included in Table~\ref{Table:isotope_shifts}. The largest effect appears naturally for the pair $^6$Li - $^{11}$Li, which amounts to $\delta \nu_{\rm SODS}^{6,11}=34(11)$~kHz.

The values reported here were carefully reanalyzed with respect to statistical and systematic uncertainties including effects that were discovered and evaluated \cite{Sanchez2009} after the original publication \cite{Sanchez06}.
Hence, the values and uncertainties do slightly deviate from those given in our earlier work. However, these changes are small and do not significantly alter the determination of nuclear charge radii and their final uncertainty. Note also that all isotope shifts listed in Table~\ref{Table:isotope_shifts} are given relative to $^6$Li, whereas they were earlier given relative to $^7$Li. The reason for this is twofold: First, the data in Table~\ref{Table:isotope_shifts} are the direct results since we have used $^6$Li as the reference isotope during the measurements. Second, as will be described in a following publication, a reevaluated nuclear charge radius of $^6$Li appears to be more reliable than the previously used reference radius of $^7$Li.

The final value of the ToPLiS measurements for the $^{6,7}$Li isotope shift agrees with the result obtained at GSI \cite{Ewald04,Ewald05} but differs from a previously reported one \cite{Bushaw2003} by about four times the combined uncertainties as listed in the last line of Table~\ref{Table:isotope_shifts}. This is attributed to unaccounted systematic errors in the prior interferometric
measurements. Possible sources of systematic errors are nonlinearities in the scan of the confocal Fabry-Perot interferometer (FPI) or changes in the effective length of the interferometer with pointing and wavelength. The latter might be caused, for example, by a wavelength
dependent phase change of the multilayer dielectric mirrors of the FPI. Moreover, the FPI was calibrated at the reference lines of Rb at 778~nm and Cs at 851~nm, while the operational wavelength for the measurement was relatively far away at 735~nm for lithium. A relative error of $2\cdot 10^{-5}$ in the length of the FPI is sufficient to explain the difference in uncertainty. The data reported here are based on beat frequency determinations and do not rely on any interferometer calibration. Please note that the interferometer scale error in \cite{Bushaw2003} is only relevant for the reported isotope shift, whereas its contribution to the $A_{3s}$ factor is only a small fraction of the previously stated statistical uncertainty and thus requires no correction.

\begin{table*}
 {\caption{\label{Table:isotope_shifts}
Measured isotope shifts $\delta \nu_{\rm IS}^{6,A}$ in the $2s \, ^2S_{1/2} \to 3s \, ^2S_{1/2}$ transition as obtained in all
beamtimes at GSI and at TRIUMF and correction for the second-order Doppler-shift $\delta \nu_{\rm SODS}^{6,A}$.  All values are in kHz. A set of measurements includes typically 15 to 30 resonances at different Ti:sapphire laser powers that were used to obtain one ac Stark shift regression line as shown, for example, in Fig.~\ref{fig:ac_stark_6li_7Li.eps}. Results of sets are listed for the on-line beamtimes at TRIUMF. The uncertainty of each set is the $1\sigma$ confidence limit of the ac Stark extrapolation. If more than one set for an isotope was obtained during a beamtime, the average is listed in addition and printed in bold. The weighted average of all measurements for one isotope is marked ``Average (ToPLiS)". Here, the GSI measurement of $^9$Li was not included for reasons discussed in the text (Section~\ref{sec:systematics}).
The measured isotope shift has to be corrected for the differential second-order Doppler shift $\delta \nu_{\rm SODS}^{6,A}$ according to Eq.~(\ref{eq:SecondOrderDoppler}). The final corrected values are marked ``Corrected (ToPLiS)". Additional systematic uncertainties of 20~kHz + $\Delta(\delta \nu_{\rm SODS}^{6,A})$ (uncertainty of the second-order Doppler shift coorrection) must be considered for the isotope shift of $^{7,8,9}$Li and of $\Delta_{\rm Syst}=85$~kHz as estimated for the isotope shift of $^{11}$Li and discussed in detail in Section~\ref{sec:systematics}. The total uncertainty of the final corrected value is the geometrical sum of the statistical and the systematic uncertainty.
}
 }
  \begin{ruledtabular}
    \begin{tabular}{lllll}
      Facility Year   & \mbox{$\delta \nu^{6,7}$} & \mbox{$\delta \nu^{6,8}$} & \mbox{$\delta \nu^{6,9}$} & \mbox{$\delta \nu^{6,11}$}  \\
      \hline
      GSI 2003        & 11\,453\,950(130) & 20\,089\,705(112) & 26\,787\,087(175)$^{\rm a}$ & \mbox{-}\\
      \hline
   TRIUMF 06/2004     & 11\,453\,973(62)  &                                   &                     & \\
      \hline
   TRIUMF 09/2004 & \bf 11\,453\,982(35)  &              &  \bf 26\,787\,217(61) & \\
        Set 1         & 11\,454\,058(63)  & 20\,089\,754(81)  & 26\,787\,363(170) & \\
        Set 2         & 11\,453\,993(56)  &                                   & 26\,787\,301(105) & \\
        Set 3         & 11\,453\,889(64)  &                                     & 26\,787\,132(82)  & \\
      \hline
   TRIUMF 10/2004 & \bf 11\,453\,990(43)&\bf 20\,089\,767(57)   &                                       & \\
        Set 1         & 11\,453\,881(89)  & 20\,089\,685(104) & 26\,787\,297(55)  & 36\,555\,210(76) \\
        Set 2         & 11\,454\,023(49)  & 20\,089\,705(157) &                   & \\
        Set 3         &                                   & 20\,089\,826(76)  &                   & \\
      \hline
   Average (ToPLiS)    & 11\,453\,981(24)  & 20\,089\,754(43)  & 26\,787\,261(41)  & 36\,555\,210(66) \\
   $\delta \nu_{\rm SODS}^{6,A}$ & 11(4) &            19(6) &             25(8) &             34(11) \\
   Corrected (ToPLiS)    & 11\,453\,970(34)  & 20\,089\,735(50)  & 26\,787\,236(50)  & 36\,555\,176(108) \\
      \hline
      PNNL \cite{Bushaw2003} & 11\,453\,734(30)  \\
    \end{tabular}
  \end{ruledtabular}
  \begin{flushleft}
      $^{\rm a}$ Not included in the total average for reasons discussed in Section~\ref{sec:systematics}.
\end{flushleft}
\end{table*}

\subsubsection{Extraction of $\delta \left \langle r_c^2 \right \rangle$ for $^{6,7}$Li}
\label{sec:delta_R2_stable}

We can now use the measured isotope shift between the stable isotopes to extract the change in the mean square nuclear charge radius. This is an important test of our technique, because the result must be consistent with corresponding values extracted from other transitions or by other means, as long as those are model-independent. Therefore, we use the calculated mass shift $\delta \nu_{\rm MS,Theory}^{6,7}$ from Table~\ref{Li_isotope} to extract by use of Eq.~(\ref{eq:delta_rc}) the change in the ms nuclear charge radii and obtain
\begin{equation}
\delta \left\langle r^{2}\right\rangle^{6,7}=-0.731(22)~{\rm fm}^2.
\end{equation}

\begin{figure}[tb]
\includegraphics[width=\columnwidth, clip=]{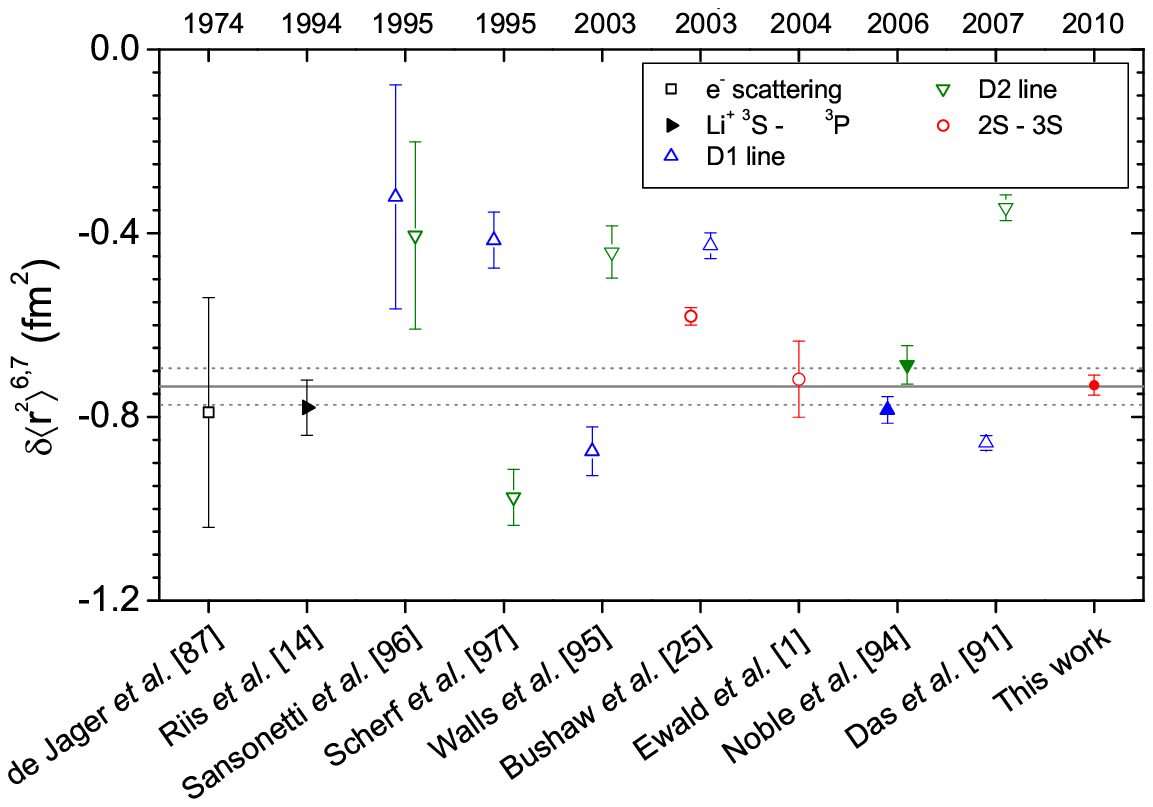}
\caption{\label{fig:delta_ms_radii67}
(Color online) Experimental results for the difference in the mean square charge radius $\delta \left\langle r^{2}\right\rangle ^{6,7}$ between the stable isotopes $^6$Li and $^7$Li. Besides the elastic electron scattering (square), all other experimental data are obtained from optical isotope shift measurements in different transitions in Li$^+$ and neutral lithium as indicated in the figure legend and discussed in the text. Experimental results are depicted in chronological order. The simple average of the consistent data from the individual atomic lithium transitions (solid symbols)
is indicated by the solid line with the respective standard deviation represented by the dashed lines.
All values and respective references are listed in Table~\ref{tab:delta_ms_radii67}. For more discussions see text.}
\end{figure}

Nuclear charge radii of the stable lithium isotopes were measured by elastic electron scattering in the 1960's and 70's
\cite{Suelzle1967,Li1971,Bumiller1972} and results are summarized in \cite{deJager1974,deVries1987}. We used the charge radii
$R_c(^7{\rm Li})=2.39(3)$~fm \cite{Suelzle1967} and $R_c(^6{\rm Li})=2.55(5)$~fm, the average of the values reported in \cite{Li1971} and \cite{Bumiller1972}. In Table~\ref{tab:delta_ms_radii67} the resulting value for $\delta \left\langle r^{2}\right\rangle^{6,7}= - 0.79(29)$~fm$^2$ is compared with optical measurements of three transitions in lithium that have been repeatedly investigated during the last decades. Besides the $2s \to 3s$ two-photon transition investigated here, isotope shifts were determined in the resonance lines $2s\;^2{\rm S}_{1/2} \to 2p\;^2{\rm P}_{1/2}$ (D$_1$ line) and $2s\;^2{\rm S}_{1/2}\to 2p\;^2{\rm P}_{3/2}$ (D$_2$ line) in neutral lithium as well as in the $1s2s \; ^3{\rm S}_1 \to 1s2p \; ^3{\rm P}_{0,1,2}$ transitions in Li$^+$. We use the recent mass shift calculations as listed in Table~\ref{tab:delta_ms_radii67} to extract $\delta \left\langle r^{2}\right\rangle^{6,7}$ from the isotope shifts in the D$_1$ and D$_2$ lines reported in the literature. The results are summarized in Table~\ref{tab:delta_ms_radii67} and also depicted in Fig.~\ref{fig:delta_ms_radii67} in chronological order. The large scatter of the data points, much larger than most of the indicated error bars, is the first striking impression. Obviously, for many measurements the systematic uncertainties were underestimated. In the following, we will shortly discuss the different experiments and discuss possible sources of additional or underestimated systematic uncertainties.

The most reliable early result is probably the measurement on Li$^+$ by Riis {\it et al.} \cite{Riis1994}. This experiment was performed using saturation spectroscopy on a fast ion beam in collinear and anticollinear geometry to obtain absolute transition frequencies in the $1s2s \; ^3{\rm S}_1 \to 1s2p \; ^3{\rm P}_{0,1,2}$ transition for the two isotopes.
The difference was then calculated to obtain the isotope shift. A crossed-beam experiment \cite{Rong1993}, where the laser crossed the ion beam under 90$^\circ$, was in contradiction for the $1s2p\; ^3{\rm P}_{1-2}$ fine structure splitting but the result by Riis {\it et al.} was confirmed in a later experiment using electrooptic modulation \cite{Wijngaarden1996}. The change in charge radius between the stable isotopes extracted from the Li$^+$ measurements \cite{Riis1994} is in very good agreement with the elastic electron scattering result.

In contrast, measurements of isotope shifts in the D$_1$ and
D$_2$ lines show a rather wide range of results, which must be attributed to a variety of uncorrected or underestimated systematic errors.  For example, a recent atomic beam measurement using diode laser resonance fluorescence spectroscopy \cite{Das2007} reports isotope shifts in the D$_1$ and D$_2$ lines,  and claims to have determined the absolute transition frequencies with accuracies of 30-60~kHz, yet the results for the two lines are internally inconsistent, if the general theories of isotope shifts and hyperfine structure are to be believed. This claimed precision is much less than 1\% of the observed linewidth, in a one-photon transition subject to residual Doppler shifts and optical pumping distortions. This becomes even more challenging  for the D$_2$ transition, where the hyperfine structure is only partially resolved. Most authors fit the observed structure by fixing the hyperfine structure interval and intensity ratios according to Clebsch-Gordan coefficients.  However, optical pumping effects can vary the component line intensities dramatically, even at excitation intensities well below saturation \cite{Bushaw1986,Cannon1988}. Further, linear polarization and magnetic shielding \cite{Das2007} do not prevent optical pumping; rather one should use vanishingly small excitation intensity and/or (pseudo) random polarization. In \cite{Walls2003,Noble2006} it was tried to account for such effects in the D$_2$ transition by simulating the effect of optical pumping. A shift of almost 2~MHz is proposed for one of the hyperfine structure components, but an uncertainty estimation for this rather large correction is not given.

There are two additional other points which must be considered carefully and may be relevant for all works on the two lithium resonance lines (D$_{1,2}$): Firstly, exact perpendicularity between the atomic beam and the laser is needed to avoid differential Doppler shifts between different isotopes. This is particularly important for light-massed isotopes and most works make specific efforts to minimize this uncertainty. Nonetheless, at 600~K a misalignment of 1~mrad from perpendicularity between laser and atomic beam leads to a differential Doppler shift of 70~kHz. Secondly, all measurements to date on the D$_{1,2}$ lines uses interferometers for essentially a wavelength measurement.  At the level of precision needed for measurements of isotope shifts in lithium, interferometric measurements are limited by systematic errors associated with pointing errors and mirror phase shifts. As discussed above, resulting scale uncertainties are most probably the reason for the deviation of our early $2s \to 3s$ measurement and may also contribute to the D$_1$ measurement. In comparison, all measurements in the $2s \to 3s$ transition reported here, are based on direct frequency measurements and do not suffer from scale errors.
There are only two sets of D$_{1,2}$ measurements that result in consistent $\left \langle \delta r^2 \right\rangle$ for both transitions. These are the measurement by Sansonetti and coworkers in 1995 \cite{Sansonetti1995}, with relatively large uncertainties, and a more recent measurement by Noble and coworkers in 2006 \cite{Noble2006} with considerably smaller uncertainty estimate. The latter group previously reported results \cite{Walls2003} using the same technique that are in relatively poor agreement: The D$_1$ isotope shift is shifted by $2 \sigma$ whereas the D$_2$ line is shifted by $4 \sigma$. In Ref.\ \cite{Noble2006} the authors disown their value for the D$_2$ and claim that the previous work had insufficient resolution.
Other measurements \cite{Scherf1996,Walls2003,Das2007} yield results that are inconsistent both internally, with respect to $\delta \left\langle r^2 \right \rangle$ values derived for the two transitions, and with each other. Hence, also the splitting isotope shift disagrees with theory.


It is obvious that an accurate determination of the isotope shifts in the D$_1$ and D$_2$ lines of lithium is very difficult. The Li$^+$ measurements \cite{Riis1994} have the clear advantage that a Doppler-free technique was used which provides the required accuracy. Similarly, measurements in the $2s \to 3s$ transition, as those reported here, are also to first-order Doppler-free and are not hampered by unresolved hyperfine splitting.

The first isotope shift measurement on the $2s\;^2{\rm S}_{1/2}\to 3s\;^2{\rm S}_{1/2}$ transition reported by some of us in \cite{Bushaw2003} was discussed above and likely suffers from a  scale error not accounted for in the original estimation of systematic uncertainties, because it deviates in the D$_1$ transition, similarly as in the two-photon transition, from more recent and more accurate measurements. The experiments on the different transitions in lithium have also been summarized and critically evaluated in \cite{Noble2009}.


In summary, some recent measurements on the D$_1$ and D$_2$ lines resulted in a conclusive picture for $\delta \left\langle r^{2}\right\rangle^{6,7}$: The optical results reported in \cite{Riis1994,Noble2006} and in this work (which includes \cite{Ewald04,Ewald05}) cover all discussed transitions and are consistent within their uncertainties. The  simple average\footnote{ Since there are obvious difficulties with the uncertainty estimations at least in some of the previously published results, we prefer not to use the uncertainty estimations for a weighted average across all transitions.} of $\delta \left\langle r^{2}\right\rangle^{6,7}$ obtained from these optical investigations is --0.734(40)~fm$^2$ and agrees with the much less accurate result obtained by elastic electron scattering. Thus, we conclude that the optical isotope shift measurements combined with atomic mass-shift calculations can provide a consistent picture for all transitions investigated so far. However, the situation in the D lines is far from being satifactory for both, the isotope shift and the spiltting isotope shift, due to the large fluctuations in the results.

\begingroup
\squeezetable
\begin{table*}
\caption{\label{tab:delta_ms_radii67}
Changes in the mean square nuclear charge radius $\delta\langle r^2 \rangle^{6,7}$ between the stable isotopes $^{6,7}$Li from elastic electron scattering and optical isotope shifts in different transitions.
Charge radii from optical isotope shifts were recalculated using the experimental value $\delta \nu^{6,7}_{\rm Exp}$ of the respective reference and the latest mass shift calculations (D$_{1,2}$ transitions from \cite{Yan2008}, $2s \to 3s$ from this work). Mass shifts $\delta \nu^{6,7}_{\rm MS}$ and field shift coefficients $C_{6,A}$ are listed for each transition only once. For the three transitions in neutral lithium the average values and the corresponding standard deviation is listed. 
The overall average of all measurements is --0.63(20)~fm$^2$, where the uncertainty is given by the standard deviation. The average of the bold printed $\delta\langle r^2 \rangle^{6,7}$ values from \cite{Riis1994,Noble2006} and this work is --0.734(40)~fm$^2$ and is indicated in Fig.~\ref{fig:delta_ms_radii67}. This subset was chosen since they give internally and mutually consistent values for $\delta\langle r^2 \rangle^{6,7}$ as discussed in the text. }
  \begin{ruledtabular}
    \begin{tabular}{lllccl}
        Transition & $\delta \nu^{6,7}_{\rm Exp}$ & $\delta \nu^{6,7}_{\rm MS}$ & $C_{6,A}$ & $\delta \left\langle r^{2}\right\rangle ^{6,7}$  & Year \& Ref. \\
                   &      (MHz)                   &    (MHz) & (MHz/fm$^2$) &    (fm$^2$)&  \\

        \hline
        Electron scattering
        &--              &                 &      & -0.79(29)  &  1972 \cite{Suelzle1967,Li1971,Bumiller1972} \\
\hline

        Li $^{+}$ $1s2s \; ^3{\rm S}_1 \to 1s2p \; ^3{\rm P}_{0}$
        & 34\,747.73(55) &  34\,740.17(3) & 9.705 & -0.78(6) & 1994  \cite{Riis1994} \\
        Li $^{+}$ $1s2s \; ^3{\rm S}_1 \to 1s2p \; ^3{\rm P}_{1}$
        & 34\,747.46(67) &  34\,739.87(3) & 9.705 & -0.78(7) & 1994  \cite{Riis1994} \\
        Li $^{+}$ $1s2s \; ^3{\rm S}_1 \to 1s2p \; ^3{\rm P}_{2}$
        & 34\,748.91(62) &  34\,742.71(3) & 9.705 & -0.64(6) & 1994  \cite{Riis1994} \\
        Weighted Average
        & --             &                 &       & -{\bf 0.735(36)} & 1994  \cite{Riis1994} \\
\hline
        Li $2s\;^2{\rm S}_{1/2}\to 2p\;^2{\rm P}_{1/2}$
        & 10\,532.9(6)     & 10\,532.111(6) & -2.457 & -0.32(24)       & 1995 \cite{Sansonetti1995} \\
        & 10\,533.13(15)   & &                       & -0.415(61)      & 1996 \cite{Scherf1996} \\
        & 10\,534.26(13)   & &                       & -0.875(53)      & 2003 \cite{Walls2003} \\
        & 10\,533.160(68)  & &                       & -0.427(28)      & 2003 \cite{Bushaw2003} \\
        & 10\,534.039(70)  & &                       & -{\bf 0.785(29)}& 2006 \cite{Noble2006} \\
        & 10\,534.215(32)  & &                       & -0.857(16)      & 2007 \cite{Das2007} \\
        Average
        & 10\,533.62(62)   & &                       & -0.61(25) \\
\hline

        Li $2s\;^2{\rm S}_{1/2}\to 2p\;^2{\rm P}_{3/2}$
        & 10\,533.5 (5)    & 10\,532.506(6) & -2.457 & -0.41(20)       & 1995 \cite{Sansonetti1995} \\
        & 10\,534.90(15)   & &                       & -0.975(61)      & 1996 \cite{Scherf1996} \\
        & 10\,533.59(14)   & &                       & -0.441(57)      & 2003 \cite{Walls2003} \\
        & 10\,534.194(104) & &                       & -{\bf 0.687(42)}& 2006 \cite{Noble2006} \\
        & 10\,533.352(68)  & &                       & -0.344(28)      & 2007 \cite{Das2007} \\
        Average
        & 10\,533.91(64)   & &                       & -0.57(26) \\
\hline
        	
        Li $2s\;^2{\rm S}_{1/2}\to 3s\;^2{\rm S}_{1/2}$
        & 11\,453.734(30)  & 11\,452.821(2)& -1.572  & -0.581(19)      & 2003 \cite{Bushaw2003} \\
        & 11\,453.95(13)	 & &                       & -0.719(83)      & 2004 \cite{Ewald04,Ewald05}$^{\rm a}$ \\
        & 11\,453.970(34)  & &                       & -{\bf 0.731(22)}& 2010 this work \\
        Average
        & 11\,453.88(11)	   & &                       & -0.673(71)

    \end{tabular}
  \end{ruledtabular}
  \begin{flushleft}
  $^{\rm a}$ Result included in the final value of this work.
  \end{flushleft}
\end{table*}
\endgroup

\subsubsection{Isotope shifts of $^{8,9}$Li}
The isotope shifts of the isotopes $^8$Li and $^9$Li relative to $^6$Li were determined similar to the measurements of the stable isotopes.  A peculiarity of the
radioactive isotope scans is the fact that ions implanted into the channeltron
detector will inevitably decay there and produce $\beta$-delayed $\alpha$ particles which will again trigger channeltron signals.  This will on average result in
more than one event per ionized atom and, thus, increase the
detection efficiency. However, one has to ensure that the additional
events are counted in the channel of that frequency applied to produce the photo ion. Otherwise, the line profile will be skewed depending
on the scanning direction and the resonance center will be shifted. To avoid this effect, the
radioactive ion beam was turned on only for 10~s for each frequency
channel and turned off thereafter for at least five half-lives of
the respective isotope (5~s for $^8$Li, 1~s for $^{9,11}$Li). The events occurring during the beam-off time were still registered and added into the corresponding frequency channel. This was possible because the
channeltron had a very small dark count rate on the order of 10~mHz
and the applied detection technique of resonance ionization mass
spectrometry is, apart from the dark count rate, practically
background-free. We also looked for events from $\beta$-decay of the $^{11}$Li daughter $^{11}$Be ($T_{1/2}=13.8$~s) and found nothing above background."

The intensity of the radioactive ion beam was monitored with a NaI detector and a plastic scintillator mounted outside the vacuum chamber but close to the carbon catcher foil, detecting the $\gamma$-rays produced by $\beta$ Bremsstrahlung in the stainless steel chamber and $\beta$-delayed neutrons.

\begin{figure*}[tb]
\includegraphics[width=\textwidth,angle=0]{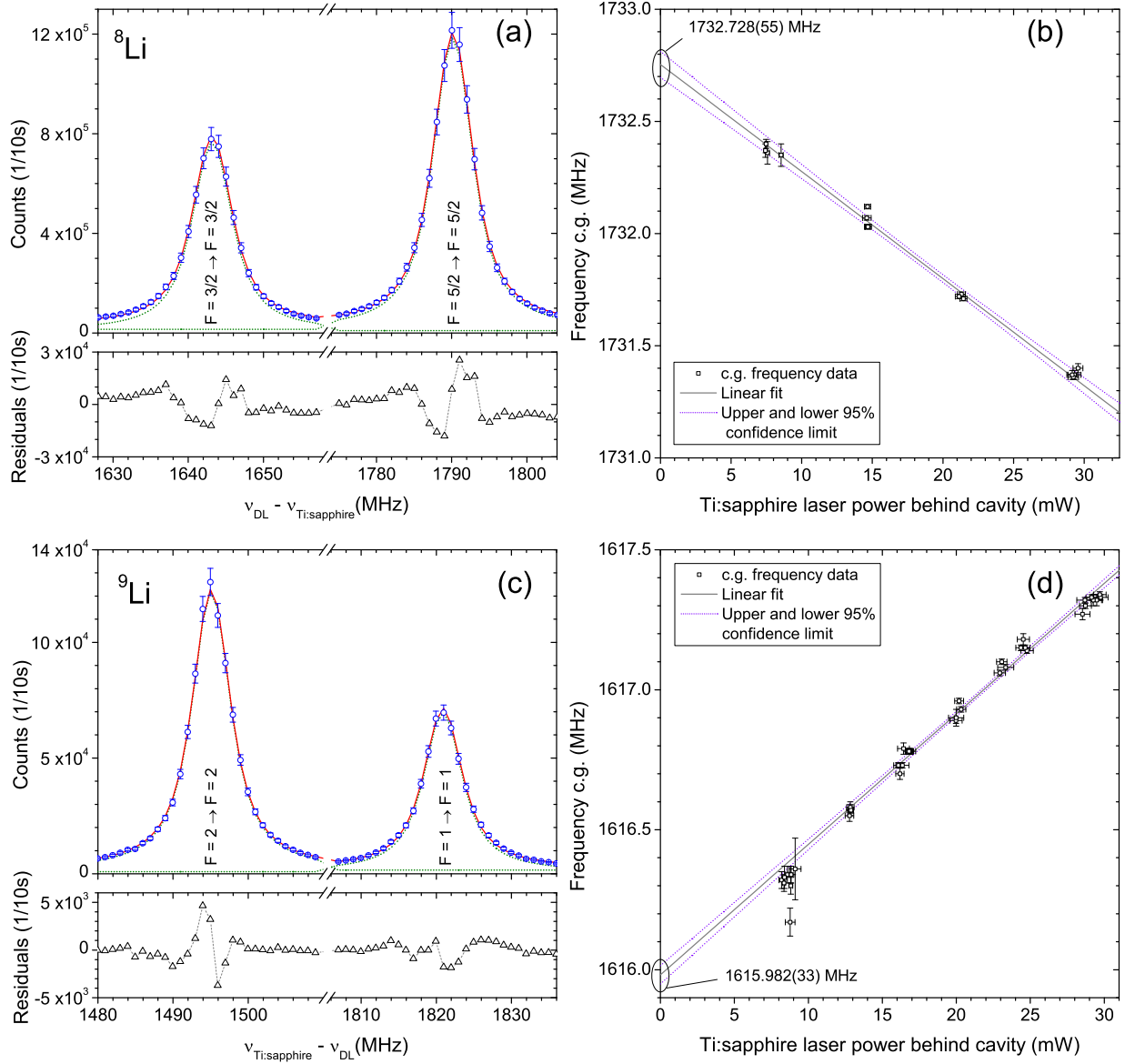}
\caption{\label{fig:li8_9_com_hv30_oct}
(Color online) Typical spectra (a,c) and ac Stark shift results (b,d) for $^{8}$Li and $^{9}$Li. Data points of the spectra are normalized to intensity variations of the ion beam and laser beam according to Eq.~(\ref{eq:norm_li11}). Voigt profiles with Doppler pedestal are fitted to the hyperfine resonances with the same procedure as used for the stable isotopes. The solid lines in (a) and (c) are the common fit results while the dashed lines are the line profiles of the individuals components. Results of ac Stark shift measurements are plotted in (b) and (d) for $^8$Li and $^9$Li, respectively. The solid line is the result of a linear regression and the dashed lines indicate the 95\% confidence band. The intercept of the regression lines with the $y$-axis is used to determine the isotope shift.}
\end{figure*}

Typical resonance profiles of $^{8}$Li and $^{9}$Li are shown in Fig.~\ref{fig:li8_9_com_hv30_oct}~(a,c). The count rate in each channel $N_{\rm Raw}$ was normalized to changing Ti:sapphire laser powers and to variations in the ion beam
intensity according to
\begin{equation}
N_{\rm Norm} = N_{\rm Raw}\cdot \left( \frac{N_{\gamma}}{\langle N_{\gamma} \rangle} \right)^{-1} \cdot \left( \frac{P_{\,\rm Ti:sapphire}}{\langle P_{\,\rm Ti:sapphire} \rangle} \right)^{-2}.
\label{eq:norm_li11}
\end{equation}
$N_{\gamma}$ is the $\gamma$-ray intensity observed with the plastic scintillator, and $P_{\,\rm Ti:sapphire}$ is the Ti:sapphire laser power recorded with the photodiode behind the resonator while recording the respective channel. $\langle N_{\gamma} \rangle$ and $\langle P_{\,\rm Ti:sapphire} \rangle$ are the average values during the complete scanning period. A correction for the $\gamma$-ray background was not necessary for $^{8,9}$Li since the count rate with the ion beam turned on
was several orders of magnitude larger than the background with the ion beam turned off. The error bars of the
data points in Fig.~\ref{fig:li8_9_com_hv30_oct} include the statistical uncertainty as
well as the propagated uncertainty from the normalization procedure.

Fitting was performed as for the stable isotopes and the c.g. was calculated from the hyperfine component positions
according to Eq.~(\ref{eq:Licg}). As in the case of the stable isotopes, a small asymmetry in the line profile is observable in the residuals. It appears that the asymmetry is inverted for $^9$Li compared to the stable isotopes and $^8$Li as is the case for the order of the hyperfine components. Both is caused by the inverted frequency scale for the heavier isotope $^9$Li. The frequency of the Ti:sapphire laser is measured relative to the iodine-locked diode laser ($\nu_{\rm DL}$). Resonance frequencies for $^{6,7,8}$Li are below the diode laser frequency while the $^{9,11}$Li resonances appear at higher frequencies.

Results of ac Stark shift measurements for $^{8,9}$Li are shown in Fig.~\ref{fig:li8_9_com_hv30_oct}~(b,d). The seemingly opposite slope of the lines is again due to the inverted frequency scale. As for the stable isotopes, the isotope shift is determined from the intercept of the regression line with the $y$-axis referenced to a similar measurement of the reference isotope $^6$Li. Measurements of $^6$Li were regulary interspersed between the ones on short-lived isotopes.
The results for the isotope shift of $^8$Li and $^9$Li are included in Table~\ref{Table:isotope_shifts}.
Four sets of measurements were obtained for each isotope during the two beamtimes at TRIUMF. The results of the different measurements and their averages are plotted in Fig.\,\ref{fig:IS_Statistics_7-8-9}. Systematic uncertainties are similar as for the stable isotopes. The comparison with the measurements from GSI ($\blacklozenge$) shows good agreement for both isotopes.

\subsubsection{Isotope Shift of $^{11}$Li}

\begin{figure}
\includegraphics[width=0.8\columnwidth, clip=]{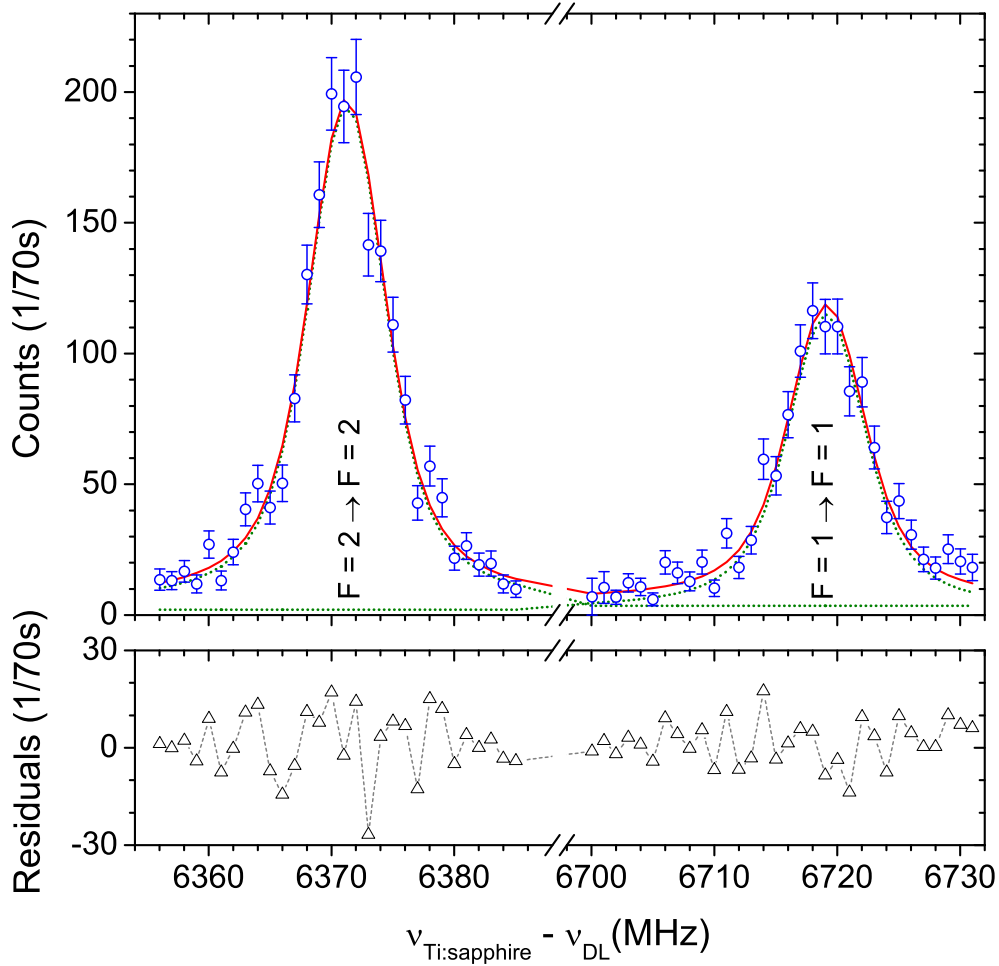}
\caption{\label{fig:li11_spectrum}
(Color online) Typical spectrum of $^{11}$Li as it was obtained in the beamtime at TRIUMF. Data points are normalized to intensity fluctuations of the ion and laser beams. The solid line is again the common fit function of the two hyperfine components while the dotted lines are the contributions of the individual hyperfine components.}
\end{figure}

The very short-lived isotope  $^{11}$Li was available at TRIUMF with a yield of about 35\,000 ions/s. A single scan of a
$^{11}$Li resonance curve was acquired within approximately 10 min. Typically 5 to 7 consecutive individual scans were added to obtain a resonance signal as shown in Fig.~\ref{fig:li11_spectrum}. The small production rate of  $^{11}$Li as compared with the other isotopes did not allow for full Ti:sapphire laser power-series measurements since it was not possible to collect sufficient statistics at low laser power within a reasonable time. Thus, a different approach was used to correct for the ac Stark shift of this isotope: All $^{11}$Li spectra were recorded at maximum laser power of the Ti:sapphire laser and the average power value during the scan was evaluated. The count rate was normalized for ion beam and laser intensity fluctuations as described above for $^{8,9}$Li, but this time the background in the $\gamma$ and neutron count rate of the plastic scintillator had to be taken into account. Therefore, the 1~s pause after each frequency step was used to determine the ion-beam independent background and
to subtract it before normalization.
The resonance signal was then fitted with the same line profile as applied to the other isotopes
and the hyperfine c.g. was determined. For reference, a complete ac Stark shift power-series measurement of $^6$Li was performed at least once per day. Then, the ac Stark shift function determined for $^6$Li
\begin{equation}
\nu(^6{\rm Li}) = \nu_0(^6{\rm Li}) + b_1^{\rm Ti:sapphire} \cdot P_{\rm Ti:sapphire}
\end{equation}
was evaluated at the average power $\langle P_{\rm Ti:sapphire} \rangle_{^{11}{\rm Li}}$ during the $^{11}$Li measurements and the isotope shift was determined according to
\begin{equation}
\delta\nu^{6,11} = \nu(^{11}{\rm Li})-\left(\nu_0(^6{\rm Li}) + b_1^{\rm Ti:sapphire} \cdot \langle P_{\rm Ti:sapphire} \rangle_{^{11}{\rm Li}} \right).
\end{equation}
This procedure requires that the ac Stark shift of both isotopes have the same slope $b_1^{\rm Ti:sapphire}$. This
is theoretically expected since all resonance frequencies in the $2s \to 3s$ transition are far-off from any
allowed E1 transition that would connect the $2s$ or $3s$ level to a
$p$ state. In fact, this was also experimentally verified for all other lithium isotopes as will be
described in the next Section.

\begin{figure}
\includegraphics[width=\columnwidth,angle=0]{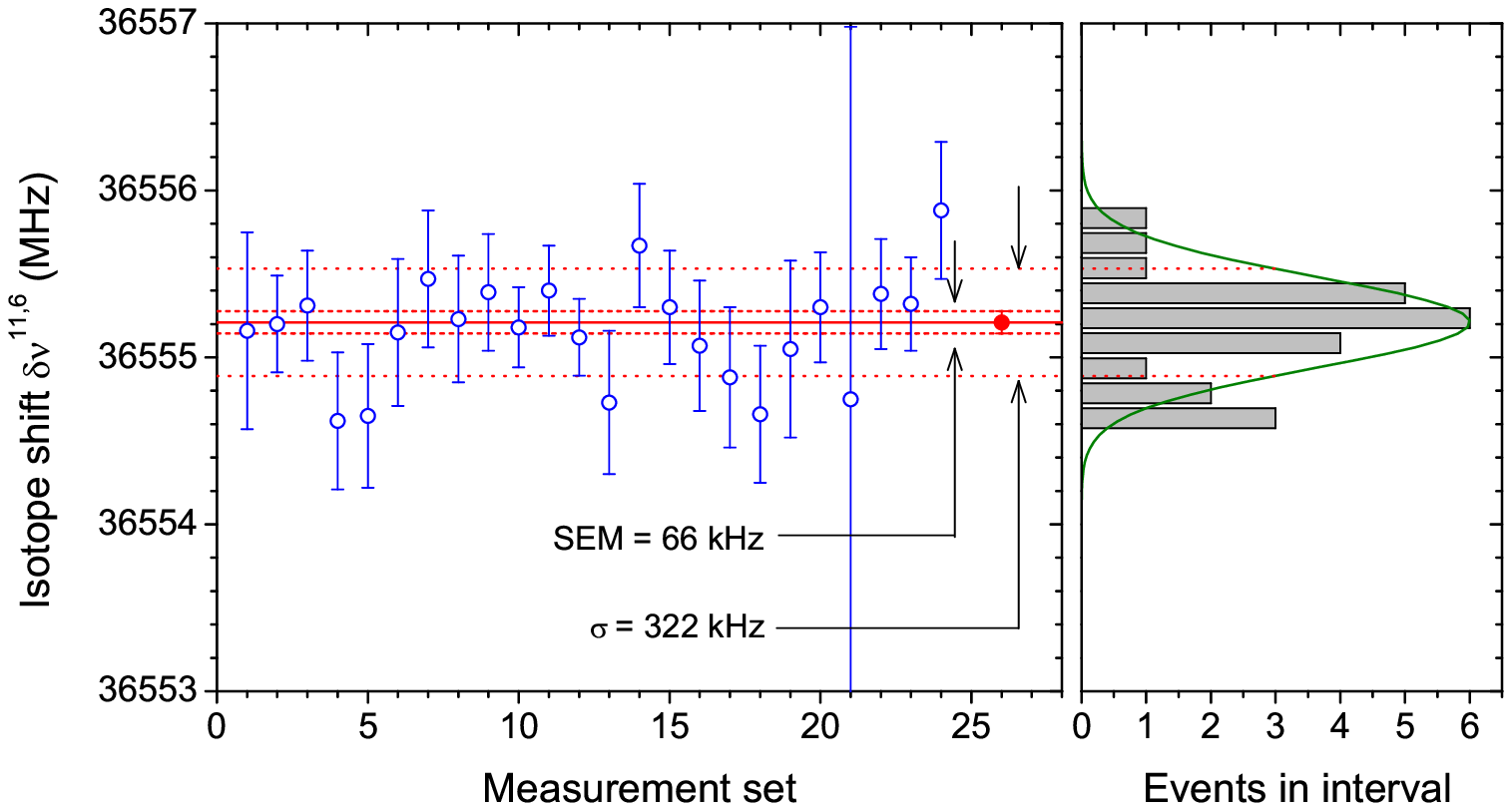}
\caption{\label{fig:li11_is_statistics}
(Color online) Distribution and the average of the $^{11}$Li - $^6$Li isotope shift (c.g.) of all 24 $^{11}$Li resonances obtained at TRIUMF. The dotted lines represent the standard deviation ($\sigma$) of the individual measurements and the dashed ones the standard error of the mean. The average value is plotted as solid circle. A histogram of the measured frequencies is depicted on the right.}
\end{figure}

In total, 24 resonances could be measured for $^{11}$Li. The distribution of the extracted isotope shift values
$\delta\nu^{6,11}$ is shown in Fig.~\ref{fig:li11_is_statistics} (open circles). The weighted average is included in
Table~\ref{Table:isotope_shifts}. The uncertainty was estimated as
the standard error of the mean as depicted in the figure. This is in accordance with the uncertainties obtained for the more abundant isotopes discussed above.

\subsection{Systematic uncertainties in the isotope shift}
\label{sec:systematics}

Several effects that can possibly affect the extracted resonance position were investigated. These include deviations from the
Lorentzian, respectively Voigtian resonance lineshape, isotope-dependent slopes or nonlinearities in the ac Stark shift
correction, and the influence of stray magnetic fields.

It was found that the most crucial effect is an isotope-dependent slope of the regression line in the ac Stark correction. This does not matter for the stable and the less exotic isotopes $^{6-9}$Li since those were all measured at different powers and extrapolated to zero laser intensity. However, the slope for $^{11}$Li could not be measured experimentally and a difference in the slope between $^{11}$Li and the reference isotope $^6$Li would directly influence the extracted isotope shift. Therefore, the slopes from all $^{6-9}$Li ac Stark shift measurements at TRIUMF were compared after the first on-line beamtime. Previous measurements at GSI showed no indication for an isotope-dependent slope in the ac Stark shift of $^{6,7,8}$Li within the level of accuracy of approximately 2~kHz/mW. However, at TRIUMF the ac Stark shift showed larger isotope-dependent trends.
\begin{figure}
\begin{center}
\includegraphics[width=\columnwidth,angle=0]{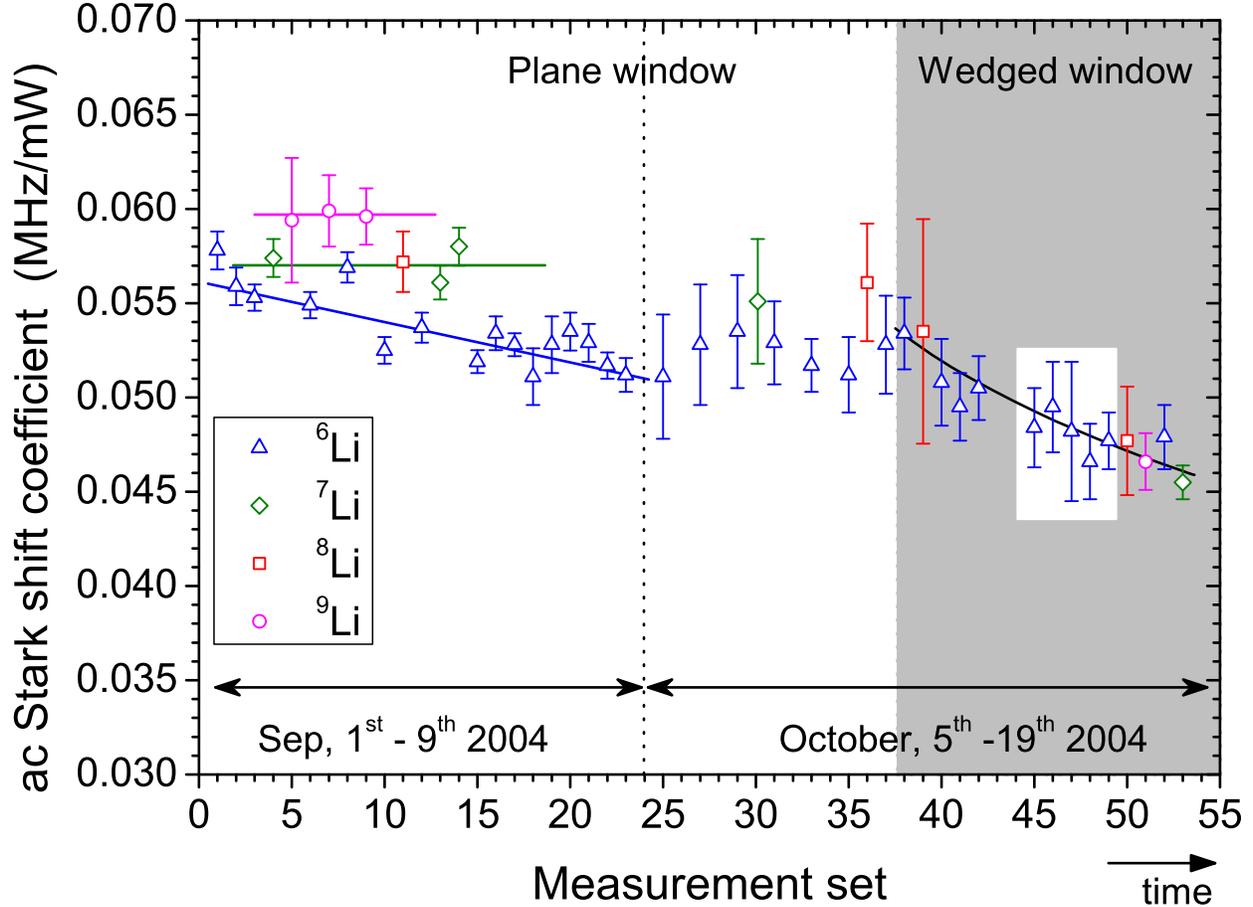}
\end{center}
\caption{\label{fig:AC-Stark_slope_coeff}
(Color online) Experimentally determined ac Stark shift coefficients $b_1^{\rm Ti:sapphire}$ for all measurement sets and all isotopes conducted at TRIUMF.
The periods of the two beamtimes are indicated at the bottom. Measurements 1-37 were performed with a plane parallel viewport window while later measurements (gray background) were carried out with a wedged window to avoid interference effects. The lines are included to guide the eye along the measurements of the respective isotope (left) and to indicate the common drift after the replacement of the viewport with a wedged window. The highlighted part on the right indicates the period in which all $^{11}$Li resonances were recorded.
}
\end{figure}
This can be seen in the left part of Fig.~\ref{fig:AC-Stark_slope_coeff} where this is indicated by the straight lines. There is a clear offset between $^7$Li and $^8$Li, and the reference isotope $^6$Li shows even a small drift. This effect
was found to be an artefact caused by interference effects in the viewport behind the optical resonator. The geometry of the vacuum chamber, the resonator and the viewport is depicted in Fig.~\ref{fig:CavityOptics}. While the entrance window was antireflection coated, the exit window behind the resonator was uncoated. The plane-parallel surfaces of the 5~mm thick window gave rise to an etalon effect, and thus the amount of light detected on the photodiodes was wavelength-dependent and therefore different from isotope to isotope since the free-spectral-range of the window-etalon was comparable
to the isotope shift between $^6$Li and $^{11}$Li. A direct
measurement of this effect was performed by comparing the amount of
light entering the vacuum chamber and leaving it. The etalon effect
was clearly observed and the plane-parallel window was replaced by a
wedged window. A second measurement after the replacement did not
show a detectable wavelength-dependent variation anymore.

Ac Stark shift coefficients that were measured after the replacement are plotted in the right part of Fig.~\ref{fig:AC-Stark_slope_coeff}. There is still a slow drift of the coefficients with time, which might
be caused by a slow degradation of the mirror reflectivity, but all
isotopes show equal coefficients within the uncertainty.
To estimate the remaining systematic uncertainty for the $^{11}$Li
measurements, the standard deviation of the slope coefficients for
the reference isotope within the periode of $^{11}$Li measurements, as
indicated in the figure, was determined ($\sigma=0.81$~kHz/mW) and
multiplied with the average Ti:sapphire laser power on the photodiode during all
measurements (30~mW). Since much less measurements were performed for $^{11}$Li than for the other isotopes, the ac Stark shift of the dye laser might not be fully included in the statistical uncertainty. Therefore, a systematic contribution was estimated by multiplying the slope of the ac Stark shift caused by the dye laser (Eq.~(\ref{eq:b_dye}), obtained from Fig.~\ref{fig:ac_stark_6li}) with the standard deviation of the dye laser power as measured on the photodiode during all $^{11}$Li measurements. This resulted in systematic uncertainties of 46~kHz for the shift induced by the intensity of the Ti:sapphire laser light and of 8~kHz by that of the dye laser. These values are listed in Table~\ref{tab:ErrorAnalysis}.

The observation of this interference effect is the reason why the $^9$Li isotope shift measurement at GSI was excluded in the final average: Insufficient statistics did not allow us to perform an ac Stark shift measurement for this isotope at GSI. Instead, a measurement as for $^{11}$Li at TRIUMF was carried out but with the plane-parallel viewport. Therefore the value might be systematically shifted. However, the reasonable agreement with the TRIUMF data within the uncertainty shows that the size of a possible systematic contribution was reasonably estimated in \cite{Ewald04}.

Recoil effects are not important for two-photon transitions, since the momenta of the two photons cancel and the velocity of the
atom stays constant during the excitation process. The differential second-order Doppler shift correction $\delta \nu_{\rm SOD}^{6,A}$ was already discussed and its uncertainty is included in the systematic uncertainty.

The influence of unresolved Zeeman splitting in the hyperfine structure has been theoretically estimated by calculating the effect of the 0.7~G stray magnetic field measured at the beamline. A large part of the induced shifts in the $2s$ and $3s$ levels cancels. The remaining part shifts the peak positions of the two hyperfine components by only about 10 kHz in opposite directions. Thus, the c.g. of the hyperfine structure is expected to stay constant. For safety, the total expected shift of $\approx 10$~kHz is included in the error budget for the isotope shift (Table~\ref{Table:isotope_shifts}) as well as for the hyperfine splitting (Table~\ref{Table:hfs_results}).

\begin{table*}
\caption{\label{tab:ErrorAnalysis} Summary of statistical and systematic uncertainties of the corrected $^{11}$Li - $^6$Li isotope shift $\delta\nu_{\rm IS}^{6,11}$. Systematic uncertainties arise from slope variations in the ac Stark shift induced by the Ti:sapphire or dye laser induced between $^{11}$Li and $^6$Li, unresolved Zeeman splitting caused by the stray magnetic fields (0.7~G), a lineshape asymmetry caused by a position-dependent ac Stark shift in the laser beam intensity profile as discussed in detail in \cite{Sanchez2009}, and from second-order Doppler shift. All systematic contributions are linearly added to $\Delta_{\rm syst}$. The total uncertainty is the geometrical sum of the statistical and the systematic contributions.}
  \begin{ruledtabular}
    \begin{tabular}{llrr}
      Effect                                            & Contribution \\
      \hline
      \multicolumn{2}{l}{Statistical uncertainty}         &         & \\
      & Standard error of the mean for 24 measurements    &         & 66~kHz \\
      \multicolumn{2}{l}{Systematic uncertainty}          &         & \\
      & ac Stark shift induced by the Ti:sapphire laser 				&  46~kHz & \\
      & ac Stark shift induced by the dye laser 					&   8~kHz & \\
      & Unresolved Zeeman splitting                       &  10~kHz & \\
      & Lineshape asymmetry                               &  10~kHz & \\
      & Second-order Doppler shift		    								&  11~kHz & \\
      & Total $\Delta_{\rm Syst}$                         &         & 85~kHz \\
      \hline
      \multicolumn{2}{l}{RMS Total}                       & & 109~kHz \\ 
    \end{tabular}
  \end{ruledtabular}
\end{table*}

Finally, the small asymmetry in the peak profile has to be considered. This was observed  much more pronounced in off-line measurements at GSI which were carried out very recently in order to determine the absolute frequency of the $2s \to 3s$ transition.
The asymmetry was attributed to the position-dependent ac Stark shift that the atoms experience when they cross the Gaussian laser beam profile \cite{Sanchez2009}. The strongest shift occurs in the beam center but all atoms that are excited and ionized in the wings of the laser beam are less ac Stark shifted and are therefore causing an asymmetry of the lineshape. This assumption was supported by lineshape simulations which are presented in detail in \cite{Sanchez2009}. There it was shown that the asymmetry leads to a small shift in the position and results in a small deviation of the ac Stark shift behavior from a straight line at very low power. This is crucial concerning the absolute transition frequency, but the isotope dependence was found to be very small ($\lesssim 10$~kHz). This value was added to the systematic uncertainty for all isotopes.

All effects contributing to the uncertainty of the $^{11}$Li - $^{6}$Li isotope shift are summarized in Table~\ref{tab:ErrorAnalysis}. Even though the systematic uncertainties are mutually independent and will
not necessarily cause a shift in the same direction, we have added them linearly to obtain the total systematic uncertainty which is then geometrically added to the statistical uncertainty. In total, a
relative uncertainty of the $^{6,11}$Li isotope shift of $3 \cdot 10^{-6}$ was obtained in this work.

For the other isotopes, power fluctuations of the dye laser are included in the statistical uncertainty given in
Table~\ref{Table:isotope_shifts}. The Zeeman splitting (10~kHz), the lineshape asymmetry (10~kHz) and the uncertainty in the second-order Doppler shift are correlated for all isotopes and have therefore been linearly added to the total systematic uncertainty. The geometric sum of the systematic and the uncorrelated statistical contribution is then taken as the final isotope shift as listed in Table~\ref{Table:isotope_shifts}.

\section{Changes in the mean square charge radius}

\begin{figure}[htb]
\includegraphics[width=\columnwidth, clip=]{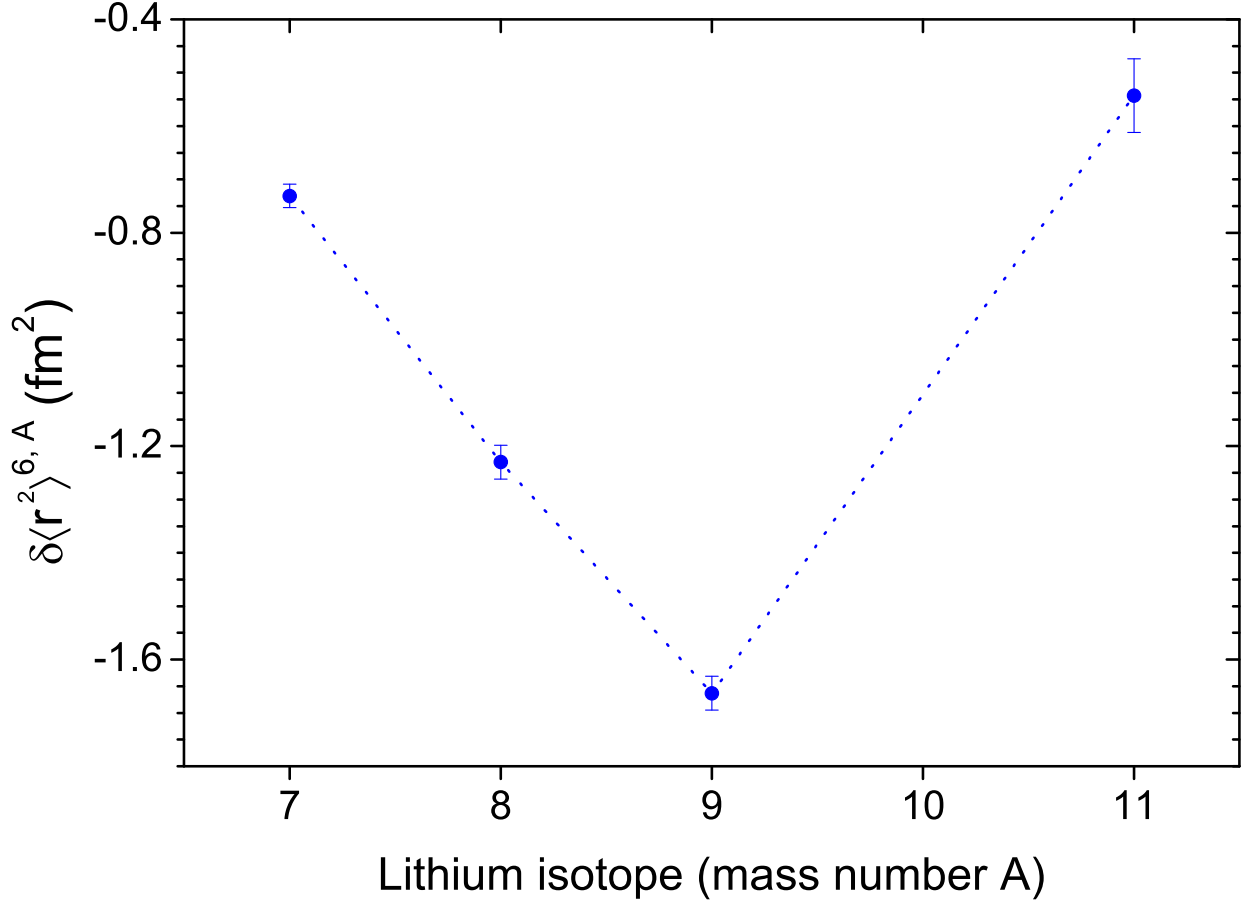}
\caption{\label{fig:msChargeRadiiChanges} (Color online) Changes $\delta\langle r^2\rangle^{6,A}$ in the mean square charge radius along the lithium isotope chain with respect to the mean square nuclear charge radius of $^6$Li. The values are obtained from the measured isotope shift (Table~\ref{Table:isotope_shifts}) and theoretical mass shift calculations (Table~\ref{Li_isotope}) using Eq.~(\ref{eq:delta_rc}).}
\end{figure}

The measured isotope shifts (Table~\ref{Table:isotope_shifts}) can now be combined with the mass shift
calculations discussed in Section~\ref{IS_theory} (Table~\ref{Li_isotope}) to extract by use of Eq.~(\ref{eq:delta_rc}) the change in the rms nuclear charge radii $\delta \left\langle r^{2}\right\rangle$ along the lithium chain of isotopes. Results are listed in Table~\ref{tab:FS-DR2} and plotted in Fig.~\ref{fig:msChargeRadiiChanges} relative to $^6$Li.
This table represents the summary of the complete analysis of the
experimental results and theoretical mass shift calculations. We
want to stress that slightly varying mass shift values along
previous publications on this topic
\cite{Sanchez06,Puchalski2006,Yan2008,Puchalski2008} are not caused
by a model-dependence of the mass shift calculations but several improvements: 
First, since the start of the experiments the accuracy of the atomic calculations could be improved by an order of magnitude; second, more accurate atomic masses became available by mass measurements at MISTRAL \cite{Bachelet2005, Bachelet2008} and TITAN \cite{Smith2008}; third, the influence of the nuclear polarizability on the transition frequency could be calculated for the first time \cite{Puchalski2006}. Hence, previous values should not be used anymore for calculating the nuclear charge radii. 

Figure \ref{fig:msChargeRadiiChanges} clearly depicts the trend in the nuclear charge radii: The size of
the proton distribution shrinks continuously from $^6$Li to $^9$Li.
As already stated in \cite{Sanchez06} this reduction is attributed
to the strong clusterization of the lithium nuclei. Clusterization
is a phenomenon that often appears in light nuclei; they tend
to behave as composed of well-defined subsystems, called
``clusters.'' Experimentally, this is reflected in the fact that $\alpha$-particles,
tritons ($t=\,^3$H), and helions ($^3$He) can be knocked out from a
number of nuclei with substantial probability. Often $\alpha$
clusterization is observed even in heavier stable nuclei along the $N=Z$
line, for example in $^{12}$C or $^{16}$O. According to nuclear
structure calculations, $^6$Li can be described as $\alpha+d$,
$^7$Li as $\alpha+t$, and the heavier isotopes as $\alpha+t+xn$ with
$x=1,2$ for $^{8,9}$Li, respectively. The consequence of
clusterization is a strong center-of-mass (cm) motion of the
clusters that leads to larger  proton distributions. This cm
motion is obviously reduced with increasing number of neutrons, hence, the nuclear charge radius decreases. This trend is reversed
when going from $^9$Li to $^{11}$Li. Here the charge radius
increases again and $^{11}$Li is similar in size to $^7$Li. There
are in general two effects that can be made responsible for this increase:
Either the $^{9}$Li core of $^{11}$Li is changed in its internal structure,
{\it i.e.} core polarization takes place, or the correlation of the
halo neutrons induces a cm motion of the $^9$Li core nucleus. 

Core polarization is related to the neutron-$^9$Li interaction and the internal structure of $^9$Li, whereas the center-of-mass motion gives insight into the neutron-neutron correlation in the three-body system. Pure three-body calculations assume the core to be inert (``frozen core" approximation) whereas more elaborated nuclear models try to take both effects into account. There is experimental evidence that the frozen core assumption is a very good approximation, however, both effects might contribute to the total change in the charge radius
between $^{9,11}$Li and it is impossible to disentangle their
contributions without consideration of further experimental data or theory.
More complete theoretical models were considered in \cite{Ewald04,Sanchez06} and will be discussed in more detail in a following publication \cite{Noertershaeuser2010}. 
\begin{table}
\caption{\label{tab:FS-DR2}Field shifts contributions $\delta\nu_{\rm FS}^{6,A}$ in the $2s\to 3s$ transition of the lithium isotopes (extracted from the measured isotope shifts in Table~\ref{Table:isotope_shifts} by subtracting the theoretically evaluated mass shift contributions listed in Table~\ref{Li_isotope}), field shift coefficients $C$ as defined by Eq.~(\ref{eq:c_coef}), and the change in the mean square nuclear charge radius $\delta\langle r^2\rangle^{6,\rm A}$ between the isotope $^A$Li and the reference isotope $^6$Li calculated using Eq.~(\ref{eq:delta_rc}).}
  \begin{ruledtabular}
    \begin{tabular}{rlll}
      $\rm ^A$Li & Field Shift (MHz) & $C$ (MHz/fm$^2$)& $\delta\langle r^2\rangle^{6,\rm A}$ (fm$^2$)\\
      \hline
      $^7$Li     & 1.149(34)         & -1.5719(16)     & -0.731(22)\\
      $^8$Li     & 1.933(51)         & -1.5719(16)     & -1.230(32)\\
      $^9$Li     & 2.615(50)         & -1.5720(16)     & -1.663(32)\\
      $^{11}$Li  & 0.852(108)        & -1.5703(16)     & -0.543(69)\\
    \end{tabular}
  \end{ruledtabular}
\end{table}

Comparison with nuclear models can best be performed if absolute
charge radii are calculated from the measured isotope shift. However, we need the charge radius of one reference isotope to fix the absolute charge radius. The value for the charge radius obtained by this procedure is model-dependent if the reference radius is obtained using a nuclear model. This is the case for the charge radii
from elastic electron scattering of lithium, because the model-independent approach gave rather large uncertainties \cite{Bumiller1972}. Since the value of $\delta \left\langle r^{2}\right\rangle ^{6,7}$ obtained in our isotope shift meaurement is slightly different from that obtained by elastic electron scattering \cite{deJager1974} (even though within uncertainties), 
the absolute charge radii differ if they are based either on the charge radius
reported for $^{6}$Li \cite{Bumiller1972,Li1971} or that reported
for $^{7}$Li \cite{Suelzle1967}. In previous publications, we have
used $r_{c}(\,^{7}\mathrm{Li)=2.39(3)}$~fm from \cite{Suelzle1967}
which is not model-independent and 
a recent analysis of the world scattering data revealed that the
charge radius of $^{6}$Li is better understood. Since this topic is of major importance
for the comparison with nuclear structure models and for conclusions
about the structure of the halo nucleus, it is discussed in more
detail in the following publication \cite{Noertershaeuser2010}.

\section{Summary}
This long writeup reports about the isotope shift measurement of
the complete lithium isotope chain from the stable isotopes up to
the exotic two-neutron halo nucleus $^{11}$Li. The theory of atomic
mass shift calculations is explained in detail and the most
accurate results for the mass shifts in the $2s\;^2{\rm S}_{1/2} \to 3s\;^2{\rm S}_{1/2}$
two-photon transition are presented. The experimental setup for
high-accuracy resonance ionization mass spectroscopy with cw lasers,
including a two-photon transition is described with all important
details to judge the quality and accuracy of the data obtained. All
sources of possible systematic uncertainties in these measurements
that are known to us are evaluated and discussed. Compared to previous publications, a small additional correction for the second-order Doppler shift is included in the isotope shift result. Combination of
the accurate measurements with the mass shift calculations allows
us to extract the changes in the nuclear charge radii along the
isotope chain spanning from the stable $^6$Li to the halo nucleus $^{11}$Li. 
Absolute nuclear charge radii will be extracted and compared with predictions by nuclear models in a following publication.

\begin{acknowledgments}
This work is supported by BMBF (contract No. 06TU203, 06TU263I, 06MZ215) and by the Helmholtz Association of German Research Centres (contract VH-NG 148). Support from the U.\ S.\ DOE Office of Science (B.\ A.\ B.\ ), NRC through TRIUMF, NSERC and SHARCnet (G.\ W.\ F.\ D.\ and Z.-C.\ Y.) is acknowledged. A.\ W.\  was supported by a Marie-Curie Fellowship of the European Community Programme IHP under contract number HPMT-CT-2000-00197. K.\ P.\ and M.\ P.\ acknowledge support by the NIST Precision Measurement Grants. We thank the target laboratory at GSI for providing the carbon foil catcher, Nikolaus Kurz, Mohammad Al-Turany, Christophor Kozhuharov (GSI) and the ISAC Computer Division at TRIUMF for support in data acquisition, Reinhard Kirchner, Haiming Wang, Frank Schmitt and Sascha Faber for contributions during the early part of this experiment, Melvin Good for help during installation of the experiment at TRIUMF, and Ren\'e Roy for providing a liquid scintillator. The role of Isao Tanihata for motivating and initiating these experiments is particulary acknowledged by the authors.
\end{acknowledgments}

\end{document}